%% file: article_rev1.tex
\documentclass[11pt]{article}

\usepackage{amsfonts}
\usepackage{amsbsy}
\usepackage{amsmath}
\usepackage{amssymb}
\usepackage{amsthm}
\usepackage[margin=1in]{geometry}
\usepackage[T1]{fontenc}
\usepackage{graphicx}
\usepackage{setspace}
\usepackage[font={footnotesize,sl}]{caption}
\usepackage{titlesec}
\usepackage{fancyhdr}
\usepackage{lastpage}
\usepackage{chngpage}
\usepackage{color, soul}
\usepackage[affil-it]{authblk}
\usepackage{float}
\usepackage{appendix}
\usepackage{fancyvrb}
\usepackage[x11names]{xcolor}
\usepackage{colortbl}
\usepackage{adjustbox}
\usepackage{booktabs,tabularx}
\usepackage[flushleft]{threeparttable}
\usepackage{threeparttablex}
\usepackage{subcaption}
\usepackage{rotating}
\usepackage{pdflscape}
\usepackage{afterpage}
\usepackage{multirow}
\usepackage{longtable}
\usepackage{footmisc}
\usepackage{accanthis}
\usepackage{cancel}
\usepackage{svg}

\usepackage{lmodern}

\usepackage{titlesec}

\DefineVerbatimEnvironment%
  {gams}{Verbatim}
  {formatcom=\color{blue},fontsize=\footnotesize,fontfamily=lmodern,xleftmargin=1cm}

\definecolor{blue}{rgb}{0,0,0.4}
\usepackage[citecolor=blue, linkcolor=blue, urlcolor=blue]{hyperref}
\hypersetup{colorlinks=true}

\usepackage{lineno}

\usepackage{natbib}
\usepackage{chapterbib}

\usepackage[version=4]{mhchem}  
\usepackage{siunitx}            

\newcommand{\DSCIM}{\texttt{DSCIM}}
\newcommand{\GIVE}{\texttt{GIVE}}
\newcommand{\MimiGIVE}{\texttt{MimiGIVE}}
\newcommand{\FrEDI}{\texttt{FrEDI}}
\newcommand{\FaIR}{\texttt{FaIR}}  

\begin{document}

\include{front_matter_rev1}

\include{main_matter_rev1}

\include{appendix_rev1}

\end{document}

%% file: front_matter_rev1.tex
\title{Economic Impacts of Climate Change in the United States: 
Integrating and Harmonizing Evidence from Recent Studies\thanks{The views expressed in this paper are those of the author(s) and do not necessarily represent those of the United States Environmental Protection Agency (EPA). No official Agency endorsement should be inferred. We thank Charles Griffiths and Michael Howerton for contributions to prior work. All errors are our own. Declarations of interest: none.}}

\author{\small{Elizabeth Kopits\textsuperscript{\textdagger}, Daniel Kraynak\textsuperscript{\textdagger}, Bryan Parthum\textsuperscript{\textdagger}, Lisa Rennels\textsuperscript{\textdaggerdbl}, David Smith\textsuperscript{\textdagger}, 
Elizabeth Spink\textsuperscript{\textdagger}, Joseph Perla\textsuperscript{\textdagger}, and Nshan Burns\textsuperscript{\textdagger\textdagger}
}}

\affil{
\small{\textdagger National Center for Environmental Economics, Environmental Protection Agency, USA

\textdaggerdbl Doerr School of Sustainability, Stanford University

\textdagger\textdagger Department of Agricultural and Resource Economics, University of California, Berkeley
}}

\date{\today}

\maketitle
\thispagestyle{plain}
\vspace{25pt}

\begin{abstract}
This paper synthesizes evidence on climate change impacts specific to U.S. populations. We develop an apples-to-apples comparison of econometric studies that empirically estimate the relationship between climate change and gross domestic product (GDP). We demonstrate that with harmonized probabilistic socioeconomic and climate inputs these papers project a narrower and lower range of 2100 GDP losses than what is reported across the published studies, yet the implied U.S.-specific social cost of greenhouse gases (SC-GHG) is still greater than the market-based damage estimates in current enumerative models. We also take a first step in integrating evidence on nonmarket damages with the GDP impacts. Our findings highlight the need for more research addressing the limitations of macroeconomic empirical studies and research on nonmarket and international spillover impacts. Further investigation of how macroeconomic and enumerative approaches can be integrated would enhance the usefulness of both strands of literature to climate policy analysis going forward.
  \\[10pt]
  \\[10pt]
\end{abstract}

\maketitle

\doublespacing

%% file: main_matter_rev1.tex
\defcitealias{ipcc2021climate}{IPCC, \citeyear{ipcc2021climate}}
\defcitealias{ipcc2021earth}{IPCC, \citeyear{ipcc2021earth}}
\defcitealias{ipcc2022policy}{IPCC, \citeyear{ipcc2022policy}}
\defcitealias{ipcc2023policy}{IPCC, \citeyear{ipcc2023policy}}
\defcitealias{national2017valuing}{NASEM, \citeyear{national2017valuing}}
\defcitealias{un2019world}{UN, \citeyear{un2019world}}
\defcitealias{census2020county}{U.S. Census, \citeyear{census2020county}}
\defcitealias{census2023national}{U.S. Census, \citeyear{census2023national}}
\defcitealias{census2024tiger}{U.S. Census, \citeyear{census2024tiger}}
\defcitealias{epa2011sab-vsl}{EPA, \citeyear{epa2011sab-vsl}}
\defcitealias{epa2023report}{EPA, \citeyear{epa2023report}}
\defcitealias{epa2024climate}{EPA, \citeyear{epa2024climate}}
\defcitealias{epa2024pattern}{EPA, \citeyear{epa2024pattern}}
\defcitealias{epa2024tdfredi}{EPA, \citeyear{epa2024tdfredi}}
\defcitealias{epa2024guidelines}{EPA, \citeyear{epa2024guidelines}}
\defcitealias{cil2023dscim}{CIL, \citeyear{cil2023dscim}}
\defcitealias{ciesin2016gridded}{CIESIN, \citeyear{ciesin2016gridded}}
\defcitealias{rff2024increasing}{RFF, \citeyear{rff2024increasing}}
\defcitealias{cbo2022long}{CBO, \citeyear{cbo2022long}} 
\defcitealias{cbo2023long}{CBO, \citeyear{cbo2023long}}
\defcitealias{USGCRP2023fifth}{USGCRP, \citeyear{USGCRP2023fifth}}


\section{Introduction}
\label{sec:introduction}

Informed development of policies to address the global externality of climate change requires a way to assess the economic consequences of changes in anthropogenic greenhouse gas (GHG) emissions---the primary driver of climate change \citepalias{ipcc2021climate, USGCRP2023fifth}. Understanding both the aggregate net impact to current and future generations and how these consequences will be distributed across regions and populations is helpful for designing climate mitigation and adaptation policies and for conducting benefit-cost analysis of a wide range of policies affecting GHG emissions.\footnote{See \citet{Kopits2025federal} for an overview of the use of social cost of greenhouse gas estimates in U.S. federal policy.} How the consequences of GHG emissions are likely to be experienced across particular regions and populations is especially challenging to estimate in the case of a global pollutant because it requires a mapping between physical effects and their ultimate economic incidence through both direct and indirect effects. For example, the impacts of GHG emissions not only have direct effects inside the borders of the United States (U.S.), but also indirect impacts from effects occurring outside of U.S. borders due to the interconnectedness of the global economy and populations (e.g., through supply chains, investments abroad, and national security). 

Despite the daunting task of modeling such a broad scope of scientific issues across a complex global landscape, researchers continue to make progress in estimating various economic impacts of GHG emissions. Models that take an enumerative or endpoint-specific approach to estimating market and nonmarket damages of GHG emissions have incorporated methodological advances in recent years, and researchers continue to look for opportunities to integrate additional impact categories as empirical studies become available. Another line of research has focused on econometrically estimating the effect of temperature and other climate variables on aggregate measures of economic outcomes, such as national or regional gross domestic product (GDP). The focus on macroeconomic indicators is often viewed as a way to estimate many market effects without the need to fully enumerate and estimate them. As both strands of the literature evolve, so too must the syntheses of their findings. In particular, it is necessary to develop direct comparisons of the results across macroeconomic econometric studies using consistent inputs and modeling frameworks, and to integrate the results of these studies with evidence from enumerative damage approaches where appropriate. 

This paper takes a step forward in synthesizing the existing evidence on the economic consequences of climate change specific to U.S. populations. We focus our efforts on developing an apples-to-apples comparison of the implications of a set of macroeconomic econometric studies. We develop a damage module based on the relationship between U.S. GDP and temperature change from each study’s empirical results and projection methods. We show the implications for both projected end-of-century U.S. GDP loss from climate change and for U.S.-specific measures of the marginal damages from GHG emissions, i.e., the social cost of carbon dioxide (\ce{CO2}), methane (\ce{CH4}), and nitrous oxide (\ce{N2O}), collectively known as the social cost of greenhouse gases (SC-GHG). We calculate U.S.-specific SC-GHG estimates under a harmonized set of U.S. socioeconomic and emissions projections, climate modeling (incorporating the latest advances in the representation of carbon feedback effects), and discounting methods, with explicit representation of key uncertainties in each of these inputs. 

Among the macroeconomic empirical studies, we find that significant progress has been made in estimating the dynamic effects of temperature change on economic growth. Studies are increasingly finding temperature changes have some persistent but not permanent effects on U.S. economic growth. Yet it is not necessary to deviate far from the assumption of permanent growth effects for the GDP impact to decrease dramatically. When evaluated using a consistent set of baseline temperature data and socioeconomic and climate inputs, which capture a range of uncertainties, the central specification of papers projecting some persistence in the effect of temperature on GDP yield end-of-century U.S. GDP loss projections ranging from 0.4 to 3.4 percent---a narrower and lower range than what is reported within the independently published studies. These results translate to U.S.-specific SC-\ce{CO2} estimates ranging from \$10 to \$65 per metric ton \ce{CO2} for 2030 emissions (under a 2 percent near-term Ramsey discount rate). The higher end of this range is derived from recent studies that use more flexible empirical methods that better allow for tracing out nonlinear dynamics in the temperature-GDP relationship and addressing serial correlation in temperature and GDP over time \citep[e.g.,][]{nath2024}. Finally, we discuss our findings in context of existing estimates of U.S. climate damages, including those based on existing enumerative models. The U.S.-specific SC-\ce{CO2} estimates from the macroeconomic models are larger than the sum of U.S. market damage estimates based on existing enumerative models. In addition, we illustrate how these GDP-based market damage estimates can be augmented with nonmarket impacts for a more comprehensive estimate of the full economic burden of climate change. Integrating evidence on two categories of nonmarket health damages (temperature- and wildfire smoke-related mortality) to the GDP-based market damage function from the macroeconomic studies increases the range of U.S.-specific SC-\ce{CO2} estimates to \$46-\$99 per metric ton of \ce{CO2} for 2030 emissions. 

There are still many categories of climate impacts omitted from this analysis. The U.S.-specific damages from some market impacts are not yet reflected, even in estimates based on macroeconomic econometric studies, as they can only account for net climate impacts on macroeconomic outcomes that have, to some extent, been experienced in the historical record and are identified by annual country-level average temperature shocks. And while these studies may capture some forms of long-run adaptation \citep{merel2021climate, merel2024sufficient}, they may not account for the full extent of long-run adaptation and the costs associated with it. Many categories of nonmarket impacts are also not yet accounted for, such as mortality and morbidity effects from many climate-mediated extreme weather events and various nonmarket impacts associated with the loss of ecosystem services, among others. Our findings highlight the need for more research on both market and nonmarket climate damages to U.S. populations, including through interactions and international spillover impacts. Further investigation of how the results of macroeconomic and enumerative approaches can be compared and combined would also enhance the usefulness of both strands of literature in policy analysis going forward.

\section{Review of Literature}
\label{sec:literature}

The economic consequences of global changes driven by GHG emissions are most often examined using integrated assessment models (IAMs) of climate change, which incorporate climate processes and economic systems into a single unified modeling framework. Climate change IAMs vary in their complexity, structure, and geographic resolution, but those used in SC-GHG estimation are generally comprised of four modules: socioeconomic, climate, damages, and discounting \citepalias{national2017valuing}. The socioeconomic module consists of jointly estimated projections of economic growth, population, and GHG emissions, which feed into the climate module to project future earth system conditions such as global temperatures, ocean acidity, and sea level rise. The damage module translates changes in these climatic conditions into physical and monetized estimates of economic damages. These economic damages represent the amount of money those experiencing the changes would be willing to pay to avoid them and can be experienced through impacts to goods and services traded in markets (e.g., changes in agricultural productivity, energy expenditures, or property damage) or nonmarket goods and services (e.g., changes in mortality and morbidity risks or ecosystem services) \citepalias{epa2023report}. Lastly, the discounting module translates the stream of undiscounted economic damages from the damage module into the present value of net damages. This four-step procedure is modeled with both baseline emissions projections and with a small additional amount (a pulse) of GHG emissions in a particular year. The SC-GHG is the per-ton difference in the present value of damages between the baseline and pulsed models from the perspective of the year of the emissions pulse.

\subsection{Modeling economic damages of climate change}
\label{sec:modeling_overview}

This paper focuses on the evidence that can inform the damage module of an SC-GHG IAM. Two methods that have produced estimates of the effect of climate change on U.S.-specific outcomes include an enumerative approach to damage function development and a more aggregated approach to econometrically estimating impacts on economy-wide outcomes.\footnote{Other methods include expert elicitation methods or other survey techniques \citep[e.g.,][]{pindyck2019social,hulshof2020willingness,moore2024synth}. Economy-wide computable general equilibrium (CGE) models are often used to explore how market-based climate impacts propagate through the economy \citep[e.g.,][]{dellink2019sectoral,takakura2019dependence} but have not generally been used on their own to develop SC-GHG estimates.} An enumerative damage module is often developed by calibrating to, or building up, disaggregated estimates of the damages resulting from various types of climate impacts. This approach to estimating damage functions typically involves spatially-explicit and sectoral- or category-specific modeling and aggregation of damages across sectors or impact categories. Three models that provide an enumerative approach to monetizing U.S.-specific climate damages are: the Greenhouse Gas Impact Value Estimator (\GIVE) \citep{rennert2022comprehensive}, the Data-driven Spatial Climate Impact Model (\DSCIM) \citepalias{cil2023dscim}, and the EPA's Framework for Evaluating Damages and Impacts (\FrEDI) \citepalias{epa2024tdfredi}.\footnote{The \GIVE{} damage module includes country-level representation of monetized damages from four climate impacts: changes in mean temperature-related mortality risk, agricultural yields for staple crops, energy related expenditures, and sea level rise induced mortality risk and physical capital loss in coastal areas \citep{rennert2022comprehensive,rennert2022social}. \DSCIM{} contains a subnational-scale, sectoral damage module that estimates net damages resulting from similar health, energy, agriculture, and coastal impact endpoints as considered in \GIVE{} \citetext{\citet{rode2021estimating,carleton2022valuing}; \citetalias{cil2023dscim}}. In addition, \DSCIM{} includes representation of net economic impacts from labor supply responses to changes in temperature, particularly in high-risk weather-exposed industries. The third model, \FrEDI{}, is a U.S. focused model that includes representation of additional impact endpoints across categories including human health, infrastructure, labor, electricity supply and demand, agriculture, and ecosystems and recreation \citepalias{epa2024tdfredi}. See Tables \ref{tab:impact_categories} and \ref{tab:impact_categories_fredi} in \ref{app:sec:add_fig_tab} and discussion in EPA \citeyearpar{epa2023report,epa2024pattern,epa2024tdfredi} for more information about each of these three models.} While these models reflect recent advances in the scientific literature on several key endpoints such as temperature-related mortality, energy, agriculture, and coastal impacts, there are many categories of climate change impacts and associated damages that are not yet or only partially represented in enumerative IAMs \citepalias{epa2023report}. Continued progress in filling data gaps and estimating the magnitude of many of these omitted impacts includes a growing body of research providing evidence on a wide range of U.S.-specific outcomes, such as heat-related kidney disease and other morbidity outcomes \citep[see, e.g.,][]{bell2024climate,yang2024degree}, human capital impacts \citep{park2021learning}, flooding impacts on mortality \citep{mueller2024sunny,lynch2025large} and drinking water  quality \citep{austin2025drinking-water}, forestry and ocean impacts \citep{baker2023projecting,bastien2026ocean-scc}, long-term impacts to coastal wetlands \citep{fant2022valua}, net impacts on outdoor recreation \citep{parthum2022winterrecreation,willwerth2023}, climate impacts on mental health \citep{obradovich2022} and violent crime \citep{harp2020crime}, and the distortionary effects of fiscal impacts \citep{barrage2023fiscal}, to name a few.  The majority of this research has not yet been converted into damage functions of the kind needed for developing U.S.-specific SC-GHG estimates.\footnote{Several exceptions include research on climate-driven wildfire-related health impacts \citep{qiu2026wildfire} and emerging research monetizing nonuse damages from biodiversity loss \citep{wingenroth2024accounting}. We explore adding both of these endpoints in the \GIVE{} model and discuss further in \ref{app:sec:td_wildfire}.}

Another line of research has focused on econometrically estimating the effect of temperature and other climate variables on more aggregate measures of economic outcomes, such as U.S. GDP (e.g., \citealp{dell2012tempera, burke2015global, acevedo2020effects, kalkuhl2020impact, kahn2021long, newell2021the, casey2023projecting, harding2025distribution, nath2024}). The focus on readily available macroeconomic measures can be appealing given the still limited scope of damage functions based on enumerative modeling and the resource intensive research needed to model each damage pathway. These studies offer a reduced-form approach to estimating the impact of changes in the climate on the combined set of market goods and services, thereby accounting for many sectors without the need to fully enumerate them. That is, measured climate-driven changes in GDP are thought to reflect the value of net impacts on goods and services traded in markets (e.g., changes in agricultural crop yields or energy use) that have an associated market price. Such measures do not reflect the value of nonmarket goods and services and thus cannot be used to provide a comprehensive accounting of net damages to U.S. citizens and residents from GHG emissions. The results of these studies have not been as widely used for recent damage module development in SC-GHG estimation. In fact, the National Academies \citeyearpar{national2017valuing} highlighted both the lack of traceability to damage pathways and the lack of accounting for nonmarket damages as reasons for recommending against this damage function approach for SC-GHG estimation. However, given the still limited scope of damage functions based on enumerative modeling and that GDP has been extensively studied as an indicator of the productivity of the economy as a whole, it is useful to conduct a closer review of the findings of this strand of the literature to inform the magnitude of market-based climate damages.

\subsection{Macroeconomic econometric studies of GDP impacts}
\label{sec:macro_damage_functions}

In our review of the recent macroeconomic econometric research on GDP impacts, we focus on studies whose results are in a form that can be combined with the other modules and components of an IAM modeling framework (e.g., country-level annual socioeconomics and mean temperature projections) to estimate U.S.-specific SC-GHGs.\footnote{Incorporating the results of studies that investigate the economic growth impacts of changes in other aspects of the temperature distribution, or weather variables measured at finer geographic and/or temporal scales, would require additional extensions to our modeling framework. This includes, for example, macroeconomic empirical studies that estimate the impact of temperature or precipitation on sectoral GDP \citep{conte2021local}, studies that use seasonal or daily temperature and precipitation data \citep{deryugina2017marginal,colacito2019temperature,kotz2021day}, and those that include measures of annual temperature variability or extremes \citep{kotz2021day,schwarz2022empirical,waidelich2024climate}. Similarly, additional modeling is needed to incorporate the results of papers investigating the effects of climate-driven changes in tropical cyclones and other natural hazards on economic growth \citep[e.g.,][]{hsiang2014causal,bakkensen2018climate}. Finally, we have also not included studies using time-series methods that make it more challenging to control for time-varying confounders, including a recent study by \citet{bilal2024} that estimates the economic effects of global temperature fluctuations and finds large statistically significant impacts on GDP.} Two prominent early papers using global panel datasets that serve as a foundation for subsequent literature looking to identify plausibly causal relationships between climate and economic growth (i.e., growth in GDP) are \citet{dell2012tempera} and \citet{burke2015global}. The source of econometric identification in these studies is fluctuations in annual mean temperature and precipitation, aggregated from gridded climate measurements to the country or region level to study its effects irrespective of other factors. \citet{dell2012tempera} estimate economic growth in a given year as a linear function of population-weighted temperature and precipitation including lagged weather variables and controlling for country-specific effects and time trends. They find a one-time \SI{1}{\celsius} increase in temperature to have statistically significant negative impacts for poor countries, but imprecise smaller impacts for rich countries.\footnote{In \citet{dell2012tempera}, a country is categorized as ``poor'' if it has below median purchasing power parity-adjusted GDP per capita in the initial year of the data.} Including multiple lags of temperature in a distributed lag (DL) model, the authors examine whether these growth impacts persist over time and find evidence of temperature changes affecting economic growth for at least 10 years. 

\citet{burke2015global} extend this framework by considering a nonlinear relationship between a country’s economic growth and temperature. They estimate economic growth as a quadratic function of temperature and find a statistically significant inverted U-shaped relationship with GDP peaking at approximately \SI{13.1}{\celsius}, with the response in rich countries not statistically different from that in poor countries. They find suggestive, but statistically ambiguous, evidence for permanent growth effects. Finally, \citet{burke2015global} develop a climate projection approach that applies their econometric results to estimate future GDP losses from climate change under a key assumption: that the long-run impact of changes in climate will be the same as the effect of short-run impacts from changes in weather. The resulting damages are estimated to be large globally, and substantial even for many high-income countries.  

\begin{figure}[ht]
    \begin{center}
        \caption{Conceptual plot of the response of gross domestic product (GDP) per capita to a temporary temperature increase}
        \label{fig:gdp_response}
        \includegraphics[width=1\linewidth]{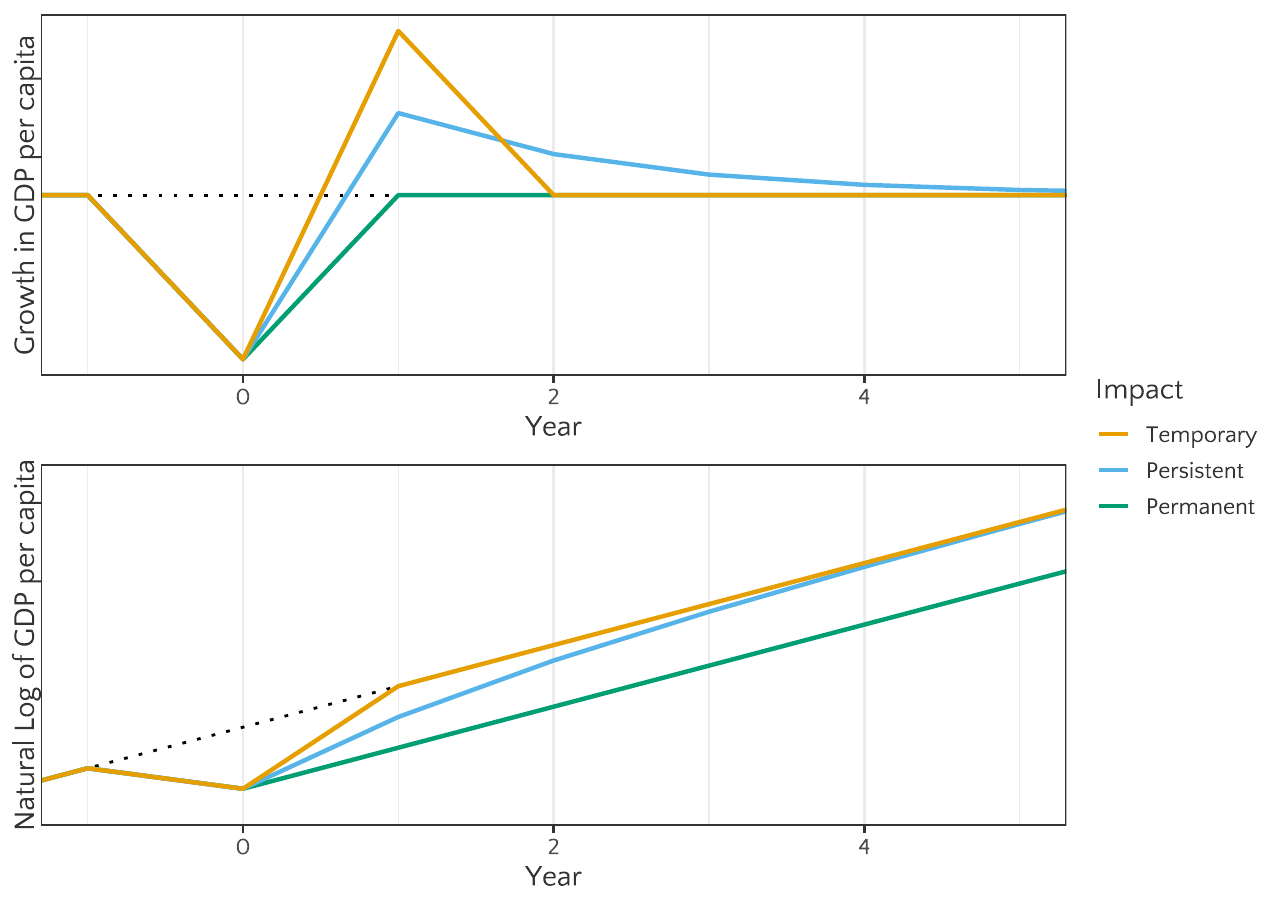}
    \end{center}
    \footnotesize
    Source: Adapted from \citet{nath2024}.
\end{figure}

Following the early findings of \citet{dell2012tempera} and \citet{burke2015global}, researchers began to further investigate the extent to which temperature changes have temporary, persistent, or permanent effects on economic growth. Understanding these dynamics has emerged as a key focus in this literature due to the important implications for estimating future damages from climate change. Figure \ref{fig:gdp_response} illustrates, for a temporary shock to temperature, what an impact on GDP growth implies for the impact on the level of GDP, and vice versa. If a temporary increase in temperature leads to a reduction in economic growth in the initial year of the change only and is then reversed (i.e., a temporary growth effect), it will still lead to a change in the level of GDP, but that level effect will dissipate quickly over time, holding all else equal. However, if a one time change in temperature causes a permanent impact on GDP growth, then the impact on the level of GDP remains over time. This is simply because there is no ``catching-up'' that occurs after the initial shock  (under a permanent increase in temperature, this case would lead to an indefinitely widening divergence between the level of GDP under climate change and without climate change.) An impact on the growth rate of GDP with some persistence but eventual convergence back to the original growth rate represents an intermediate case.\footnote{The temporary and permanent responses illustrated in Figure \ref{fig:gdp_response} are often called ``levels'' and ``growth'' effects, respectively, in macroeconomic econometric studies of the relationship between temperature and GDP. However, there is inconsistency in terminology used across the literature.}

Some studies have focused on examining the robustness of the \citet{burke2015global} results. For example, \citet{newell2021the} conduct a large cross-validation exercise in which they explore model uncertainty by estimating 800 plausible specifications of the relationship between temperature and GDP growth and test the out-of-sample performance of the models. They find that models estimating a relationship between temperature and economic growth exhibit significant model uncertainty leading to a wide range of forecasted climate impacts by the end of the century. They argue that their results do not support a statistically significant marginal effect of temperature on global GDP growth and emphasize that specifications estimating a relationship between temperature and GDP levels (i.e., a temporary effect on GDP growth) generally find a more robust and much narrower range of GDP losses by the end of the century. 

\citet{kalkuhl2020impact} also find evidence in support of temperature fluctuations only affecting the level of economic output using different sources of data. They measure economic activity using data from a variety of sources with increased resolution and detail of subnational regions, or gross regional product (GRP). In their preferred specification, a one-year distributed lag (DL) model with an interaction between the lagged level of temperature and the change in temperature, the authors find a strong nonlinear relationship between area-weighted annual temperature and GRP, but do not find the level of temperature to affect economic growth when the annual change in temperature is also included in the model. They interpret these findings to be consistent with permanent changes in temperature affecting the level of GDP but not the long-run growth rate of the economy. Aside from differences in the underlying data used, their projected damages are larger than those of \citet{newell2021the} in part because their results imply a lower optimal temperature, above which climate damages accrue.

Several recent papers have further advanced the examination of the extent of GDP effects using more flexible empirical methods and/or different measures of temperature variation that can better address common econometric challenges that arise in the analysis of temperature-GDP relationships. For example, \citet{acevedo2020effects} examine the effects of weather on a large set of outcome variables using the local projections (LP) method \citep{jorda2005} to trace the impulse response function of GDP per capita to a change in population-weighted temperature and precipitation. The LP method is more flexible than a DL model because it imposes fewer restrictions on the dynamic process by directly estimating the cumulative impact of a current shock in each future horizon period.\footnote{By estimating a collection of projections local to each forecast horizon, the authors can interpret the estimated effects in the 0-horizon regression as the contemporaneous impact and the estimated effects in subsequent regressions as reflecting the impact $h$ years after the shock. The LP method produces less biased (but higher variance) forecasts at intermediate and long horizons than vector autoregression methods because it does not constrain the shape of the impulse response functions and is thus less sensitive to misspecification \citep{li2024local}.} In their main specification, \citet{acevedo2020effects} estimate both the contemporaneous and medium term (7-year) change in per capita GDP as a quadratic function of temperature and precipitation, controlling for a one-year lag of the dependent and weather variables and country and time fixed effects. The inclusion of the lagged variables, importantly, helps to address serial correlation in temperature and GDP growth over time that could lead to biased estimates. They find statistically significant negative effects of temperature on per capita output and that these effects continue for at least 7 years. However, they do not interpret these findings as evidence of permanent growth effects because they are unable to reject that the contemporaneous and medium-term effects on output are identical.\footnote{Using the same methodology, \citet{acevedo2020effects} also estimate the impact of temperature on channels of impacts and find the largest impacts on crop production and the agricultural sector. They also find impacts on manufacturing but find no impacts on the services sector. The paper also estimates the impact of temperature increases on some of the basic determinants of GDP. The authors find that temperature reduces labor productivity in heat exposed industries but find no impact on labor productivity in non-heat exposed industries. Temperature also appears to have persistent impacts on investment. \citet{acevedo2020effects} further find that temperature increases led to persistent increases in infant mortality and decreases in the Human Development Index.} 

\citet{kahn2021long} use a panel autoregressive distributed lag (ARDL) model\footnote{An ARDL model is a type of a DL model that includes lags of the dependent variable (autoregressive) and lags of an explanatory variable (distributed lag).} with four lags. They attempt to address the econometric challenges with trended variables by focusing on a country’s deviation in temperature relative to a moving historical average, rather than levels or squares of temperature. The authors argue that this measure allows for a more explicit modeling of changes in the distribution of weather patterns and an implicit model of adaptation. The length of time over which their moving average is calculated is the assumed amount of time for economies to adapt to changes in climate (either 20, 30, or 40 years), and thus the authors view the GDP impacts of the deviation from this moving average as the relevant temperature shock after accounting for assumed adaptation. However, the authors examine the evolution of the temperature sensitivity coefficient and find that it becomes larger in absolute value over time. While adaptation is expected to diminish the effect of temperature on economic growth over time, the authors note that this analysis does not imply that adaptation has not occurred historically, but that the assumptions about adaptation in their paper may not reflect the current state of the world. The authors derive climate projections that incorporate the full long-run implications of their ARDL estimates, including assumed adaptation, under parametric assumptions on the distribution of future deviations of temperature from its historical norm. 

\citet{casey2023projecting} further advance the examination of the extent of GDP growth effects using macroeconomic growth theory and exploring the mechanisms through which climate impacts on GDP growth could occur. The authors focus on the effect of climate variables on total factor productivity (TFP) to help distinguish between temporary and persistent impacts of climate change on economic output. More precisely, they argue that even a one-time effect of a change in temperature on the level of GDP is likely to manifest over several time periods due to an endogenous capital response, while climate change effects on TFP ought to be more easily distinguishable as temporary or persistent. Under certain assumptions, growth theory implies that temporary effects on TFP lead to persistent but relatively short-lived effects on GDP, while persistent effects on TFP lead to longer-term growth impacts on GDP because TFP is the key driver of the long-run growth rate of output. Guided by this reasoning, the paper empirically investigates how temperature shocks affect TFP. In regressions allowing for potential impacts both on the level and growth rate of TFP, \citet{casey2023projecting} find temperature has a persistent effect on the level of TFP, but not its rate of growth. This result implies that the impact of a change in temperature on GDP growth may still persist for several years (through temperature impacts on TFP and TFP’s impact on investment and capital accumulation) but will not lead to permanent long-run growth impacts. Model-based climate change impacts on economic output are derived from reduced-form projections of country-level TFP using their empirically estimated coefficients. 

\citet{harding2025distribution} use a reduced-form approach to align their empirical analysis more closely with theoretical models and methods used in the broader empirical macroeconomic growth literature by accounting for growth convergence when estimating the relationship between temperature and country-level economic growth. Similar to \citet{casey2023projecting}, they argue that under neoclassical macroeconomic growth theory, the only way for climate change to have permanent effects on economic growth is if it permanently affects the determinants of long-run economic growth, such as the rate of innovation. By contrast, non-permanent climate-induced changes in productivity can only have a temporary effect on economic output because the economy will eventually converge back to the steady-state long-run growth rate. They empirically estimate this speed of convergence by regressing GDP growth on one lag of GDP in addition to the weather variables, including the same controls for time trends and country fixed effects as in \citet{burke2015global} and \citet{newell2021the}. In their central specification, the authors find significant support for short-run effects on economic growth, but no statistically significant evidence of permanent long-term growth impacts. The authors emphasize that while including the convergence term has little impact on the estimated effects of weather variables on GDP growth, accounting for the convergence effect is important when projecting long-run damages from climate change; otherwise, any estimated effects of climate on growth, by construction, will be permanent. 

\citet{nath2024} also take models of economic growth as a starting point and argue that temperature can have persistent but not permanent growth effects because fundamental drivers of growth, specifically technological change, link countries’ growth rates together. They present a range of evidence to support that international technology spillovers prevent countries from differing entirely in growth as global temperatures change. They then use country-level panel data to empirically estimate the dynamic effects of temperature on GDP. Like other recent macroeconomic econometric papers discussed above, \citet{nath2024} control for lagged GDP growth, but they also address econometric difficulties in estimating a dynamic causal relationship between GDP and temperature in additional ways. First, they emphasize that because temperature itself is serially correlated, it is important to include lags of both temperature and GDP. Second, they examine the impact of a temperature shock\footnote{\citet{nath2024} define the shock to temperature as the innovation in a nonlinear auto-regressive model of country temperature, a common practice in analysis of macroeconomic data which isolates current shocks from their relationship to past shocks in the presence of serial correlation. Specifically, they construct the temperature shock as the residual from a regression of temperature on lagged values of temperature and lagged temperature interacted with mean temperature.}, rather than temperature itself, and explore an alternative specification to capture nonlinearities in the way that the temperature shock affects GDP, relative to the quadratic terms typical in many previous studies. Finding that a state-dependent model outperforms a quadratic function of temperature variables, in their main specification, \citet{nath2024} interact the temperature shock variable with mean baseline temperature to investigate whether a shock to temperature has different effects on GDP depending on the country’s mean historical temperature. Finally, the authors estimate a flexible impulse response function of GDP to temperature shocks using the LP approach, similar to \citet{acevedo2020effects}, while putting additional emphasis on the importance of accounting for the persistence in temperature when using their empirical results to project the effects of future increases in temperature on GDP. For instance, for a moderate temperature country (\SI{15}{\celsius}), the paper finds persistent effects of temperature shocks on GDP. The persistence in the effect of temperature shocks on GDP is partly driven by persistence in temperature \citep[see][Figure 6b]{nath2024}. For such a moderate temperature country, \citet{nath2024} find 9 percent of a temperature shock persists 9 years later. \citet{nath2024} account for this persistence in temperature by using a cumulative response ratio of GDP to temperature, defined as the ratio of the cumulative response of GDP to the cumulative response of temperature to a temperature shock, which accounts for the dynamic impact of the initial shock and the continuing impacts driven by its persistence. 

Table \ref{tab:summary_macroecon} presents a summary of the empirical methods used in the main specification of the econometric papers discussed above. While each of these papers provides results of many alternative specifications and robustness checks, Table \ref{tab:summary_macroecon} focuses on the authors’ stated ``preferred'', ``main'', or ``central'' specification, or the specification that is used for climate damage projections in the paper if no preferred specification is stated. The last column of Table \ref{tab:summary_macroecon} qualitatively classifies what each paper projects about the impact of temperature on U.S. GDP growth. The papers employing DL models to estimate long-run impacts project either permanent or temporary impacts on GDP growth. By contrast, the papers using more flexible estimation approaches and incorporating lagged dependent variables project some amount of persistence in the impact on GDP growth. 

\begin{landscape}
    \begin{table}[ht]
        \caption{Summary of methods in recent macroeconomic econometric studies}
        \label{tab:summary_macroecon}
        \centering
        \begin{threeparttable}
            \footnotesize
                \begin{tabular}{lcccccc}
                    \toprule \toprule
                    \textbf{Study} & \textbf{Econometric} & \textbf{Data} & \textbf{Weather} & \textbf{Weather} & \textbf{Lags} & \textbf{Projected Temperature} \\ 
                    \textbf{(Empirical Specification)} & \textbf{Model\tnote{$\alpha$}} &     & \textbf{Variables} & \textbf{Variables} & \textbf{Included} & \textbf{Impact on} \\ 
                                   &    &    &    & \textbf{Specification} &     & \textbf{GDP Level} \\ 
                    \midrule
\citet{dell2012tempera}  & DL         & Country & Level & Linear  & Weather & N/A    \\
(Table 3, col 4)         & (10 lags)  &         &       &         &         &        \\
\midrule
\citet{burke2015global}    & DL       & Country & Level & Non-linear  & None & Permanent \\
(Ext. Data Table 1, col 1, & (0 lags) &         &       & (quadratic) &      &           \\
``base''/``main'')           &          &         &       &             &      &           \\
\midrule
\citet{kalkuhl2020impact} & DL      & Sub-national & Level, first & Non-linear           & Weather & Mix of \\
(Table 4, col 5,          & (1 lag) &              & difference   & interactions)        &         & Temporary and Permanent \\
``preferred'')            &         &              &              & (level \& difference &         &           \\
                          &         &              &              & interactions)        &         &           \\
\midrule
\citet{newell2021the}                   & DL       & Country & First      & Non-linear  & None & Temporary \\
(Levels version of                      & (0 lags) &         & difference & (quadratic) &      &           \\
\citet{burke2015global})\tnote{$\beta$} &          &         &            &             &      &           \\
\midrule
\cite{acevedo2020effects} & LP           & Country & Level & Non-linear  & Weather,   & Temporary \\
(Table 1, col 5,           & (0 horizons) &         &       & (quadratic) & GDP growth &           \\
``baseline''/ ``main'')     &              &         &       &             &            &           \\
\midrule
\citet{kahn2021long}     & ARDL     & Country & Absolute deviation & Linear & Weather,   & Persistent \\
(Table 2, Spec 2,        & (4 lags) &         & from historical    &        & GDP growth &            \\
m=30(b), ``preferred'') &          &         & moving average     &        &            &            \\
\midrule
\citet{casey2023projecting} & AR (1 lag) of & Country & First      & Non-linear  & TFP growth & Persistent \\
(Table 1, col 2)            & TFP growth    &         & difference & (quadratic) &            &            \\
\midrule
\citet{harding2025distribution}  & Convergence              & Country & Level & Non-linear  & Natural Log & Persistent \\
(Table 1, col 6,              & equation\tnote{$\delta$} &         &       & (quadratic) & of GDP      &            \\
``central'')               &                          &         &       &             &             &            \\
\midrule
\citet{nath2024}    & LP           & Country & Shock and & Non-linear      & Weather, weather      & Persistent \\
(Full dynamics with & (9 horizons) &         & historical mean\tnote{$\eta$}         & (shock and mean & interacted with their &            \\
FE, Figure 6c)     &              &         &                         &  interactions)  & mean, GDP growth      &            \\
\bottomrule
                \end{tabular}
            \begin{tablenotes} 
                \footnotesize
                \item This table summarizes each study's stated ``preferred'', ``main'', or ``central'' specification, or the specification that is used for climate damage projections in the paper if no preferred specification is stated.
                \item[$\alpha$] DL - Distributed Lags, AR - Auto-regressive, LP - Local Projection. 
                \item[$\beta$] Levels only specification analogous to main \citet{burke2015global} specification; See Table 2, GDP Level effect estimation with year fixed effects and 2-degree time trend, in \citet{newell2021the}. 
                \item[$\delta$] The convergence equation \citep{acemoglu2008introduction} is similar to an AR model but controls for lagged GDP rather than lagged GDP growth. 
                \item[$\eta$] The temperature shock variable is the residual in an equation of temperature as a function of lags of temperature and lagged temperature interacted with mean temperature.
            \end{tablenotes}
        \end{threeparttable}
    \end{table}
\end{landscape}

\subsection{Comparing the estimated GDP-temperature response function across studies}
\label{sec:compare_response_functions}

One way to show how the macroeconomic econometric studies compare in their modeling of the impact of temperature changes on economic growth is using a generalized single empirical framework that accommodates the frameworks used in each paper. Equation \ref{eq:general_main_matter} illustrates how the main specification from each paper listed in Table \ref{tab:summary_macroecon} can be written in, or related to, the ARDL form, where studies differ in how they transform the temperature variable, how they account for state dependence or nonlinearity, and how they treat persistence. Specifically, the change in log GDP per capita over time can be written as:

\begin{equation}\label{eq:general_main_matter}
    \Delta y_{i,t} =
        -\sum_{l=1}^{L} \lambda_{l}\, y_{i,t-l}
        + \sum_{l=0}^{M} \Big[ \beta_{1l}\, \tau_{i,t-l}
        + \beta_{2l}\, \tau_{i,t-l}\, S_{i,t-l} \Big]
        + \Omega_{it}
        + \epsilon_{i,t},
\end{equation}

\noindent where $y$ is log GDP per capita, $\tau$ is the temperature term, $i$ indexes the panel unit and $t$ the time period, $L$ determines the autoregressive order, $M$ determines the distributed-lag order of the climate terms, and $\lambda_{l}, \beta_{1l},\beta_{2l}$ are their coefficients. The climate regressors are built from two objects, $\tau$ and a heterogeneity term, $S$. The term $\Omega_{it}$ represents a sum of controls that typically includes at least country and year fixed-effects. 
\newcommand{\That}{\hat{T}} 
\newcommand{\Tbar}{\bar{T}} 
The measured temperature in region $i$ in time period $t$, $T_{it}$, enters as part of $\tau_{i,t-l} = T_{i,t-l} -\That_{i,t-l}$. Depending on how the benchmark $\That$ is specified, temperature can enter in levels ($\That = 0$, as in \citet{burke2015global}, \citet{harding2025distribution}, and \citet{acevedo2020effects}), first differences ($\That_{i,t-l} = T_{i,t-l-1}$, as in \citet{newell2021the} and \citet{kalkuhl2020impact}), a moving-average norm ($\That_{i,t-l} =\Tbar_{i,t-l}$, as in \citet{kahn2021long}), or a temperature shock (where $\That$ is a fitted value from a filtering regression, as in \citet{nath2024}). The heterogeneity term $S$ introduces state dependence or nonlinearity. For example, setting $S = T$ gives a quadratic effect, whereas setting $S =\Tbar_i$, a fixed baseline such as a country’s mean historical climate, gives a state-dependent effect that scales with location. A group indicator gives group heterogeneity. Setting $S = 0$ gives an effect that is linear in $\tau$. \ref{app:sec:common_framework} shows in more detail how the preferred specification of each study relates to the common model (1) via the choices of $\That$ and $S$, the lag and autoregressive order parameters $L,M$, and other related modeling choices.  

Using this common framework, we compare studies through the dynamic multiplier, $\Psi_h$, the response of the level of the outcome $h$ periods ahead to a temperature shock in year $t$.
\begin{equation}\label{eq:psi}
    \Psi_h \;=\; \frac{\partial\, y_{i,t+h}}{\partial\, T_{i,t}}
\end{equation}

\noindent Because $\tau$ is itself a function of $T$, we define the multiplier with respect to the underlying temperature and differentiate through $\tau$. In the multipliers, dynamic terms may arise from the lags of temperature and autoregressive structure. 

\begin{figure}[ht]
    \begin{center}
        \caption{Response of gross domestic product (GDP) per capita to a temporary marginal temperature (\SI{}{\celsius}) increase at U.S. average temperature in 2100 (\SI{16.3}{\celsius})}
        \label{fig:gdp_response_temp}
        \includegraphics[width=1.0\linewidth]{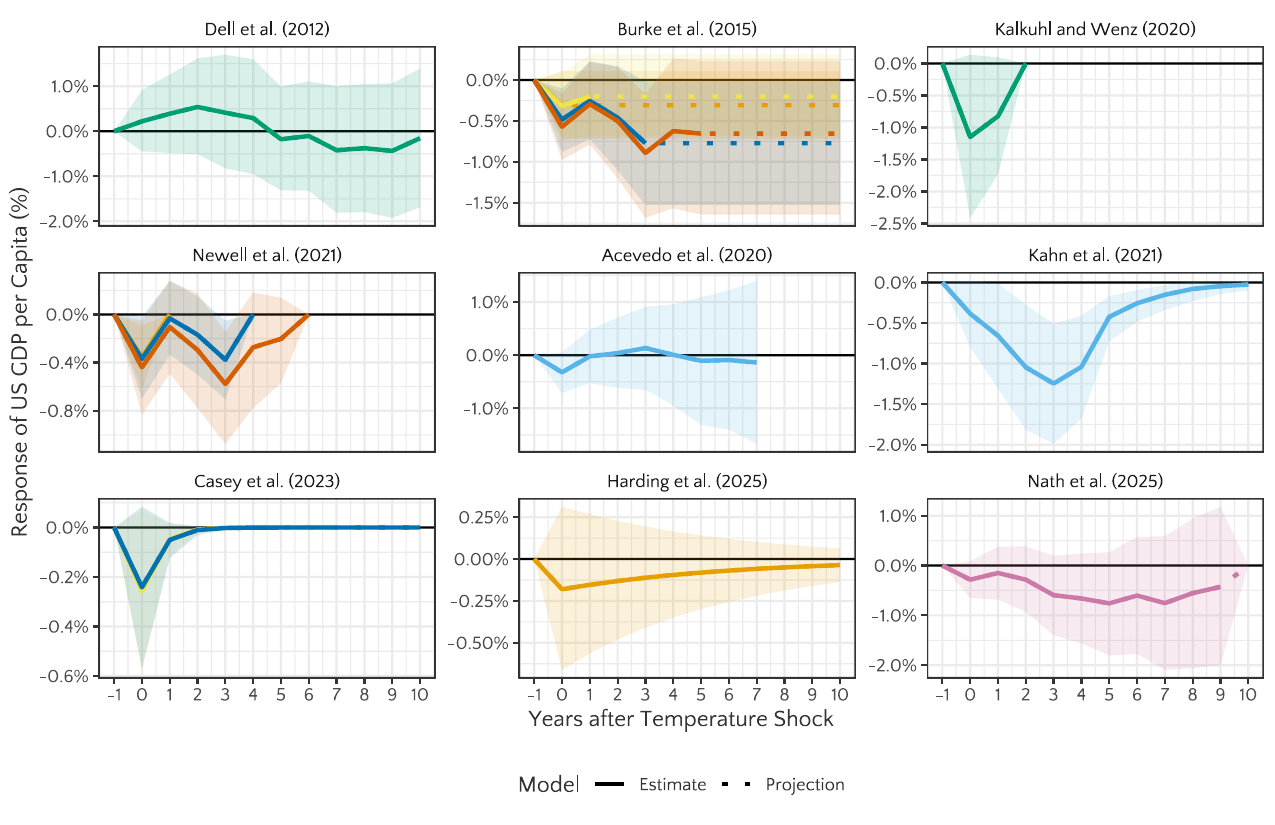}
    \end{center}
    \footnotesize
    This figure shows the response from the main specification for each paper (as described in Table \ref{tab:summary_macroecon}) as well as some additional lags and horizons explored in the case of \citet{burke2015global,acevedo2020effects,newell2021the}. Specifically, the \citet{burke2015global} panel shows the estimated response under the 0, 1, 3, and 5 lag specifications from the published paper; the \citet{,newell2021the} panel shows the 0, 3, and 5 lag specifications; and the \citet{acevedo2020effects} panel shows the 0 and 7 horizon specifications. For \citet{casey2023projecting}, this figure shows the response of both total factor productivity (Table 1, col 2) and GDP per capita (Table 2, col 4). These responses are similar enough that they overlap in the figure. The dotted line shows the implicit assumption about the response beyond the estimate lags or horizons based on the projections in the papers. The shaded ribbon represents the econometric uncertainty (95\% confidence interval) from the corresponding specification in the paper.
\end{figure}

The estimated dynamic multiplier from each study is provided in \ref{app:sec:common_framework} and  illustrated in Figure \ref{fig:gdp_response_temp} which traces out the estimated GDP-temperature response function over time based on the econometric results of the main specification of each paper discussed above, evaluated at a projected U.S. average temperature for 2100 (\SI{16.3}{\celsius}).\footnote{This 2100 U.S. temperature projection is consistent with the U.S. temperature projections presented in Figure \ref{fig:usmst_wo_feedbacks} and described in \ref{app:sec:td_feedback}. That is, it is based on the average of 10,000 Monte Carlo simulations of \FaIR{} 1.6.2, inclusive of two additional carbon feedbacks based on \citet{dietz2021economic}, using the RFF-SP emission projections as inputs and \FaIR{} 1.6.2 uncertainty to recover the increase of \SI{2.67}{\celsius} in U.S. temperature relative to 1980-2010 average and a 1980-2010 baseline U.S. temperature level of \SI{13.62}{\celsius} \citep{burke2015global}.} Specifically, the figure displays the GDP response to a temporary marginal shock in temperature ($^\circ\mathrm{C}$)  which is reversed in the following year. Figure \ref{fig:gdp_response_temp} shows how the temperature shock affects GDP contemporaneously and each year following the shock. For papers using DL methods and the temperature level as regressors \citep{dell2012tempera,burke2015global}, the cumulative GDP response is the sum of the marginal effects of the lags of temperature, up to the longest lag. For papers using DL methods and temperature change as regressors \citep{newell2021the}, the cumulative GDP response is simply equal to the marginal effects of the lags, while \cite{kalkuhl2020impact} is a hybrid between the two.\footnote{Papers that estimate the marginal effect of the change in temperature on the change in the natural log of GDP are estimating a first-differenced equation of the marginal effect of temperature on the natural log of GDP. Therefore, the parameters can be directly used to estimate the response of GDP.} For the papers using LP methods \citep{acevedo2020effects,nath2024}, the GDP response is the marginal effects of the lag of the temperature variables for each regression. For papers using ARDL or growth convergence methods, the cumulative GDP response is the sum of the marginal effects estimated using the dynamic multipliers (temperature level in \citet{harding2025distribution}) and the marginal effects (change in temperature in \citet{kahn2021long}), computed by using the ARDL or growth convergence coefficients and repeated substitution of the lags of GDP \citep{harding2025distribution} or GDP growth \citep{kahn2021long}. Figure \ref{fig:gdp_response_temp} displays each study's response function over a 10 year time horizon without making any additional changes. That is, no further adjustments were made to account for the difference between an impulse response function estimated using local projections and serially correlated data versus a response function estimated using a distributed lag model. The figure also does not reflect how the temperature variable specification in \citet{kahn2021long}, which implicitly incorporates an assumed 30-year adaptation period, dampens the impacts of a permanent temperature shock over time. This dampening is better illustrated in Figure \ref{fig:impulse_perm} in \ref{app:sec:add_fig_tab} which shows how the response function of each study evolves over time following a permanent marginal temperature shock.

Because the econometric specifications vary along several dimensions, the dynamic multipliers are comparable in a way that raw regression coefficients from each paper are not. One additional margin of difference that can be compared across the papers is the underlying data used for estimation. To understand the importance of the specific GDP, temperature, and precipitation datasets for the estimates, we re-estimated the main specification of each paper using the underlying dataset from each of the other papers. We present the results of this exercise in Figure \ref{fig:gdp_response_temp2} using a format analogous to Figure \ref{fig:gdp_response_temp}. Each panel in the figure displays the estimated impulse response function for the main specification of the labeled paper when using the underlying dataset from each of the other papers. Encouragingly, re-estimating each specification on different datasets used in the literature yielded similarly shaped response functions that were generally well-covered by the confidence intervals estimated in the original papers. This may be unsurprising given that there is substantial overlap in the underlying data sources used (indeed several papers simply build on the original \cite{burke2015global} dataset). 

To further investigate the robustness of results to the underlying data used in estimation, we used the dataset from from \cite{nath2024}, a newer dataset with broad data coverage relative to the other papers, to estimate the response functions of all of the main specifications. Figure \ref{fig:gdp_response_temp_nath} presents the result estimated impulse responses, which suggest large impacts estimated in \cite{nath2024} and \cite{kahn2021long}. In spite of this apparent result, the results of the rest of our modeling indicate that even these harmonized estimates do not tell the full story of how each paper projects climate damages. We rely on the main specification of each paper as estimated on the paper's own dataset going forward, but to carry forward the idea of sensitivity to estimation data, we supplement our central estimates of the U.S.-specific SC-\ce{CO2} with estimates where the central specification from each paper is generated using only the dataset from \citet{nath2024}.

\section{Methods}
\label{sec:methods}

While nearly all the macroeconomic econometric papers reviewed above\footnote{While \citet{dell2012tempera} developed an early framework for examining the relationship between changes temperature and precipitation and changes in a country’s economic performance, they do not carry their framework through to project future damages, stating: ``\emph{Given uncertainty over adaptation, international spillovers, technical change, and other issues, the estimates here—driven primarily from short-run fluctuations in temperature—alone cannot provide precise predictions about the estimated impacts of future climate change.}'' Therefore, we also do not extend their framework into a damage function.} present some projections of U.S. GDP impacts out to 2100 based on their estimated GDP-temperature relationships, our review found that the studies vary in the underlying temperature data used, assumed socioeconomic and climate scenarios, and scope of their projections. Without a consistent framework to compare these studies, inferences comparing the magnitude of projected GDP losses across these papers can be misleading. The papers use different aggregation methods and weighting schemes to recover country-level temperatures and have different assumptions on the shape of the path of warming. For example, most studies apply some form of population weighting to their temperature data using a single year of gridded population data and different baseline years, while \citet{kalkuhl2020impact} apply area weighting at the U.S. state level (or subregion). Similarly, the authors differ in the ways they select the baseline temperature from which to develop their temperature projections under various climate change scenarios going forward. Several papers follow the approach of \citet{burke2015global} and use the mean country-level temperature observed over 1980-2010, while others use temperature from a single recent year as a starting point (2005, 2014, and 2010 in case of \citet{acevedo2020effects}, \citet{kahn2021long} and \citet{casey2023projecting}, respectively), or the mean temperature over a shorter, more recent time period (2015-2019) in the case of \citet{kalkuhl2020impact}. Moreover, when projecting future climate impacts, even under the same emissions scenario, the papers use a variety of emissions-consistent temperature projections based on different global climate models. For example, while the papers provide projections under similar climate change scenarios (including with an RCP8.5 radiative forcing scenario), we found that authors’ temperature projections consistent with RCP8.5 are drawn from a variety of data sources, projections are made with different base years or measurements of baseline temperatures, or with different treatment of precipitation effects. These and other differences in data inputs (see Tables \ref{tab:macro_summary_temp}, \ref{tab:macro_summary_precip},  and \ref{tab:macro_summary_socio} in \ref{app:sec:td_macro} for a complete listing) make an apples-to-apples comparison difficult based on the projections presented in each paper (or based on the U.S.-specific results obtained from each paper’s replication code).    

We synthesize and compare the evidence on U.S.-specific climate damages across the macroeconomic studies by placing them within a consistent modeling framework capable of providing U.S.-specific SC-GHG estimates and show what each study's empirical results imply for both end-of-century U.S. GDP losses and for U.S.-specific SC-GHG estimates under the same harmonized U.S. socioeconomic and emissions inputs, climate modeling, and discounting methods. We conduct our analysis using the \MimiGIVE{} modeling platform that can easily accommodate modifications to the various components, or modules, of IAM-based SC-GHG estimation: socioeconomics, climate, damages, and discounting.\footnote{\MimiGIVE{} \citep{rennert2022comprehensive} is written using the Julia programming language and \GIVE{} is built on the \texttt{Mimi.jl} platform, an open-source package for constructing modular integrated assessment models (\url{www.mimiframework.org} \citep{anthoff2024mimi})}

We construct the damage module to be consistent with the econometric results of the main specification for each paper based on the underlying temperature and GDP dataset the authors used in their estimation. The approach to implement the damage projections was taken directly from each paper and did not rely on the response functions presented for comparison above. Each paper had a unique methodology for damage projections that aligned with the econometric specification and assumptions on the implications for long-term projections. 

The socioeconomic and climate inputs needed for estimating U.S.-specific SC-GHG are based on \citet{rennert2022social} and the methodology used in \citet{rennert2022comprehensive} and EPA \citeyearpar{epa2023report}. We rely on U.S. population and economic growth trajectories that were taken from an internally consistent set of probabilistic projections of population, GDP, and GHG emissions (\ce{CO2}, \ce{CH4}, and \ce{N2O}) (hereafter collectively referred to as the RFF-SPs) developed under the Resources for the Future Social Cost of Carbon Initiative \citep{rennert2022comprehensive}. The RFF-SPs were developed using a mix of statistical and expert elicitation techniques, as recommended by the National Academies \citeyearpar{national2017valuing} See \ref{app:sec:td_discounting} for additional details on the RFF-SPs.

Global mean surface temperature (GMST) is estimated based on the RFF-SP probabilistic global emissions projections used as inputs to the Finite amplitude Impulse Response (\FaIR) climate model \citep{millar2017a,smith2018fair}, a widely used peer-reviewed Earth system model recommended by the National Academies \citeyearpar{national2017valuing} for SC-GHG estimation. For damage functions that require U.S. temperature inputs to calculate U.S.-specific impacts, the GMST projections from \FaIR{} 1.6.2 are downscaled to U.S. temperature using a pattern-scaling approach based on continental U.S. data from global climate models. The temperature projections also reflect updates to partially incorporate the impact of Earth system feedback effects that are not accounted for in the 1.6.2 version of the \FaIR{} climate model \textemdash permafrost thaw and the dieback of the Amazon rainforest \textemdash based on the peer-reviewed \citet{dietz2021economic} modeling of these feedbacks. See \ref{app:sec:td_feedback} for more details the on earth system feedbacks considered in this study.

The projected future streams of economic damages are discounted following the methodology used in EPA \citeyearpar{epa2023report}, using the stochastic discounting approach from \citet{newell2022a}. This approach is based on the Ramsey formula \citep{ramsey1982a} and therefore provides consistency with changes in economic growth and allows the risk-adjusted discount rate to vary over time. The associated risk-free discount rates are dynamic and calibrated to empirical evidence of the interest rate term structure. This calibration uses a near term risk-free discount rate of 2 percent based on multiple lines of evidence on observed real market interest rates \citetext{\citealp{giglio2014very,drupp2018discount,pindyck2019social,bauer2020interest,howard2020wisdom,giglio2021}; \citetalias{cbo2022long}; \citealp{bauer2023rising}; \citetalias{cbo2023long}}. The stochastic discounting approach provides internal consistency within the modeling and a more complete accounting of uncertainty. This approach is consistent with economic theory \citep{arrow2013determining,cropper2014declining} and the National Academies’ \citeyearpar{national2017valuing} recommendation to employ a more structural approach to discounting that explicitly recognizes the relationship between economic growth and discounting uncertainty. See \ref{app:sec:td_discounting} for more discussion of each of the socioeconomics and climate inputs and discounting methods.

\section{Results}
\label{sec:results}

\begin{figure}[htbp]
    \begin{center}
        \caption{Climate damages as a fraction of U.S. GDP in 2100 due to a change in annual global mean surface temperature under each of the 8 macroeconomic econometric damage functions}
        \label{fig:macro_climate_damages}
        \includegraphics[width=1\linewidth]{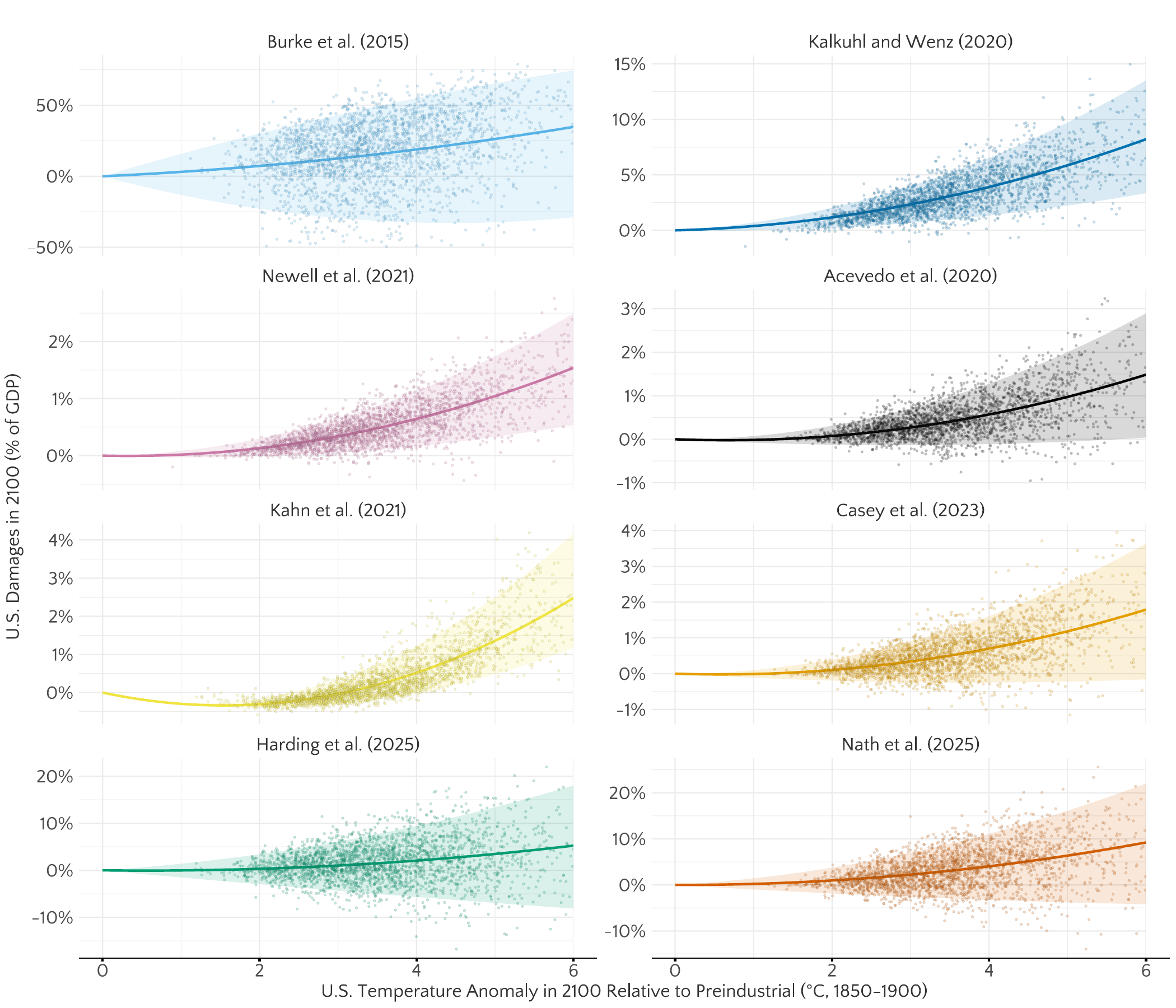}
    \end{center}
    \footnotesize
    GDP loss functions are generated using estimated damages in 2100 (points) and regressing on temperature and temperature squared at the mean (solid line) and quantile regressions at the 5\textsuperscript{th} and 95\textsuperscript{th} percentiles (shaded ribbons). 2,500 of the 10,000 points for each module are randomly selected to simplify the presentation of damages. The IPCC \citeyearpar{ipcc2021climate} notes that present day global mean surface temperatures in the year 2020 are around \SI{1.1}{\celsius} above preindustrial (1850-1900) levels. Estimated damages in 2100 shown in this figure are based on each study’s stated ``preferred'', ``main'', or ``central'' empirical specification of the temperature-GDP relationship and/or the method used to project climate damages. Each damage function requires a measure of current climate or baseline temperature; in all panels in this figure we use country-level population-weighted mean temperatures from 1980 to 2010 drawn from \citet{burke2015global} for this purpose. See \ref{app:sec:td_macro} and Figure \ref{fig:macro_gdp_loss} for more details.
\end{figure}

The constructed GDP damage component when integrated into the \MimiGIVE{} framework provides baseline projections of U.S. GDP damages. Figure \ref{fig:macro_climate_damages} displays the estimated losses to U.S. GDP from climate change in 2100 based on each macroeconomic econometric paper as a function of U.S. temperature change. The points represent a random subset of the Monte Carlo simulation (2,500 of the 10,000 simulations) with consistent draws of U.S. RFF-SP socioeconomic projections and \FaIR{} 1.6.2 climate module parameters across all studies. Also added to each panel in the figure is a reduced-form damage function based on the point estimates. Specifically, the U.S.-specific damage functions shown here are generated using estimated GDP losses in 2100 (the points) regressed on temperature and temperature squared in the year 2100 at the mean (solid line), and quantile regressions at the 5\textsuperscript{th} and 95\textsuperscript{th} percentiles (shaded ribbons). Figure \ref{fig:avg_macro_climate_damages} in \ref{app:sec:add_fig_tab} presents the solid lines (means) from Figure \ref{fig:macro_climate_damages} in a single figure to allow for an easier comparison across studies.  Note these reduced-form damage functions are provided for illustrative purposes only. They are not being used to generate the U.S.-specific SC-\ce{CO2} estimates we present below (or the other SC-GHGs presented in \ref{app:sec:add_fig_tab}).

Figure \ref{fig:macro_climate_damages}, along with the companion Figure \ref{fig:macro_summary} in \ref{app:sec:td_macro}, shows notable differences between the damage functions in their implied average climate damages for U.S. GDP in 2100. \citet{kahn2021long} state that their model implicitly incorporates assumed adaptation over a 30-year horizon, which yields the lowest average damages to U.S. GDP in 2100 of 0.4 percent. The papers finding only contemporaneous effects of temperature on the level of GDP, \citet{newell2021the} and \citet{acevedo2020effects}, also have lower average damages of 0.5 percent in 2100. \citet{casey2023projecting} find some persistence in the impact of temperature on U.S. TFP (19 percent), and results from this damage component show slightly higher average annual damages in 2100 of 0.6 percent due to their modeling of the persistence of investment, capital, and output reductions following temperature changes. The damage component from \citet{harding2025distribution}, who find higher level of persistence (85 percent)\footnote{The persistence parameter is one minus the convergence parameter estimated in the paper.} of temperature on U.S. GDP, shows larger average damages of 1.8 percent in 2100. Even with this high level of persistence, the structural restrictions of this approach imply that the contemporaneous effect dissipates. When a more flexible damage function is estimated using local projections as in \citet{nath2024}, we find damages increase to 3.4 percent in 2100. This is the result of increasing damages in the short term and continued persistence in the medium term. Our estimates using the \citet{kalkuhl2020impact} damage component are also higher and equal 3.4 percent in 2100. This paper finds much higher contemporaneous damages to U.S. GDP owing to the much lower global optimal temperature than other papers of \SI{5.4}{\celsius}, compared to \SI{13.0}{\celsius} in \citet{nath2024}, \SI{13.6}{\celsius} in \citet{harding2025distribution}, and \SI{13}{\celsius} in \citet{acevedo2020effects}. The projected damages based on the \citet{burke2015global} damage component remain an extreme outlier, estimated to reach 16.5 percent of GDP in 2100, which illustrates the consequences of permanent growth effects of temperature shocks. Overall, our analysis with harmonized uncertain socioeconomic and climate inputs projects a lower and narrower range of average 2100 U.S. GDP losses of 0.4 - 3.4 percent (or 16.5 percent if including Burke) than the range derived from each paper's replication code of 1.6 - 13.8 percent (or 36.3 percent if including Burke), under their implementation of an RCP 8.5 scenario, as shown in column 1 of Table \ref{tab:us_gdppc_loss}.\footnote{To validate that the damage component provides an accurate representation of each paper’s findings, we confirmed that the U.S. damage projections based on our constructed damage function under an SSP 5/RCP 8.5 scenario through 2100 approximated the results from the U.S. damage projections in the replication code for the paper. While this SSP/RCP scenario is not used in the estimates we present in the main text, it is one that is generally available for each paper and assures consistency of the damage component with the paper results. See \ref{app:sec:td_macro} for a detailed description and results of this multi-step validation exercise.}

Table \ref{tab:macroecon_climate_damages} presents the resulting U.S.-specific SC-\ce{CO2} estimates for 2030 emissions when we combine the constructed GDP damage component based on each paper’s main specification with the harmonized U.S.-specific socioeconomic and emissions inputs, climate modeling, and discounting methods. Analogous U.S.-specific SC-GHG results for 2030 emissions of \ce{CH4} and \ce{N2O} are presented in Table \ref{tab:impact_categories_us} in  \ref{app:sec:add_fig_tab}.\footnote{All reported estimates are certainty-equivalent SC-GHGs consistent with the approach of EPA \citeyearpar{epa2023report}.} As expected, the early studies that project permanent GDP growth effects from an increase in temperature \citep{burke2015global} result in SC-\ce{CO2} estimates that are more than an order of magnitude larger than those projecting only temporary or some intermediate persistence in the impacts on GDP growth. Among the studies projecting some persistence in U.S. GDP growth effects, the implied U.S.-specific SC-\ce{CO2} for 2030 emissions ranges from \$10 to \$65 per metric ton \ce{CO2} (under a 2 percent near-term Ramsey discounting rate). For this class of papers, one of the more recent studies that incorporates many recent advancements in the relevant literature, \citet{nath2024}, is at the high end of the range of estimates.\footnote{We find the results presented in Table \ref{tab:macroecon_climate_damages} to be robust to the dataset used in the empirically estimated temperature-GDP relationship underlying the GDP damage component in the MimiGIVE modeling. Figure \ref{fig:nath_dataset_comparison} displays the U.S.-specific SC-\ce{CO2} when all studies are re-estimated using the underlying dataset from \citet{nath2024}.}

\begin{table}[htbp]
    \caption{Evidence on U.S. climate damages from recent macroeconomic econometric studies}
    \label{tab:macroecon_climate_damages}
    \centering
    \begin{threeparttable}
        \footnotesize
            \begin{tabular}{lcc}
                \toprule \toprule
				\textbf{Study\tnote{$\alpha$}} & \textbf{Impact} &  \textbf{U.S.-specific SC-\ce{CO2}} \\
                & \textbf{Represented} & (2030 emissions, 2\% near-term Ramsey \\
                & & discount rate, 2020\$/mt\ce{CO2})\\
                \midrule		
\citet{burke2015global}     & Permanent U.S. GDP growth effects	 & \$1000 \\
\midrule
\citet{kalkuhl2020impact}   & Temporary U.S. GDP growth effects	 & \$55  \\
\citet{newell2021the}       & Temporary U.S. GDP growth effects	 & \$10  \\
\citet{acevedo2020effects}  & Temporary U.S. GDP growth effects	 & \$10  \\
\midrule
\citet{kahn2021long}        & Persistent U.S. GDP growth effects & \$10  \\
\citet{casey2023projecting} & Persistent U.S. GDP growth effects & \$14  \\
\citet{harding2025distribution}  & Persistent U.S. GDP growth effects & \$38  \\
\citet{nath2024}            & Persistent U.S. GDP growth effects & \$65  \\
\bottomrule
            \end{tabular}
            \begin{tablenotes} 		
                \footnotesize
				\item Modeling was conducted in a modified \MimiGIVE{} framework: U.S. socioeconomic and emissions inputs are taken from RFF-SPs \citep{rennert2022social}; climate module temperature projections are based on \FaIR{} 1.6.2, accounting for carbon feedback effects from permafrost thaw and Amazon dieback based on \citet{dietz2021economic} modeling of these feedbacks, and GMST projections are downscaled using a pattern-scaling approach (in which CONUS GCMs are paired with \FaIR{} parameter sets); future damages are discounted using Ramsey-based discounting following \citet{newell2022a} updated to use U.S. growth rates and recalibrated $\rho$ and $\eta$ parameters. See \ref{app:sec:td_discounting} and \ref{app:sec:td_feedback} for details on each of these inputs. Due to sampling variation in the random draws of a Monte Carlo simulation, values presented in this table are rounded to the nearest dollar. 
				\item[$\alpha$] Results shown in this table are based on each study's stated ``preferred'', ``main'', or ``central'' empirical specification of the temperature-GDP relationship and/or the method used to project climate damages. See \ref{app:sec:td_macro} for more details. 
			\end{tablenotes}
	\end{threeparttable}
\end{table}

A few common themes emerge from our analysis of these macroeconomic empirical studies. First, the impacts of temperature on GDP are likely to be nonlinear - this implies that temperature changes cause greater damages in hotter countries and greater damages as countries get hotter. Second, temperature shocks appear to have a persistent impact on a country’s GDP. Empirical methods that allow for persistence are useful in tracing out the dynamic response of GDP to changes in temperature \citep{acevedo2020effects,kahn2021long,casey2023projecting,harding2025distribution,nath2024}. One potential mechanism for this persistence is through impacts on capital investment \citep{casey2023projecting}. Third, temperature shocks are themselves persistent, which partially explains the persistence of temperature’s effect on GDP \citep{acevedo2020effects,nath2024}. Not accounting for the persistence of temperature can overestimate the dynamic impact of changes in current temperature on GDP. Recent literature has attempted to control for these using the lag of temperature as a control \citep{acevedo2020effects,nath2024} and/or by filtering out the serial correlation of temperature to produce innovations to temperature \citep{nath2024}. Despite controls that reduce temperature persistence, some level of persistence still appears present---additional care is necessary to account for temperature persistence and fully identify the persistence of GDP to temperature shocks \citep{nath2024}.

\section{Discussion}
\label{sec:discussion}

In addition to producing a consistent set of U.S.-specific SC-GHG estimates from the macroeconomic studies reviewed above, we take an initial step in placing our findings in context of existing estimates of U.S. climate damages. Given that U.S.-specific SC-GHG estimates derived from macroeconomic econometric damage functions should capture all market damages from annual average temperature changes that affect U.S. GDP (inclusive of interactions and adjustment costs), the U.S-specific SC-\ce{CO2} estimates presented in Table \ref{tab:macroecon_climate_damages} ostensibly offer a more complete picture of market-based damages than currently represented in enumerative models. Our findings bear this out. Even the lower end of the range of estimates in Table \ref{tab:macroecon_climate_damages} exceed the market damages based on the enumerative models discussed in Section \ref{sec:modeling_overview}. For example, in \FrEDI, market damages comprise approximately \$7 of the \$36/mt\ce{CO2} in 2030, and in \DSCIM{}, market damages comprise approximately \$4 of the \$21/mt\ce{CO2} in 2030. In \GIVE{}, market damages comprise approximately \$2 of the \$24 /mt\ce{CO2} in 2030. See \ref{app:sec:add_fig_tab} for more results from the enumerative models. This finding further suggests that market-based damages are not fully represented in existing enumerative models.

This does not imply that these macroeconomic studies offer a fully comprehensive account of market-based climate damages, and the authors across these studies acknowledge limitations of their approaches. First, the damage functions only include market damages observed in the historical record. Some possible changes to the climate that have not been experienced, such as large-scale sea level rise and Earth system feedback effects, are not reflected in the impulse response functions from the econometric studies but are likely to cause climate damages if they occur. Similarly, empirically estimated damage functions account for adaptation only to the extent that adaptation has occurred historically. Future projections of climate damages based on these studies generally assume, implicitly or explicitly, that future adaptation occurs at a rate consistent with adaptation observed in the historical record, which may differ from adaptation to an enduring, anticipated change in climate. Relatedly, expenditures on adaptation to and mitigation of climate change (such as on coastal protection or air purification) may temporarily increase economic output, but are unlikely to increase welfare, or even output in the long run, because they they reallocate resources away from other productive activities. Finally, the macroeconomic studies reflect only impacts associated with the annual averages of climate variables (or shocks derived from those averages) at the geographic resolution used in the studies (often the country or state-equivalent level). Extreme weather events, such as extreme heat waves, extreme precipitation, and extreme wind may not be well captured by country-level temperature shocks \citep{bilal2024}, and the impact of weather variability may not be well captured by annual average temperature measures \citep{kotz2021day}. 

Another well known limitation of damage estimates based on the macroeconomic econometric studies is their inability to account for the value of nonmarket impacts of climate change. This provides an opportunity to begin to combine or integrate the GDP-based market damage estimates above with results from enumerative modeling of nonmarket damages. Such an integration allows us to ``go beyond GDP'' in the sense of \citet{jones2016beyond} in a structured framework built upon credible underlying studies, as long as care is taken (1) to avoid the potential for double counting of damages, and (2) to account for changing incomes and the relative scarcity of market and nonmarket goods (as highlighted in \citet{drupp2018limits} and \citet{drupp2021relative}) when monetizing nonmarket damages and discounting in this context. 

On the issue of double counting, it is clear that the macroeconomic empirically-based damage functions cannot be added to the full damage module of enumerative models such as \GIVE, \DSCIM, or \FrEDI{} without some double counting as each of these models also contain representation of at least a few types of market damages that are likely to be captured in GDP measures (e.g., from net changes in crop yields, energy consumption, and labor productivity). However, climate damages experienced through nonmarket pathways such as changes in net mortality rates, welfare resulting from household production, the value of time spent outside work, and changes in ecosystem services (including those provided by biodiversity) are not directly included in GDP calculations. SC-GHG estimates derived from macroeconomic econometric studies only reflect nonmarket damages to the extent that these damages have consequences for productivity or other components of GDP. It is therefore reasonable to combine one or more nonmarket damage functions from enumerative models with the GDP-based market damage functions from the macroeconomic studies without major concerns of double counting.\footnote{Precedent for this type of combination within a damage module of an IAM used for SC-GHG estimation can be found in the PAGE model \citep{yumashev2019climate,yumashev2020page}. Others have also noted the importance of augmenting GDP impact estimates with nonmarket impacts for a comprehensive estimate of the full economic burden of climate change \citep{hsiang2026climate}, and there is a broader relevant literature that suggests nonmarket factors (e.g., leisure, lower inequality, and higher life expectancy) and GDP are additive when deriving social welfare \citep{jones2016beyond}.} 

In \ref{app:sec:add_fig_tab} Table \ref{tab:combined_evidence}, we take a first step in illustrating the impact on U.S.-specific SC-GHG estimates from including two categories of nonmarket damages with the GDP-based market damage function from several of the macroeconomic studies reviewed above. Specifically, we combine the heat- and cold-related mortality damage function from \citet{cromar2022global}\footnote{We use the mortality damage function in \GIVE{} \citep{rennert2022comprehensive} due to modeling constraints in incorporating other health damage functions (\citet{carleton2022valuing} from \DSCIM) into the \MimiGIVE{} framework.} and wildfire smoke-related mortality damages \citep{qiu2026wildfire} with the U.S. GDP damage functions derived from recent macroeconomic studies that project some persistence in the effect of temperature on GDP growth over time. This includes the subset of the studies presented in Table \ref{tab:summary_macroecon} that use more flexible empirical methods that better allow for tracing out nonlinear dynamics in the temperature-GDP relationship and addressing serial correlation in temperature and GDP over time. Importantly, in this integrated modeling exercise, we do not simply add GDP and mortality impacts estimated separately, but integrate both impacts into the full Monte Carlo estimation procedure. This allows us to account for feedbacks from estimated GDP-based damages to monetization of health damages and to the discounting module.\footnote{See Section \ref{sec:combine_details} and \ref{app:sec:drupp_derivation} for more discussion.} From this integration exercise, we find U.S.-specific SC-\ce{CO2} estimates ranging from \$46 to \$99 per metric ton \ce{CO2} for 2030 emissions (under a 2 percent near-term Ramsey discount rate).\footnote{Accounting for the feedbacks from estimated GDP-based damages to monetization of health damages within our integrated modeling framework acts to slightly reduce monetized health damages and increase discounted damages, all else equal. As such, the specification with the highest market damages, \citet{nath2024}, has the lowest health damages when accounting for these feedbacks.} It is important to note that there are other categories of nonmarket damages from enumerative models that could be integrated in a similar manner, such as nonuse value of biodiversity loss \citep{wingenroth2024accounting}. In addition, as discussed above, GDP impulse response functions from the macroeconomic econometric studies are only reflective of impacts that occur due to year-to-year temperature variation observed in the historical record. It is likely that some types of market-based damages based on enumerative studies (e.g., from large-scale sea level rise or other slow to manifest changes) could be also be integrated in a similar manner without concerns about double counting. We leave this as an area for future research.

\section{Conclusion}

The primary contribution of this paper is to provide an apples-to-apples comparison of recent macroeconomic studies that empirically estimate the relationship between temperature and GDP. Studies that use more flexible empirical methods and address common econometric challenges find temperature changes to have some persistent but nonpermanent effects on economic growth. We demonstrate that the empirical results from the central specifications of these papers --- once placed within a consistent modeling framework with a comprehensive probabilistic consideration of key uncertainties --- produces a narrower and lower range of end of century GDP loss projections than what is reported across the published studies. Among the studies projecting only temporary or some intermediate persistence in the impacts on GDP growth, our harmonized analysis projects average 2100 U.S. GDP losses of 0.4 - 3.4 percent of U.S. GDP. The implied U.S. SC-\ce{CO2} for 2030 emissions (for market damages only) ranges from \$10 to \$65 per metric ton \ce{CO2}, or about \$30 on average (under a 2 percent near-term Ramsey discounting rate). Even the low end of this range is greater than the market damage estimates from current enumerative SC-GHG modeling. However, these results do not account fully for adaptation to an anticipated, persistent change in the climate distribution and its costs, for market-based damages not observed in the historical record, or for nonmarket damages of climate change. Incorporating evidence on two categories of nonmarket health damages (temperature- and wildfire smoke-related mortality) with the GDP-based market damage function from the macroeconomic studies increases the range of U.S.-specific SC-\ce{CO2} estimates to \$46-\$99 per metric ton of \ce{CO2} for 2030 emissions. 

Our findings point to several next steps in developing an improved understanding of how the consequences of GHG emissions are likely to be experienced by U.S. populations. First, additional examination of the mechanisms underlying estimated GDP-temperature relationships is needed in order to better unpack what is reflected in the results of the macroeconomic empirical studies and to more fully compare and combine them with evidence from enumerative damage approaches. Coupled with additional enumerative modeling of market based damages, this can expand the policy usefulness of both strands of the literature. Not only aggregate measures of climate consequences but also information on the source and incidence of these consequences is helpful to policy makers ---for example, when thinking about where and what types of adaptation investments in the U.S. are most critical. Relatedly, it is unclear how well the macroeconomic studies can fully account for adaptation to anticipated climate change. Empirically based estimation of revealed adaptation must rely on what has been observed in the historical record, and it remains challenging to project how the ability of societies to adapt to climate change and how the costs of adaptation will change at higher levels of warming. Similarly, results from macroeconomic studies may not reflect the market-based damages of slow-to-manifest changes such as expected large-scale sea level rise, earth system feedbacks, or other damages that have not occurred historically.  

Second, additional research is needed on various categories of nonmarket impacts. As shown in this paper, integrating climate-driven temperature- and wildfire smoke-related health impacts alone with GDP-based damages causes the average implied U.S. SC-\ce{CO2} across the studies analyzed to more than double. As noted in this paper, emerging research on biodiversity loss can also already be incorporated into an existing enumerative model, and there may be additional opportunities to convert existing impact-specific studies (e.g., existing estimates of temperature-driven impacts to some morbidity endpoints) into damage functions of the kind needed for U.S.-specific SC-GHG estimation. In other areas, new research is needed (e.g., damages associated with the loss of ecosystem services and impacts to water resources, among others). 

Finally, of relevance to both steps above, more research is needed to quantify how climate impacts occurring outside of U.S. borders affect U.S. interests. Due to the interconnectedness of the global economy and populations, climate impacts occurring abroad may impact U.S. citizens and residents and assets located in other countries as well as lead to significant supply chain disruptions that may affect local economies. Geopolitical impacts from climate change such as political unrest, war, humanitarian crises, and increased global migration are also likely to affect U.S. populations. More research is needed on these key omitted damage categories and the spillover effects on U.S. populations from climate damages abroad to develop a more comprehensive estimate of the economic damages from climate change to U.S. populations.

\newpage
\clearpage
\singlespacing
\bibliographystyle{jaere}
\bibliography{references}

\resetlinenumber

%% file: appendix_rev1.tex
\appendix
\setcounter{page}{1}                                
\setcounter{footnote}{0} 	                        

\counterwithin*{figure}{section}
\counterwithin*{table}{section}
\counterwithin*{equation}{section}

\renewcommand{\thesection}{Appendix \Alph{section}}               
\renewcommand{\thesubsection}{\Alph{section}.\arabic{subsection}} 
\renewcommand{\thefigure}{\Alph{section}.\arabic{figure}}         
\renewcommand{\thetable}{\Alph{section}.\arabic{table}}           
\renewcommand{\theequation}{\Alph{section}.\arabic{equation}}     

\defcitealias{ipcc2021climate}{IPCC, \citeyear{ipcc2021climate}}
\defcitealias{ipcc2021earth}{IPCC, \citeyear{ipcc2021earth}}
\defcitealias{ipcc2022policy}{IPCC, \citeyear{ipcc2022policy}}
\defcitealias{ipcc2023policy}{IPCC, \citeyear{ipcc2023policy}}
\defcitealias{national2017valuing}{NASEM, \citeyear{national2017valuing}}
\defcitealias{un2019world}{UN, \citeyear{un2019world}}
\defcitealias{census2020county}{U.S. Census, \citeyear{census2020county}}
\defcitealias{census2023national}{U.S. Census, \citeyear{census2023national}}
\defcitealias{census2024tiger}{U.S. Census, \citeyear{census2024tiger}}
\defcitealias{epa2011sab-vsl}{EPA, \citeyear{epa2011sab-vsl}}
\defcitealias{epa2023report}{EPA, \citeyear{epa2023report}}
\defcitealias{epa2024climate}{EPA, \citeyear{epa2024climate}}
\defcitealias{epa2024pattern}{EPA, \citeyear{epa2024pattern}}
\defcitealias{epa2024tdfredi}{EPA, \citeyear{epa2024tdfredi}}
\defcitealias{epa2024guidelines}{EPA, \citeyear{epa2024guidelines}}
\defcitealias{cil2024estimating}{CIL, \citeyear{cil2024estimating}}
\defcitealias{cil2023dscim}{CIL, \citeyear{cil2023dscim}}
\defcitealias{ciesin2016gridded}{CIESIN, \citeyear{ciesin2016gridded}}
\defcitealias{rff2024increasing}{RFF, \citeyear{rff2024increasing}}
\defcitealias{cbo2022long}{CBO, \citeyear{cbo2022long}} 
\defcitealias{cbo2023long}{CBO, \citeyear{cbo2023long}}
\defcitealias{USGCRP2023fifth}{USGCRP, \citeyear{USGCRP2023fifth}}

\begin{center}
\large
\textbf{Appendix} \\

\vspace{0.15in}

\normalsize
\textsc{Economic Impacts of Climate Change in the United States:
Integrating and Harmonizing Evidence from Recent Studies} \\
\vspace{0.15in}

Elizabeth Kopits, Daniel Kraynak, Bryan Parthum, Lisa Rennels, David Smith, 
Elizabeth Spink, Joseph Perla, and Nshan Burns \\

\vspace{0.15in}

\end{center}

\doublespacing

\section{Additional Tables and Figures}
\label{app:sec:add_fig_tab}

\begin{table}[htbp]
    \caption{Impacts represented in \GIVE{} and \DSCIM{} Integrated Assessment Models}
    \label{tab:impact_categories}
    \centering
    \begin{threeparttable}
        \footnotesize
        \begin{tabular}{lcc}
            \toprule \toprule
            \textbf{Impact}   & \textbf{\DSCIM{} \citepalias{cil2023dscim}} & \textbf{\GIVE{} \citep{rennert2022comprehensive}} \\
            \textbf{Category} & & \\
            \midrule	
Temperature     & Heat- and cold-related mortality  & Heat- and cold-related mortality \\
Mortality       & \citep{carleton2022valuing}       & \citep{cromar2022global} \\
\midrule
Energy & Expenditures for electricity and other & Expenditures for space heating and  \\
       & direct fuel consumption                & cooling in buildings \\
       & \citep{rode2021estimating}                   & \citep{clarke2018effects}  \\
\midrule
Agriculture & Production impacts for six crops: maize, rice, & Welfare changes from temperature driven \\
            & wheat, soybeans, sorghum, and cassava          & changes in production of four crops: maize, \\
            & \citep{hultgren2022estimating}                      & rice, wheat, and soybeans \\
            &                                                & \citep{moore2017new} \\
\midrule
Coastal & Impacts of SLR as realized through inundation,     & Impacts of SLR as realized through inundation, \\
        & migration, protection, dry and wetland loss;       & migration, protection, dry and wetland loss; \\
        & mortality and physical capital loss from SLR       & mortality and physical capital loss from SLR \\
        & (\citet{kopp2016temp}, \citet{garner2021ipcc}  for SLR; & (\citet{wong2017} for SLR;  \\
        & \citet{diaz2016estimating}, \citet{depsky2023dscim} for damages)     &  \citet{diaz2016estimating} for damages) \\
\midrule
Labor & Labor disutility costs from labor supply & \\
      & responses to increased temperature       & \\
      & \citep{rode2022labor}                      & \\
\bottomrule       
        \end{tabular}
        \begin{tablenotes} 				
            \item Source: Adapted from EPA \citeyearpar{epa2023report}. 
        \end{tablenotes}
	\end{threeparttable}
\end{table}

\begin{table}[htbp]
    \caption{Impact categories in \FrEDI, version 4.1}
    \label{tab:impact_categories_fredi}
    \centering
    \begin{threeparttable}
        \begin{adjustbox}{width=\textwidth}
            \begin{tabular}{llc}
                \toprule \toprule
                \textbf{Impact}   & \multicolumn{1}{c}{\textbf{\FrEDI{}\citepalias{epa2024tdfredi}}} & \textbf{Underlying study} \\
                \textbf{Category} & & \\
                \midrule
\multirow[c]{8}{*}{Health} & Heat- and cold-related mortality & \citet{cromar2022global} \\
\cmidrule{2-3}
& Climate driven \ce{O3} and PM-related mortality  & \citet{fann2021} \\
\cmidrule{2-3}
& Wildfire PM exposure mortality and morbidity impacts & \citet{neumann2021estim}\\
& (hospitalization costs and lost productivity) and some suppression costs & \\
\cmidrule{2-3}
& Southwest region dust mortality and morbidity impacts & \citet{achakulwisut2019effects} \\
\cmidrule{2-3}
& Valley fever mortality and morbidity impacts & \citet{gorris2021economic} \\
\cmidrule{2-3}
& Suicide incidence & \citet{Belova2022projecting} \\
\cmidrule{2-3}
& Vibriosis mortality and morbidity impacts & \citet{sheahan2022examining} \\
\midrule
\multirow[c]{8}{*}{Energy} & Electric power sector costs (e.g., capital, fuel, variable operation and & \citet{mcfarland2015impacts} \\
& maintenance (O\&M), and fixed O\&M costs) for heating and cooling & \\
& (demand) and required capacity expansion (supply)  & \\
\cmidrule{2-3}
& Stress to electricity transmission and distribution infrastructure & \citet{fant2020climate} \\
&  (e.g., leading to system failure, changes in infrastructure & \\
& health/lifespan, efficiency/capacity).  & \\
& Replacement or repair costs, O\&M costs, costs from efficiency or & \\
& capacity losses, and interruption costs to consumers &  \\
\midrule
\multirow[c]{4}{*}{Agriculture} & Effects of changes in temperature, precipitation, and  & \citet{schlenker2009nonlinear}; \\
& \ce{CO2} fertilization on yields of cotton, maize, &  \citet{mcgrath2013regional}  \\
& soybean, and wheat. & \citet{hsiang2017estimating}, citing\\
& & \citet{hsiang2013climate} \\
\midrule
\multirow[c]{15}{*}{\parbox{2cm}{Coastal and Infrastructure}} & Coastal Properties --- Costs related to armoring, elevation, nourishment, & \citet{neumann2021climate}; \\
& structure repair, and abandonment (including storm surge impacts) & \citet{lorie2020modeling} \\
\cmidrule{2-3}
& High Tide Flooding --- Coastal property damage/adaptation costs, traffic & \citet{fant2021mere}\\
& delays, and adaptation costs due to high tide flooding & \\
\cmidrule{2-3}
& Hurricane Wind Damage – Property damage from hurricane winds  &  \citet{dinan2016potential,dinan2017project}; \\
& & \citet{marsooli2019climate} \\
\cmidrule{2-3}
& Urban Drainage --- Costs of proactive urban drainage infrastructure & \citet{price2016calibrated}; \\
& adaptation (construction costs and annual maintenance costs of & \citet{neumann2015climate} \\
& temporary storage or infiltration to manage runoff volumes)  & \\
\cmidrule{2-3}
& Inland flooding --- Residential property damages from riverine flooding & \citet{wobus2019,wobus2021} \\
\cmidrule{2-3}
& Rail impacts – Costs of replacing tracks and costs of delays & \citet{neumann2021climate}, citing \\
& associated with temperature-induced track buckling. & \citet{chinowsky2019impacts} \\
\cmidrule{2-3}
& Road impacts – Temperature-driven damage to paved and unpaved & \citet{neumann2021climate}, citing \\
& road surfaces including road repair costs and road user costs (travel &  \citet{neumann2015climate}\\
& time delays and vehicle operating costs). & \\
\midrule
\multirow{2}{*}{Labor} & Impacts on hours worked in weather-exposed industries & \citet{neidell2021temper} \\
& (e.g., agriculture, construction, manufacturing) & \\
\midrule
\multirow[c]{9}{*}{\parbox{2cm}{Ecosystems and Recreation}} & Fisheries impacts --- value of change in weight of landings due to & \\
& changes in thermally available habitat (no additional adaptation) (16 US & \citet{moore2021est}; \\
& fisheries that account for 56\% of US commercial fishing revenues) & \citet{morley2018project}\\
\cmidrule{2-3}
& Water quality impacts – recreation-related WTP for changes in water & \citet{fant2017climate} with \\
& quality & \citet{boehlert2015climate}; \\
& & \citet{yen2016}\\
\cmidrule{2-3}
& Winter recreation impacts – Lost snowmobiling, alpine skiing, and & \citet{wobus2017}\\
& cross-country skiing revenues (some adaptation captured – i.e., snow & \\
& blowing), only some regions & \\
\midrule
\multirow{3}{*}{Crime} & Property (robbery, burglary, larceny, and motor vehicle theft) and & \citet{hsiang2017estimating}, citing \\
& violent (murder, rape, and assault) crime impacts from temperature & \citet{Jacob489,heaton2010hidden}; \\
& change & \citet{ranson2014crime} \\
\bottomrule
            \end{tabular}
        \end{adjustbox}
	\end{threeparttable}
\end{table}

\begin{table}[htbp]
    \caption{U.S.-specific SC-\ce{CO2} estimates from enumerative models, 2030 emissions, 2\% Ramsey discount rate}
    \label{tab:impacts_bottom_up}
    \centering
    \begin{threeparttable}
\begin{tabular}{lcr}

\toprule \toprule
\textbf{Model}   & \multicolumn{2}{c}{\textbf{U.S.-specific SC-\ce{CO2}}} \\
                 & \multicolumn{2}{c}{(\$/mt\ce{CO2}, 2020\$)} \\

\midrule		
\textbf{DSCIM}                                         & \$21 & \\
\hspace{0.5cm} Heat- and cold-related mortality &      & 16    \\
\hspace{0.5cm} Energy                           &      & -0.3  \\
\hspace{0.5cm} Agriculture                      &      & 1     \\
\hspace{0.5cm} Coastal                          &      & 1     \\
\hspace{0.5cm} Labor                            &      & 2     \\

\midrule	
\textbf{GIVE}                                          & \$24 & \\
\hspace{0.5cm} Heat- and cold-related mortality &      & 21    \\
\hspace{0.5cm} Energy                           &      & 1  \\
\hspace{0.5cm} Agriculture                      &      & 2     \\
\hspace{0.5cm} Coastal                          &      & 0.4     \\

\midrule	
\textbf{FrEDI}                                         & \$36 & \\

\midrule\midrule	
\textbf{Additional Impacts}                            & \$19 & \\
\hspace{0.5cm} Wildfire smoke-related mortality \citep{qiu2026wildfire}          &  & 18 \\
\hspace{0.5cm} Nonuse value of biodiversity loss \citep{wingenroth2024accounting} &  & 1  \\

\bottomrule
\end{tabular}
        \begin{tablenotes} 		
            \footnotesize
            \item U.S. socioeconomic and emissions inputs taken from RFF-SPs \citep{rennert2022social}; climate module temperature projections are based on \FaIR{} 1.6.2 and account for carbon feedback effects from permafrost thaw and Amazon dieback based on \citet{dietz2021economic} modeling of these feedbacks; discounting follows \citet{newell2022a} updated to use U.S. growth with recalibrated $\rho$ and $\eta$ parameters. To calculate the temperature-related mortality damages in \GIVE{} (based on \citet{cromar2022global}), GMST projections from \FaIR{} 1.6.2 are downscaled to U.S. temperature using a pattern-scaling approach (in which CONUS GCMs are paired with \FaIR{} parameter sets). This downscaling approach, a historical period validation exercise of the approach, and resulting U.S. temperature projections are described in detail in \ref{app:sec:td_discounting}. Implementation of \citet{qiu2026wildfire} and \citet{wingenroth2024accounting} is described in detail in \ref{app:sec:td_wildfire}.
        \end{tablenotes}
    \end{threeparttable}
\end{table}

\begin{table}[htbp]
    \caption{U.S.-specific SC-\ce{CH4} and SC-\ce{N2O}, 2030 emissions, 2\% near-term Ramsey discount rate}
    \label{tab:impact_categories_us}
    \centering
    \begin{threeparttable}
    \small
        \begin{tabular}{llcc}
            \toprule \toprule
            \textbf{Model} & \textbf{Impacts represented} & \textbf{U.S. impact-} & \textbf{U.S. impact-} \\
             & & \textbf{specific} & \textbf{specific} \\
             & & \textbf{SC-\ce{CH4}} & \textbf{SC-\ce{N2O}} \\
             & & (2020\$/mt\ce{CH4}) & (2020\$/mt\ce{N2O}) \\
            \midrule	
\DSCIM                & Health energy, agriculture, coastal, labor 	 & \$130 & \$6,000 \\
& (see Table \ref{tab:impact_categories}) & & \\
\GIVE                 & Health\tnote{$\alpha$} energy, agriculture, coastal 		 & \$400 & \$7,700 \\
& (see Table \ref{tab:impact_categories}) & & \\
\FrEDI\tnote{$\beta$} & Health energy, agriculture, coastal, labor, 	 & \$590 & \$11,000 \\
& others (see Table \ref{tab:impact_categories_fredi}) & & \\
\midrule
\citet{qiu2024mortality}\tnote{$\alpha$}    & Mortality risk changes from climate-driven  & \$190 & \$5,200 \\
                                            & wildfire \ce{PM_{2.5}} exposure (CONUS only) & & \\
\citet{wingenroth2024accounting} & nonuse value from biodiversity loss	 & \$4 & \$180 \\
\midrule
\citet{burke2015global}\tnote{$\alpha$}     & Permanent U.S. GDP growth effects	 & \$17,000 & \$340,000 \\
\midrule
\citet{kalkuhl2020impact}\tnote{$\alpha$}   & Temporary U.S. GDP growth effects	 & \$850 & \$17,000  \\
\citet{newell2021the}\tnote{$\alpha$}       & Temporary U.S. GDP growth effects	 & \$140 & \$3,100  \\
\citet{acevedo2020effects}\tnote{$\alpha$}  & Temporary U.S. GDP growth effects	 & \$120 & \$3,000  \\
\midrule
\citet{kahn2021long}\tnote{$\alpha$}        & Persistent U.S. GDP growth effects & \$220 & \$3,300  \\
\citet{casey2023projecting}\tnote{$\alpha$} & Persistent U.S. GDP growth effects & \$170 & \$4,100  \\
\citet{harding2025distribution}\tnote{$\alpha$}  & Persistent U.S. GDP growth effects & \$450 & \$11,000  \\
\citet{nath2024}\tnote{$\alpha$}            & Persistent U.S. GDP growth effects & \$900 & \$20,000  \\
\bottomrule
        \end{tabular}
        \begin{tablenotes} 				
            \item[] U.S. socioeconomic and emissions inputs taken from RFF-SPs \citep{rennert2022social}; climate module temperature projections account for carbon feedback effects from permafrost thaw and Amazon dieback based on \citet{dietz2021economic} modeling of these feedbacks; discounting module follows \citet{newell2022a} with Ramsey parameters, $\rho$ and $\eta$, recalibrated to the U.S. growth rate of consumption. The health damages represented in \GIVE{} do not account for non-climate mediated effects of \ce{CH4} and \ce{N2O} emissions experienced by U.S. populations. For example, they do not capture the monetized increase in U.S. respiratory-related human mortality risk from the ozone produced from a marginal pulse of \ce{CH4} emissions (which recent research estimates to be on the order of \$320/mt\ce{CH4} for methane emissions in 2030 \citep{mcduffie2023social} or the U.S. health risks from stratospheric ozone destruction from \ce{N2O} emissions \citep{kanter2021improving}.
            \item[$\alpha$] To calculate the U.S. temperature inputs needed for these damage functions, GMST projections from \FaIR{} 1.6.2 are downscaled to U.S. temperature using a pattern-scaling approach (in which CONUS GCMs are paired with \FaIR{} parameter sets). This downscaling approach, a historical period validation exercise of the approach, and resulting U.S. temperature projections are described in detail in \ref{app:sec:td_discounting}.
            \item[$\beta$] \FrEDI{} results presented in this table are based on \FrEDI{} version 3.4 and do not yet account for the carbon feedback effects from permafrost thaw and Amazon dieback.
        \end{tablenotes}
	\end{threeparttable}
\end{table}

\begin{table}[htbp]
    \caption{Combined evidence on U.S. nonmarket health damages and U.S. GDP-based market damages from \ce{CO2}, \ce{CH4} and \ce{N2O} emissions}
    \label{tab:combined_evidence}
    \centering
    \begin{threeparttable}
            \begin{tabular}{lccc}
                \toprule \toprule
                \textbf{Panel A} & \multicolumn{3}{c}{\textbf{U.S. impact-specific SC-\ce{CO2}}} \\
                & \multicolumn{3}{c}{(2030 emissions, 2\% near-term Ramsey} \\
                & \multicolumn{3}{c}{discount rate, 2020\$/t\ce{CO2})} \\
                \cmidrule{2-4}
				  \textbf{Source of GDP-based \hspace{4em}} & \textbf{GDP-based Market} & \textbf{Health Nonmarket \hspace{2em}} & \textbf{Total\tnote{$\delta$}} \\
				   \textbf{Market Damages\tnote{$\alpha$}}& \textbf{Damages} & \textbf{Damages} &  \\
				  & & (\citet{cromar2022global}\tnote{$\beta$}, &  \\
                 & &  \citet{qiu2026wildfire}) &  \\
                \midrule	
\cite{acevedo2020effects}	& \$10 & \$37 & \$47 \\
\citet{kahn2021long}        & \$10 & \$37 & \$46 \\
\citet{casey2023projecting}	& \$14 & \$36 & \$50 \\
\citet{harding2025distribution}	& \$38 & \$35 & \$73 \\
\citet{nath2024}	        & \$65 & \$34 & \$99 \\
\midrule
 \\
                \textbf{Panel B} & \multicolumn{3}{c}{\textbf{U.S. impact-specific SC-\ce{CH4}}} \\
                & \multicolumn{3}{c}{(2030 emissions, 2\% near-term Ramsey} \\
                & \multicolumn{3}{c}{discount rate, 2020\$/mt\ce{CH4})} \\
                \cmidrule{2-4}
				  \textbf{Source of GDP-based} & \textbf{GDP-based Market} & \textbf{Health Nonmarket} & \textbf{Total\tnote{$\delta$}} \\
				   \textbf{Market Damages\tnote{$\alpha$}}& \textbf{Damages} & \textbf{Damages} &  \\
				  & & (\citet{cromar2022global}\tnote{$\beta$}, &  \\
                 & &  \citet{qiu2026wildfire}) &  \\
                \midrule	
\cite{acevedo2020effects}	& \$120 & \$530 & \$650 \\
\citet{kahn2021long}        & \$220 & \$530 & \$740 \\
\citet{casey2023projecting}	& \$170 & \$520 & \$690 \\
\citet{harding2025distribution}	& \$450 & \$510 & \$960 \\
\citet{nath2024}	        & \$900 & \$500 & \$1,400 \\
\midrule
 \\
                \textbf{Panel C} & \multicolumn{3}{c}{\textbf{U.S. impact-specific SC-\ce{N2O}}} \\
                & \multicolumn{3}{c}{(2030 emissions, 2\% near-term Ramsey} \\
                & \multicolumn{3}{c}{discount rate, 2020\$/mt\ce{N2O})} \\
                \cmidrule{2-4}
				  \textbf{Source of GDP-based} & \textbf{GDP-based Market} & \textbf{Health Nonmarket} & \textbf{Total\tnote{$\delta$}} \\
				   \textbf{Market Damages\tnote{$\alpha$}}& \textbf{Damages} & \textbf{Damages} &  \\
				  & & (\citet{cromar2022global}\tnote{$\beta$}, &  \\
                 & &  \citet{qiu2026wildfire}) &  \\
                \midrule	
\cite{acevedo2020effects}	& \$3,000  & \$11,000 & \$14,000 \\
\citet{kahn2021long}        & \$3,300  & \$11,000 & \$15,000 \\
\citet{casey2023projecting}	& \$4,100  & \$11,000 & \$5,000 \\
\citet{harding2025distribution}	& \$11,000  & \$11,000 & \$22,000 \\
\citet{nath2024}	        & \$20,000 & \$10,000 & \$30,000 \\
\bottomrule
            \end{tabular}
            \begin{tablenotes}
                \footnotesize
                \item Modeling was conducted in a modified \MimiGIVE{} framework: U.S. socioeconomic and emissions inputs are taken from RFF-SPs \citep{rennert2022social}; climate module temperature projections account for carbon feedback effects from permafrost thaw and Amazon dieback based on \citet{dietz2021economic} modeling of these feedbacks; U.S. temperature projections based on \FaIR{} 1.6.2 GMST projections are downscaled using a pattern-scaling approach (in which CONUS GCMs are paired with \FaIR{} parameter sets); and uses Ramsey-based discounting following \citet{newell2022a} updated to use U.S. growth rates and recalibrated $\rho$ and $\eta$ parameters. Modeling accounts for feedbacks from estimated GDP-based damages to monetization of health damages and to the discounting module. Due to sampling variation in the random draws of a Monte Carlo simulation, values presented in this table are rounded to the nearest dollar.
				\item[$\alpha$] Results shown in this table are based on each study's stated ``preferred'', ``main'', or ``central'' empirical specification of the temperature-GDP relationship and/or the method used to project climate damages. See \ref{app:sec:td_macro} for more details. 
                \item[$\beta$] Temperature-related health damages shown in this table are based on the mortality damage function in \GIVE{} \citep{rennert2022comprehensive} due to modeling constraints in incorporating other health damage functions (e.g., \citet{carleton2022valuing} from \DSCIM) into the \MimiGIVE{} framework.
                \item[$\delta$] Columns are individually rounded to two significant digits and, as such, may not add to be equal to the total column. Unrounded results are available in the paper's public repository. 
			\end{tablenotes}
	\end{threeparttable}
\end{table}

\begin{figure}[htbp]
    \begin{center}
        \caption{Response of gross domestic product (GDP) to a permanent marginal temperature (\SI{}{\celsius}) increase at U.S. average temperature in 2100 (\SI{16.3}{\celsius})}
        \label{fig:impulse_perm}
        \includegraphics[width=1.0\linewidth]{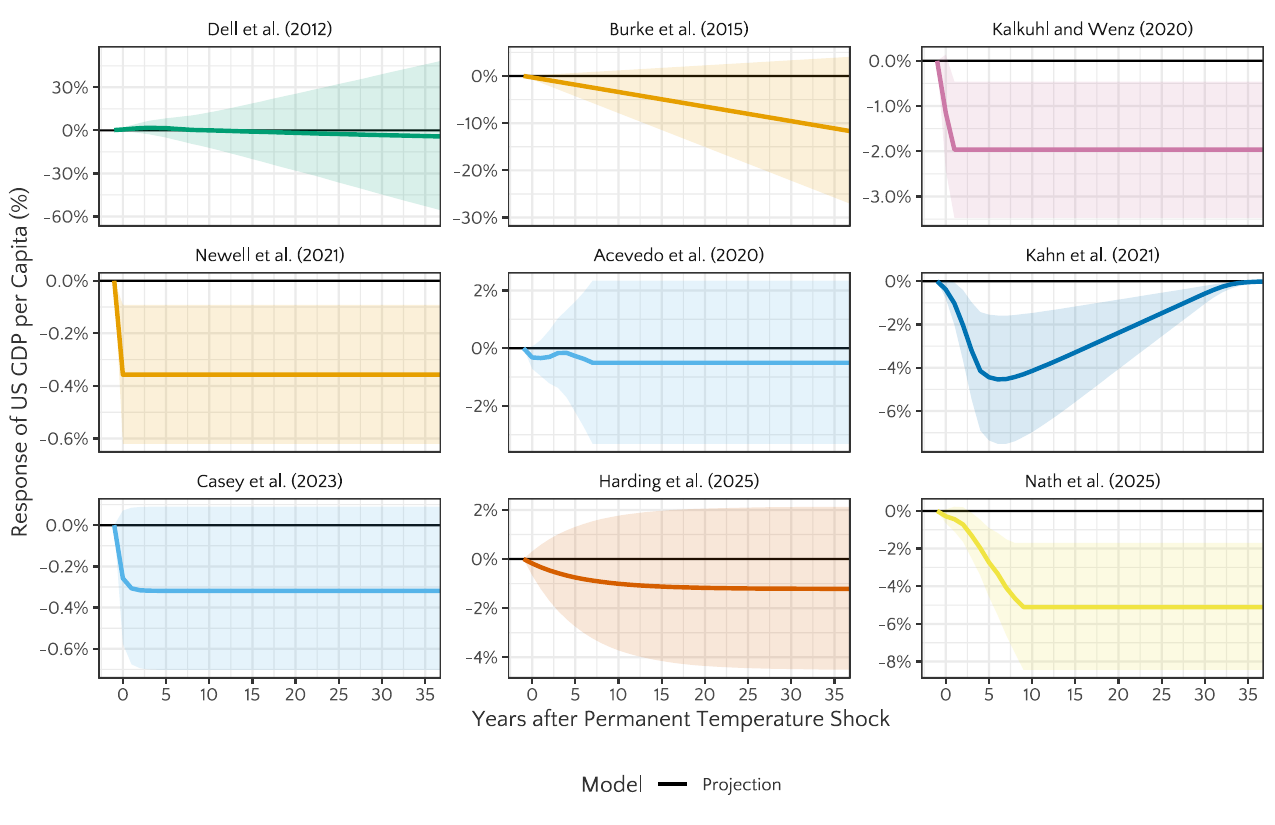}
    \end{center}
    \footnotesize
    This figure displays the estimated GDP
response to a permanent marginal shock in temperature. As in Figure \ref{fig:gdp_response_temp}, this figure traces out the estimated GDP-temperature response function over time based on the econometric results of the main specification of each paper (as described in Table \ref{tab:summary_macroecon}). The shaded ribbon represents the econometric uncertainty (95\% confidence interval) from the corresponding specification in the paper.
\end{figure}

\begin{figure}[htbp]
    \begin{center}
        \caption{Response of gross domestic product (GDP) to a temporary marginal temperature (\SI{}{\celsius}) increase at U.S. average temperature in 2100 (\SI{16.3}{\celsius}) for the dataset from each study}
        \label{fig:gdp_response_temp2}
        \includegraphics[width=1.0\linewidth]{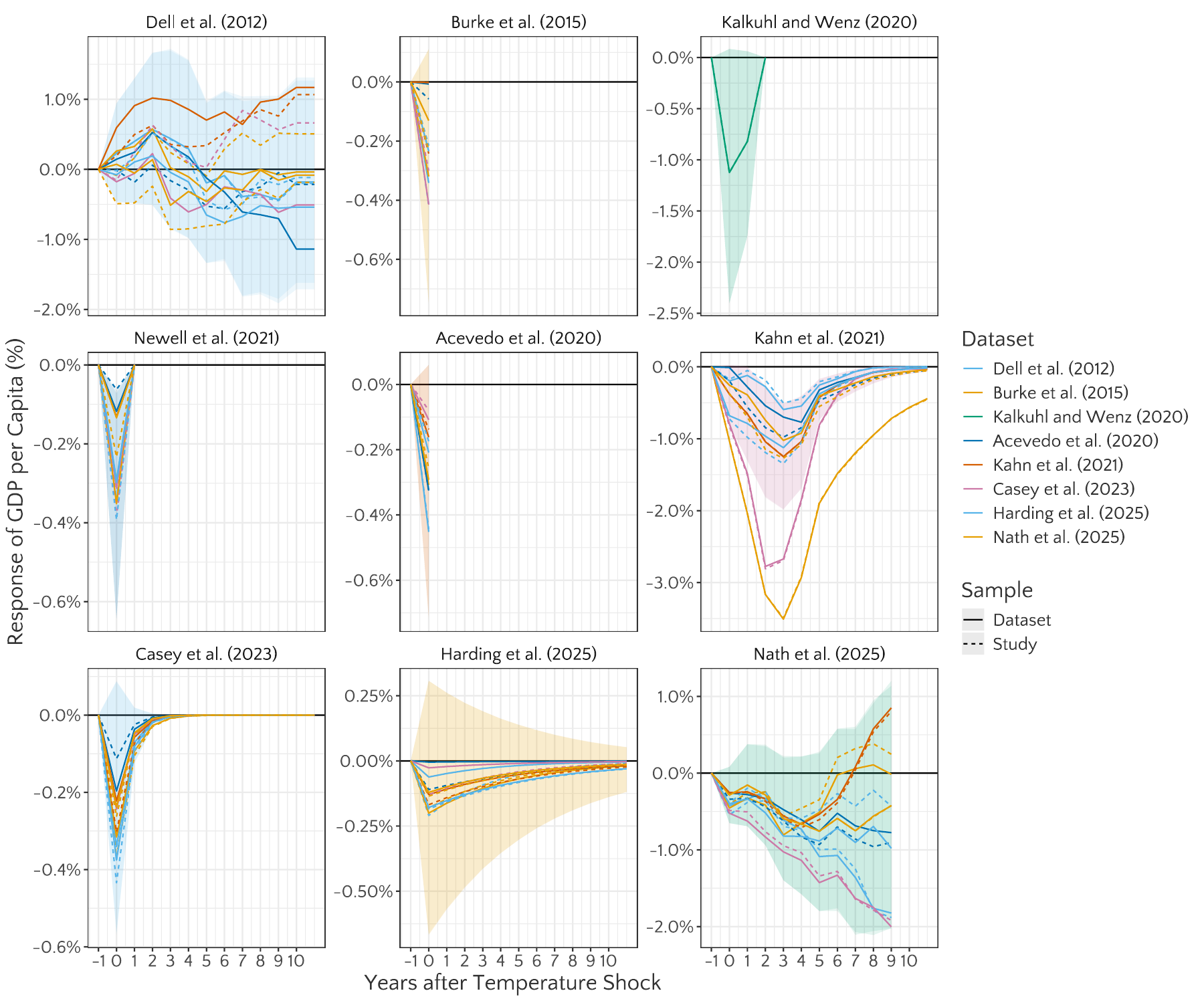}
    \end{center}
    \footnotesize
    This figure shows the estimated response from the main specification for each paper (as described in Table \ref{tab:summary_macroecon}) for the main specification of the labeled paper when using the underlying dataset from each of the other papers. The Dataset legend indicates the source of the underlying dataset used in the re-estimation. The Sample legend indicates whether all available data from the Dataset study was used (Dataset) or if the sample was restricted to match the countries and years covered in the labeled study's original dataset (Study).The shaded ribbon represents the econometric uncertainty (95\% confidence interval) from the main specification in the paper using the paper's original dataset.
\end{figure}

\begin{figure}[htbp]
    \begin{center}
        \caption{Response of gross domestic product (GDP) to a temporary marginal temperature (\SI{}{\celsius}) increase at U.S. average temperature in 2100 (\SI{16.3}{\celsius}) using \citet{nath2024} dataset}
        \label{fig:gdp_response_temp_nath}
        \includegraphics[width=1.0\linewidth]{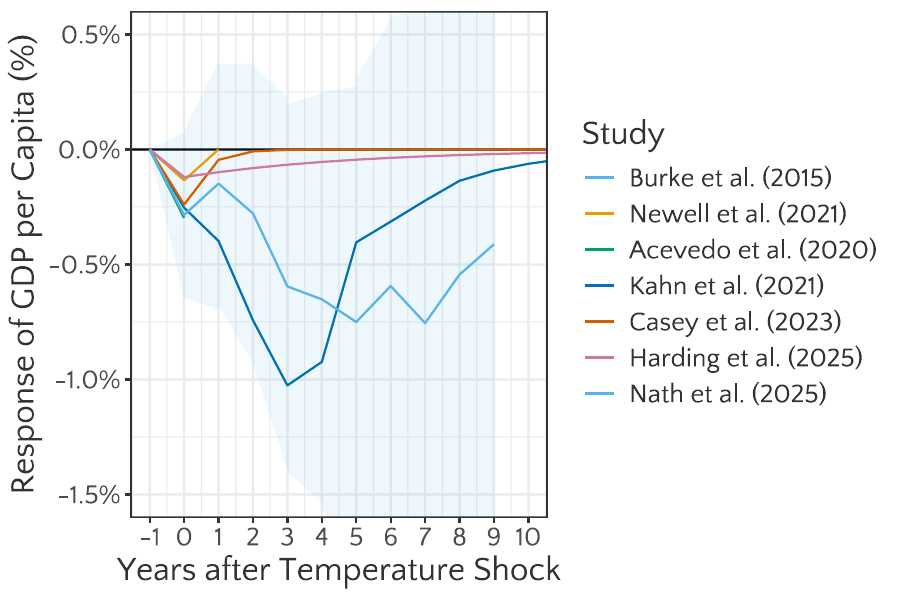}
    \end{center}
    \footnotesize
    This figure shows the response from the main specification for each paper (as described in Table \ref{tab:summary_macroecon}), but estimated on the \citet{nath2024} dataset. The shaded ribbon represents the econometric uncertainty (95\% confidence interval) from the corresponding specification in the paper.
\end{figure}

\begin{figure}[htbp]
    \begin{center}
        \caption{Average climate damages as a fraction of U.S. GDP in 2100 due to a change in annual global mean surface temperature under each of the 8 macroeconomic econometric damage functions}
        \label{fig:avg_macro_climate_damages}
        \includegraphics[width=1\linewidth]{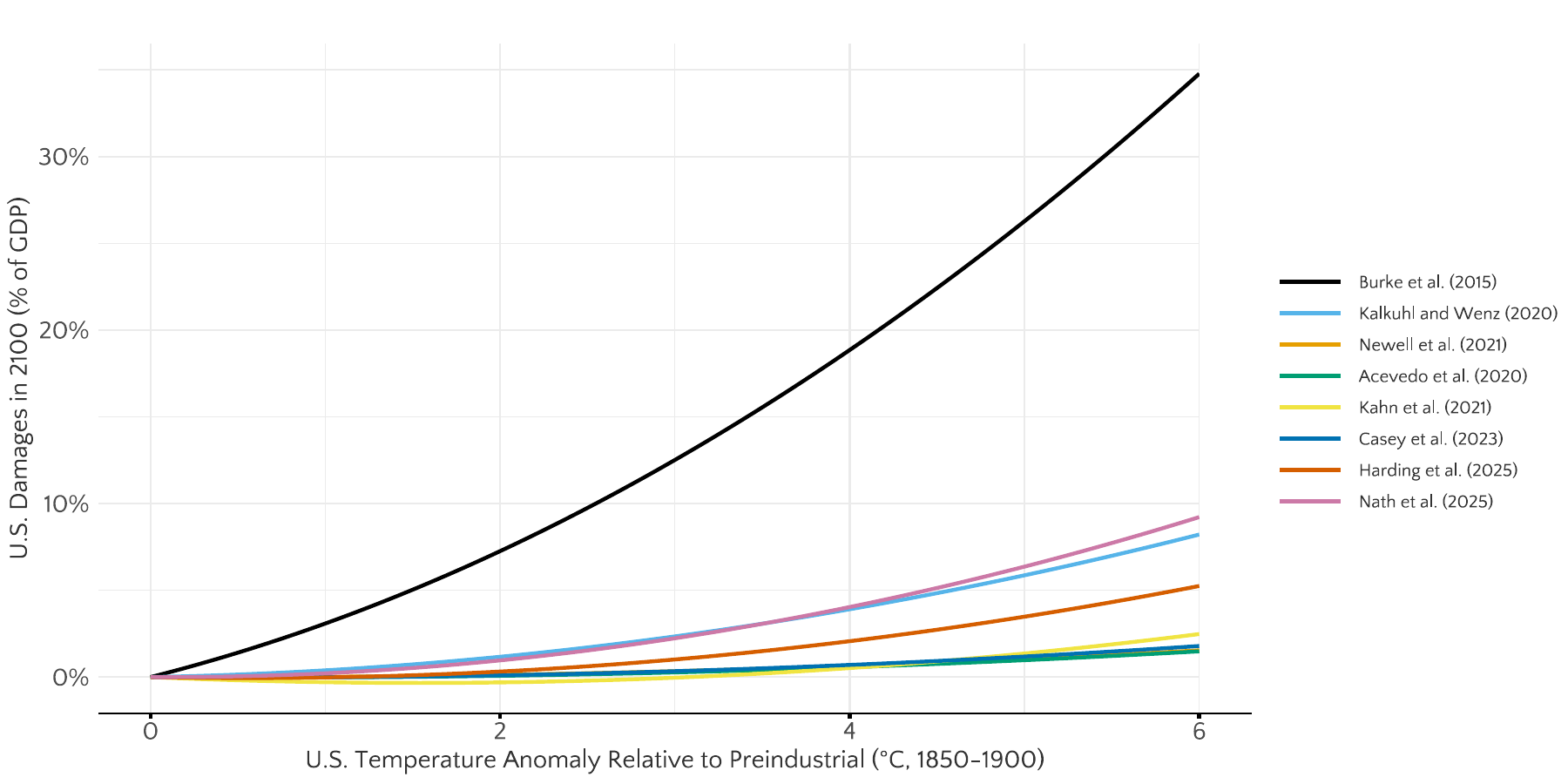}
    \end{center}
    \footnotesize
   Average (mean) GDP loss functions are generated using estimated damages in 2100 for the 10,000 Monte Carlo simulations and regressing on temperature and temperature squared (solid lines). This figure combines the mean lines from Figure \ref{fig:macro_climate_damages} into a single plot. The IPCC \citeyearpar{ipcc2021climate} notes that present day global mean surface temperatures in the year 2020 are around \SI{1.1}{\celsius} above preindustrial (1850-1900) levels. Estimated damages in 2100 shown in this figure are based on each study's stated ``preferred'', ``main'', or ``central'' empirical specification of the temperature-GDP relationship and/or the method used to project climate damages. Each damage function requires a measure of current climate or baseline temperature; in this figure we use country-level population-weighted mean temperatures from 1980 to 2010 drawn from \citet{burke2015global} for this purpose.
\end{figure}

\begin{figure}[htbp]
    \begin{center}
        \caption{Evidence on U.S. market damages from macroeconomic studies: sensitivity to underlying data used in empirical estimation}
        \label{fig:nath_dataset_comparison}
        \includegraphics[width=1\linewidth]{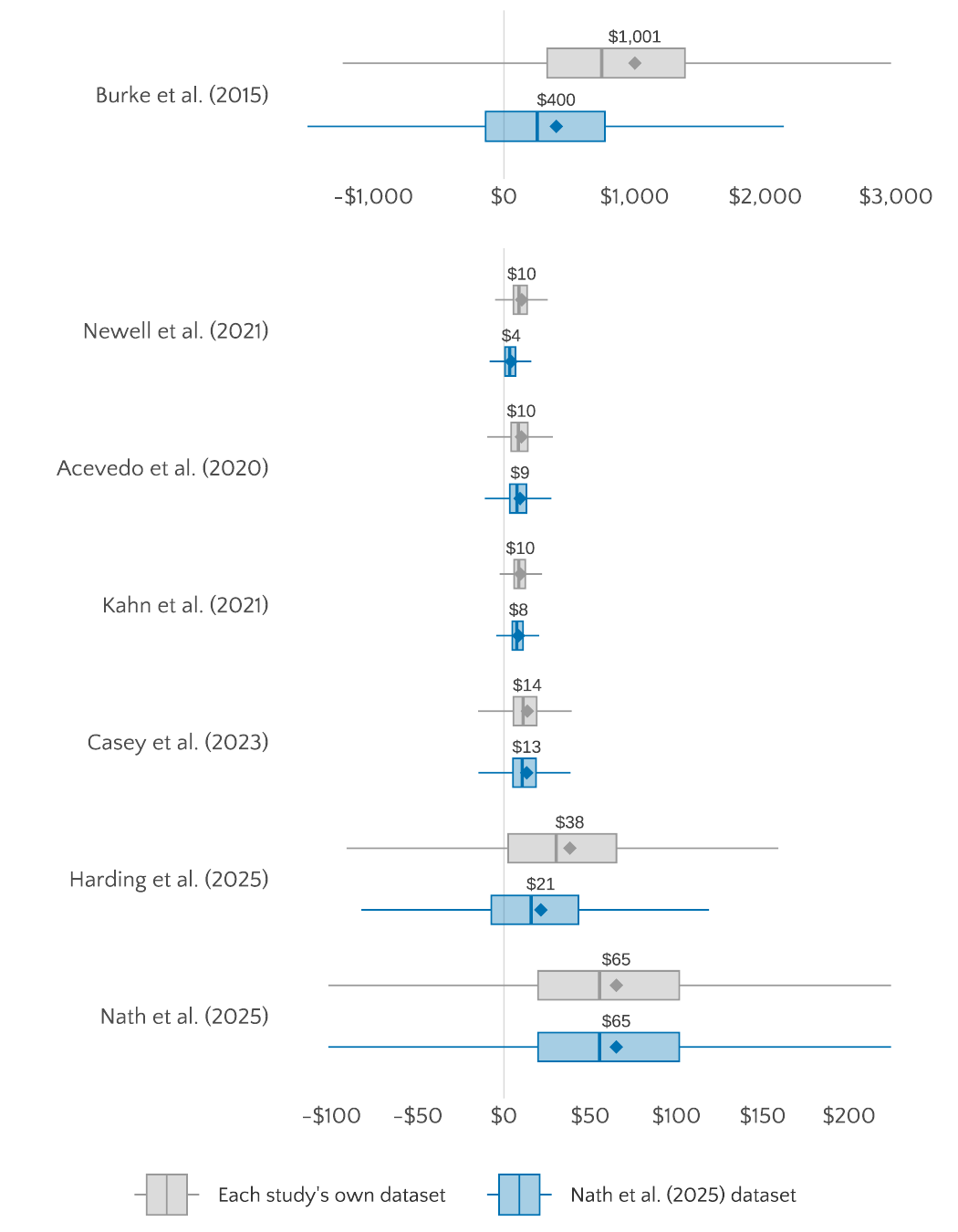}
    \end{center}
    \footnotesize
   This figure shows the mean U.S.-specific SC-\ce{CO2} (2030 emissions, 2 percent near-term Ramsey discount rate, 2020\$/t\ce{CO2}) and confidence interval of the discounted marginal damages under the main specification from each paper when re-estimated using the underlying dataset from \citep{nath2024}. 
\end{figure}

\newpage

\section{A Common Empirical Framework for Comparing the Macroeconomic Econometric Studies}
\label{app:sec:common_framework}

This Appendix illustrates that many specifications in macroeconomic econometric studies investigating the temperature-economic growth relationship can be written in, or related to, the autoregressive distributed-lag (ARDL) model. First, we introduce the common model and all notation. Second, discuss how the main specification from each paper listed in Table 1 is nested within this ARDL model based on how they transform the temperature variable, how they account for state dependence or nonlinearity, and how they treat persistence. 

\subsection{The common model}\label{sec:model}

Let the outcome $y$ be log GDP per capita and $T$ temperature. Write the first difference $\Delta y_{i,t} = y_{i,t} - y_{i,t-1}$, where $i$ indexes the panel unit and $t$ the period. Many specifications in the literature can be written in, or related to, the autoregressive distributed-lag (ARDL) form
\begin{equation}\label{eq:general}
    \Delta y_{i,t} =
        -\sum_{l=1}^{L} \lambda_{l}\, y_{i,t-l}
        + \sum_{l=0}^{M} \Big[ \beta_{1l}\, \tau_{i,t-l}
        + \beta_{2l}\, \tau_{i,t-l}\, S_{i,t-l} \Big]
        + \Omega_{it}
        + \epsilon_{i,t},
\end{equation}
where $L$ is the autoregressive order, $M$ is the distributed-lag order of the climate terms, and $\beta_{1l},\beta_{2l}$ are their coefficients. The climate regressors are built from two objects, $\tau$ and $S$. The term $\Omega_{it}$ represents a sum of controls that typically includes country and year fixed-effects.

\paragraph{The transformation $\tau$.} The temperature term enters as $\tau_{i,t-l} = T_{i,t-l} -
\That_{i,t-l}$, and studies differ in the benchmark $\That$:
\begin{itemize}
  \item $\That_{i,t-l} = 0$, $\tau$ is the level of temperature;
  \item $\That_{i,t-l} = T_{i,t-l-1}$, $\tau$ is the first difference;
  \item $\That_{i,t-l} = \Tbar_{i,t-l}$, a moving-average norm, $\tau$ is the deviation from that norm;
  \item $\That$ a fitted value from a filtering regression, $\tau$ is the residual, or shock.
\end{itemize}

\paragraph{The heterogeneity term $S$.} $S_{i,t-l}$ introduces state dependence or nonlinearity.
Setting $S = T$ gives a quadratic effect. Setting $S = \Tbar_i$, a fixed baseline such as a
country's mean climate, gives a state-dependent effect. A group
indicator gives group heterogeneity. Setting $S = 0$ gives an effect that is linear in $\tau$.

\paragraph{Dynamic Multiplier.} We compare studies through the dynamic multiplier on levels, the response of the level of the outcome $h$ periods ahead to a temperature shock today.
\begin{equation}\label{eq:psi}
    \Psi_h \;=\; \frac{\partial\, y_{i,t+h}}{\partial\, T_{i,t}}
\end{equation}

\noindent For studies that estimate the dynamic multipliers on growth, we estimate the dynamic multipliers on levels by summing over the dynamic multipliers on growth. 

\begin{equation}\label{eq:Psi}
    \Psi_h \;=\; \sum_{j=0}^h \psi_j \;=\; \sum_{j=0}^h \frac{\partial\, \Delta y_{i,t+j}}{\partial\, T_{i,t}}
\end{equation}

\noindent Because $\tau$ is itself a function of $T$, we define the multiplier with respect to the underlying temperature and differentiate through $\tau$. In the multipliers, dynamic terms may arise from the lags of temperature and autoregressive structure.

\subsection{The studies}\label{sec:studies}

The studies differ in how they transform the temperature variable, how they encode heterogeneity, and how they treat persistence. Below, we relate the specifications of each study to the common model (1) via the choices of $\That$ and $S$, lag and autoregressive orders, and the resulting multiplier on levels $\Psi_h$.

\paragraph{\cite{dell2012tempera}}: Here $\That = 0$, so temperature enters in levels, $S = D_i^{poor}$. With no autoregressive terms, \eqref{eq:general} is a distributed lag in $\tau$. 
\begin{equation}\label{eq:dell}
    \Delta y_{i,t} =
         \sum_{l=0}^{M} \left[ \beta_{1l}T_{i,t-l}  + \beta_{2l}T_{i,t-l}D_i^{poor}\right]
        + \Omega_{it}
        + \epsilon_{i,t},
\end{equation}
Because $y$ enters in first differences, a shock changes the growth rate, and the cumulative effect on the level of output for a rich country at horizon $h$ is the sum of the per-period marginal effects out to $h$:
\begin{equation}\label{eq:dell_mult}
    \Psi_h \;=\; \sum_{l=0}^{h}\beta_{1l},
    \qquad h = 0,1,\dots,M .
\end{equation}
The preferred specification, $M = 10$. The dynamic multiplier at $M=10$ is the long-run multiplier.  

\paragraph{\cite{burke2015global}}: Here $\That = 0$, so temperature enters in levels, $S = T$ gives a quadratic
effect, and there is no autoregressive term. 
\begin{equation}\label{eq:burke}
    \Delta y_{i,t} =
         \sum_{l=0}^{M} \Big[ \beta_{1l} T_{i,t-l}
        + \beta_{2l} T^2_{i,t-l} \Big]
        + \Omega_{it}
        + \epsilon_{i,t},
\end{equation}
The cumulative effect on the level of output at horizon $h$ is the sum of the per-period marginal effects out to $h$:
\begin{equation}\label{eq:burke_mult}
    \Psi_h \;=\; \sum_{l=0}^{h}\left(\beta_{1l} + 2\beta_{2l}T_{\text{US}}\right),
    \qquad h = 0,1,\dots,M .
\end{equation}
The preferred specification is contemporaneous, $M=0$, so $\Psi_h = \beta_10 + 2\beta_20 T_{\text{US}}$ for all $h$. This one-time growth effect leaves a permanent, parallel gap between the level of GDP per capita with and without a temporary shock.

\paragraph{\cite{newell2021the}}: Newell et al. show that whether temperature acts on the growth rate or the level of GDP is a restriction on the distributed-lag coefficients of the level of temperature: a level effect requires that they sum to zero. Under a level effect, the long-run growth effect of a shock is zero. Starting with equation \ref{eq:burke}, and applying the restriction on the coefficients the distributed lag can be written in first-differences:
\begin{equation}\label{eq:newell-fd}
    \Delta y_{i,t} \;=\; \sum_{l=0}^{M-1}\big[\alpha_{1l}\, \Delta T_{i,t-l}
    + \alpha_{2l}\, \Delta\big(T^2_{i,t-l}\big)\big] 
    + \Omega_{it}
    + \epsilon_{i,t},
\end{equation}
This is equivalent to a levels regression specified in first-differences. The differenced square is $\Delta(T^2)$, the change in temperature squared. For the preferred zero-lag quadratic the multiplier is the constant levels marginal effect,
\begin{equation}\label{eq:newell-psi}
    \Psi_h \;=\; \alpha_{10} + 2\alpha_{20}\, T_{\text{US}}, \qquad \text{only for } h=0 \text{, else } 0 .
\end{equation}
This is a genuine one-period level impact.

\paragraph{\cite{harding2025distribution}}: Similar to the Burke et al. specification, $\That = 0$, so temperature enters in levels, and $S = T$. Here, there is the addition of an autoregressive term.
\begin{equation}
    \Delta y_{i,t} =
        -\sum_{l=1}^{L} \lambda_{l}\, y_{i,t-l}
        + \sum_{l=0}^{M} \Big[ \beta_{1l}\, T_{i,t-l}
        + \beta_{2l} T^2_{i,t-l} \Big]
        + \Omega_{it}
        + \epsilon_{i,t},
\end{equation}
A single autoregressive term on the lagged income \emph{level} can function as a ``convergence" term, where $\lambda$ dictates the speed of convergence. With $M=0$ in the preferred specification, a shock has contemporaneous impact effect $\beta_{10} + 2\beta_{20} T$, after which convergence returns income toward its steady-state path, so
\begin{equation}\label{eq:harding-Psi}
    \Psi_h \;=\; \big(\beta_1 + 2\beta_2\, T_{\text{US}}\big)\,(1-\lambda)^{h}.
\end{equation}
The contribution of the paper is to explore the consequences of the convergence term, which bounds long-run damages of a temperature shock. Because convergence pulls growth back to its steady state, the resulting impact on the level of GDP per capita is persistent but not permanent.

\paragraph{\cite{kahn2021long}}: Here $\That = \Tbar_m$, an $m$-year moving-average norm, so temperature enters as a deviation from that norm, and $S = 0$, so the effect is linear in $\tau$. In \eqref{eq:general} the autoregressive terms act on lagged levels, $-\sum_l \lambda_l y_{i,t-l}$. If their coefficients sum to zero, then there exist equivalent autoregressive coefficients $\phi_1,\dots,\phi_{L-1}$ on lagged growth, which yields a standard Autoregressive Distributed Lag (ARDL) specification. 
\begin{equation}
    \Delta y_{i,t} =
        \sum_{l=1}^{L} \phi_{l}\, \Delta y_{i,t-l}
        + \sum_{l=0}^{M} \beta_{1l}\Delta \Big| \, T_{i,t-l} - \Tbar_m \Big |
        + \Omega_{it}
        + \epsilon_{i,t},
\end{equation}
The response follows the recursion
\begin{equation}\label{eq:kahn-psi}
    \Psi_0 = \beta_0, \qquad
    \Psi_h = \beta_{1h} + \sum_{l=1}^{\min(L,h)} \phi_l\, \Psi_{h-l}.
\end{equation}
Kahn et al.\ (their Eq.~1) build the deviation as $\tau = T - \Tbar_m$ with $\Tbar_m = \tfrac1m\sum_{\ell=1}^{m} T_{t-\ell}$, scaled by $2/(m+1)$. The preferred model sets $m=30$ and $L=M=4$.

\paragraph{\cite{acevedo2020effects}}: Here $\That = 0$, so temperature enters in levels, $S = T$, and the model is estimated by local projections. 
\begin{equation}\label{eq:acevedo}
    \Delta_h y_{i,t+h} =
         \phi_{1}\, \Delta y_{i,t-1}
        +  \beta_{1}^h\, T_{i,t}
        + \beta_{2}^h\, T^2_{i,t}
        + \Omega_{it}
        + \epsilon_{i,t},
\end{equation}
Local projections estimate the multiplier directly, running a separate regression at each horizon of the cumulative change $\Delta_h y_{i,t+h} = y_{i,t+h} - y_{i,t-1}$ on the period-$t$ temperature and controls, so the horizon profile is read off the horizon-specific coefficients rather than derived from a lag structure. Conditioned on one lag of the dependent variable, the directly estimated multipliers are:
\begin{equation}\label{eq:acevedo-psi}
    \Psi_h \;=\; \beta_1^{h} + 2\beta_2^{h}\, T_{\text{US}}, \qquad h = 0,\dots,7,
\end{equation}

\paragraph{\cite{nath2024}}: Here $\That$ is a fitted temperature from an autoregressive model of temperature, so $\tau$ is a filtered innovation, and $S = \Tbar_i$, the country-mean temperature. The model is estimated by local projections. 
\begin{equation}\label{eq:nath}
    \Delta_h y_{i,t+h} =
        \sum_{l=1}^{L} \phi_{l}\, \Delta y_{i,t-l}
        + \beta_{1}^h\, \tau_{i,t}
        + \beta_{2}^h\, \tau_{i,t}\, \Tbar_i 
        + \Omega_{it}
        + \epsilon_{i,t},
\end{equation}
Because the interaction is with a fixed baseline, the multiplier is linear in the state,
\begin{equation}\label{eq:nath-psi}
    \Psi_h \;=\; \beta_1^{h} + \beta_2^{h}\, \Tbar_{US}, \qquad h = 0,\dots,9,
\end{equation}

\paragraph{\cite{casey2023projecting}}: Here the outcome GDP per capita is replaced by total factor productivity calculated according to a simple aggregate production function. In the preferred specification, Casey et al. include terms with $\That = 0$ so that temperature enters in levels, and $\That = T_{t-1}$, so temperature enters in
first difference each with their own quadratic term through $S$. The specification can be thought of as a hybrid between the convergence equation of \cite{harding2025distribution} and the ARDL of \cite{kahn2021long}, with one autoregressive lag (in differences) rather than four. The preferred specification keeps only the level channel, and the inclusion of the AR term causes the contemporaneous effect $\beta_1 + 2\beta_2 T$ to decay geometrically,
\begin{equation}\label{eq:casey-psi}
    \Psi^{TFP}_h \;=\; \big(\beta_1 + 2\beta_2\, T_{\text{US}}\big)\,\phi_1^{h}.
\end{equation}

\paragraph{\cite{kalkuhl2020impact}}: Here $\That= T_{i,t-l}$ and $S = T_{i,t-1}$. The lags of the first difference in temperature is interacted with a single lag of temperature. No autoregressive terms are included, and the lag order parameter is $M=1$, so the temperature change and one lagged temperature change are included along with their interactions. In estimation, one temperature lag is included, but for projection it is excluded, so we exclude it here.

\begin{equation}\label{eq:kalkuhl}
    \Delta y_{i,t} =
        \sum_{l=0}^{M} \Big[ \beta_{1l}\, \Delta T_{i,t-l}
        + \beta_{2l}\, \Delta T_{i,t-l}\, T_{i,t-1} \Big]
        + \Omega_{it}
        + \epsilon_{i,t},
\end{equation}
This estimating equation is a hybrid between a pure-growth (temperature in levels) and a pure-levels (temperature in first differences), achieved through interaction terms between changes and levels of temperature. Due to this, recovering the dynamic multiplier requires carefully estimating the dynamic multipliers for each horizon. The dynamic multipliers on the growth rate are:   

\begin{equation}
    \begin{aligned}
        \psi_0 \;&=\; \beta_{10} + \beta_{20} T_{t-1} \\
        \psi_1 \;&=\; \beta_{11} + \beta_{21} T_{t}  -\beta_{10} - \beta_{20} T_{t} + \beta_{20} \Delta T_{t+1} + \beta_{21} \Delta T_{t}\\ 
        \psi_2 \;&=\; -\beta_{11} - \beta_{21} T_{t+1}
     \end{aligned}
\end{equation}
Summing over the the dynamic multipliers for the growth rates up to each horizon, the dynamic multipliers on levels are: 
\begin{equation}
    \begin{aligned}
        \Psi_0 \;&=\; \beta_{10} + \beta_{20} T_{t-1}   \\
        \Psi_1 \;&=\; \beta_{11} + \beta_{21} T_{t}  + \beta_{20} \left(\Delta T_{t+1} -  \Delta T_{t} \right ) + \beta_{21} \Delta T_{t} \\
        \Psi_2 \;&=\; \left(\beta_{20} - \beta_{21}\right) \left(\Delta T_{t+1} -  \Delta T_{t} \right )
    \end{aligned}
\end{equation}
Assuming a constant change in temperature $ \Delta T_{t+1} =  \Delta T_{t} $, including no change in temperature, then $\Psi_h = 0 \text{ for all } h>1$ and a temporary temperature shock has only a temporary impact on the level of GDP. To simplify the presentation in the figures we assume no change in the background temperature. ($\Delta T_{t}=0 \text{ for all } t$)

\newpage

\section{Technical Documentation for U.S. Socioeconomic and Climate Projections and U.S. Ramsey-based Discounting}
\label{app:sec:td_discounting}

\subsection{U.S. socioeconomic projections} 
\label{us_socio_projections}

\begin{figure}[htbp]
    \begin{center}
        \caption{Projections of U.S. population from the RFF-SPs and SSPs, 1950-2300}
        \label{fig:project_population}
        \includegraphics[width=1\linewidth]{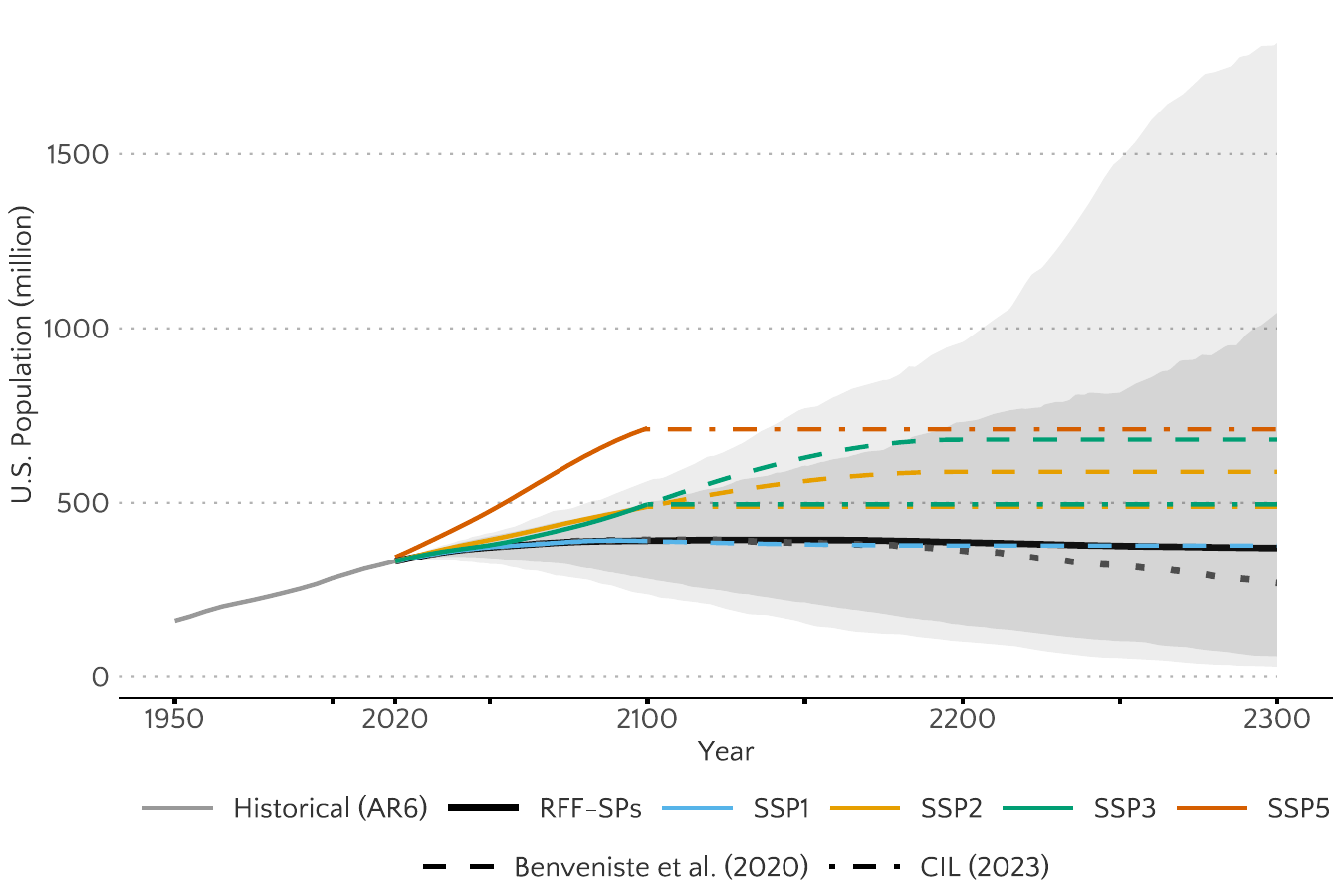}
    \end{center}
    \footnotesize
    RFF-SP probabilistic projections based on RFF-SPs \citep{rennert2022social}. Black lines represent the mean (solid) and median (dotted) along with the 5\textsuperscript{th} to 95\textsuperscript{th} (dark shade) and 1\textsuperscript{st} to 99\textsuperscript{th} (light shade) percentile ranges. Historical data from \citet{benveniste2020effect} using UN World Population Prospects 2019 \citepalias{un2019world}. SSP1, SSP2, and SSP3 data through 2100 from \citet{benveniste2020effect} using population growth rates from the International Institute for Applied Systems Analysis (IIASA) SSP Database \citep{riahi2017the}. SSP5 data through 2100 are from the IIASA database \citep{riahi2017the}. SSPs beyond 2100 (dashed) are based on two recent extrapolation methods: \citet{benveniste2020effect} and CIL \citeyearpar{cil2023dscim}.
\end{figure}

\begin{figure}[htbp]
    \begin{center}
        \caption{Projections of U.S. growth in income from the RFF-SPs and SSPs, 1950-2300}
        \label{fig:project_income}
        \centering
        \includegraphics[width=1\linewidth]{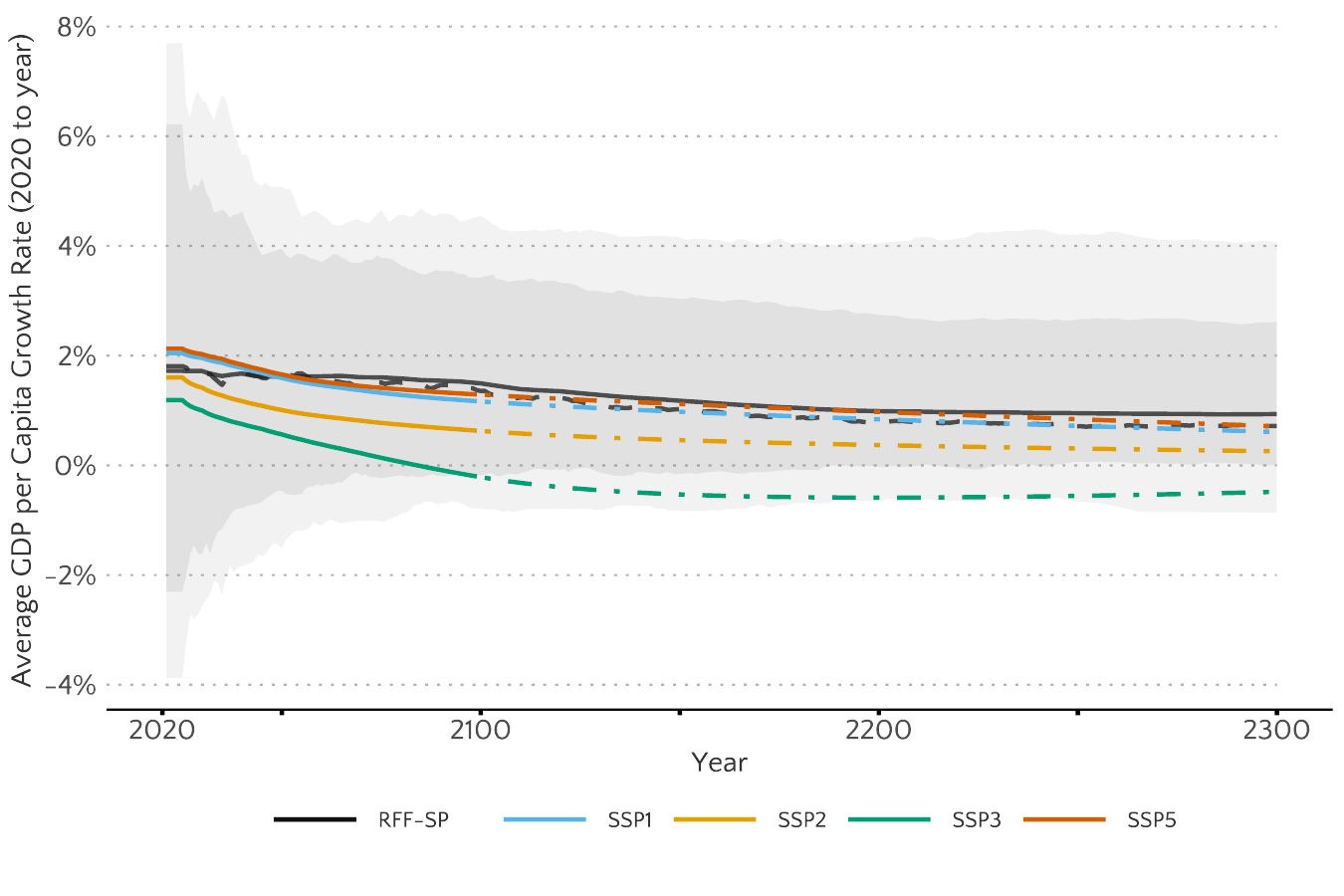}
    \end{center}
    \footnotesize
    RFF-SP probabilistic projections based on RFF-SPs \citep{rennert2022social}. Black lines represent the mean (solid) and median (dotted) along with the 5\textsuperscript{th} to 95\textsuperscript{th} (dark shade) and 1\textsuperscript{st} to 99\textsuperscript{th} (light shade) percentile ranges. Historical data from \citet{benveniste2020effect} using World Bank WDI \cite{wdi2024}. SSP data through 2100 from International Institute for Applied Systems Analysis (IIASA) SSP Database's OECD Env-Growth model \citep{riahi2017the}. SSPs beyond 2100 (dashed) are based on two recent extrapolation methods: \citet{benveniste2020effect} and CIL \citeyearpar{cil2023dscim}. The growth rates (and the uncertainty bounds around the RFF-SPs) are plotted in a time-averaged manner to accurately present the underlying year-on-year correlations that exist within each scenario.
\end{figure}

Figure \ref{fig:project_population} and Figure \ref{fig:project_income} present the RFF-SP probabilistic projections of population and economic growth for the U.S. through the year 2300. These figures also include a comparison to the deterministic projections from the database of Shared Socioeconomic Pathways (SSPs) for years 2020 to 2100. The SSPs have been used in IPCC reports and other applications.\footnote{Figures \ref{fig:project_population} and \ref{fig:project_income} contain all Tier 1 SSPs from IPCC AR6. Tier 2 scenarios, such as SSP4, were not considered.} The SSP projections presented in the figures for years beyond 2100 are based on two extrapolation methods recently used in the literature—\citet{benveniste2020effect} for SSP1, SSP2, and SSP3 (dashed lines), and CIL \citeyear{cil2023dscim} for SSP2, SSP3, and SSP5 (dashed-dotted lines)---illustrating the sensitivity to various extrapolation assumptions.

Figure \ref{fig:project_population} presents the mean (black solid line) and median (black dotted line) of the RFF-SP population projections, which follow a fairly flat trajectory through 2100, in-line with SSP1, peaking at 392 million people in the year 2150. This is followed by a slow decline to under 370 million by 2300. SSP2 and SSP3 follow the upper tail of the RFF-SPs through 2100 and, depending on the extrapolation method, drop within the 99\textsuperscript{th} or 95\textsuperscript{th} percentile of the RFF-SP distribution by 2300. While the SSP-based projections shown in Figure \ref{fig:project_population} generally fall within or near the range of the RFF-SP probabilistic distribution for U.S. population, they are limited in providing a comparison to the full RFF-SP distribution. The SSPs were intentionally developed to reflect a range of reasonably likely scenarios corresponding to different storylines rather than a more comprehensive range of plausible scenarios like the RFF-SPs. Furthermore, the SSP-based projections are sensitive to the extrapolation method used. For example, the SSP3 projections displayed in Figure \ref{fig:project_population} show U.S. population in 2300 rising to about 500 million under the CIL \citeyear{cil2023dscim} extrapolation, and 680 million under the \citet{benveniste2020effect} extrapolation. 

Figure \ref{fig:project_income} presents the mean (black solid line) and median (black dashed line) U.S. economic growth projections from the RFF-SPs along with comparisons to the SSPs in AR6.\footnote{The growth rates (and the uncertainty bounds around the RFF-SPs) shown in Figure \ref{fig:project_income} are plotted in a time-averaged manner to accurately present the underlying year-on-year correlations that exist within each scenario/storyline.} The mean (black solid line) economic growth rates start at 1.7 percent in 2021, slowly decline to 1.5 percent between 2030 and 2100 and then continue to decline through-out the next century. The mean economic growth rate levels off again after 2200 at 1 percent. The RFF-SP U.S. economic growth projections are lower but most consistent with SSP1 and SSP5 scenarios. All the SSP-based projections displayed in Figure \ref{fig:project_income} lie within the long-run RFF-SP distribution. 

\subsection{U.S. temperature projections} 
\label{us_temp_projections}

The damage module of many climate change integrated assessment models (IAMs) takes global mean surface temperature (GMST) as an input to translate changes in climate into changes in physical impacts, such as sea level rise (SLR) and, then, to associated monetized damages, or in some cases directly into monetized economic outcomes (e.g., changes in GDP). Other damage functions, however, are estimated using measures of local mean surface temperature (LMST). In this section we describe our process for transforming GMST into LMST for the U.S., and in \ref{app:sec:td_feedback} we present the resulting U.S. temperature paths underlying our estimations.

Reduced complexity climate models, such as the \FaIR{} model, typically only provide GMST and other spatially aggregated climate variables (e.g., global ocean pH, or global SLR) as outputs. More computationally intensive earth system models (ESMs) or GCMs can provide more spatially resolute climate projections, such as LMST, but are less suited to fit the needs of climate IAMs that use many, sometimes tens of thousands, of possible emissions scenarios in a probabilistic setting. However, approximating LMST from GMST can be done using a pattern-scaling approach that pairs the spatial patterns of ESMs/GCMs with the GMST output from reduced complexity models \citetext{\citealp{kravitz2017exploring,lynch2017an,kravitz2022pangeo}; \citetalias{epa2024pattern}}.\footnote{Another possible approach is that of \citet{tebaldi2020emulating}. They show that a time-shift approach to emulating earth futures and climate indices could, in some cases, outperform that of simple pattern scaling.}

The \FaIR{} model provides explicit representation of many uncertainties, including effective heat capacities of the surface and deep ocean layers, deep ocean heat uptake efficacy, the integrated airborne fraction of \ce{CO2} over the pre-industrial period, the strength of different forcing agents, a climate feedback term, and other key model parameters \citep{millar2017a,smith2018fair}. From over 1 million initial pairings of 44 random parameters in the model, \FaIR{} developers selected 2,237 unique parameter set pairings through a model calibration process as part of the IPCC AR6 \citeyearpar{ipcc2021climate,ipcc2021earth}. These parameter sets, each consisting of a unique set of values for the 44 random parameters, are calibrated to match historical observations, future GCM responses, and IPCC determined probabilities for transient climate response (TCR) and equilibrium climate sensitivity (ECS) model parameters. TCR and ECS are emergent properties in \FaIR{} and are not directly recoverable from the model. 

To recover the spatial patterns of LMST from ESMs/GCMs with GMST, we pair patterns from ESMs/GCMs with \FaIR{} parameter sets that are more similar in their climate response. That is, a hotter ESM/GCM pattern is paired with a hotter \FaIR{} parameter set, and vice versa. Because TCR and ECS are not directly recoverable from \FaIR{}, one way to pair the spatial patterns of ESMs/GCMs with \FaIR{} parameter sets is to examine the relative hotness of the \FaIR{} model runs with the relative hotness of the ESMs/GCMs. The relative hotness of the 2,237 \FaIR{} parameter sets were ranked based on the projected GMST in 2100 that resulted from running the model using SSP2-RCP4.5 storyline scenario (SSP245). 

To rank the relative ``hotness'' of ESMs/GCMs, we use the TCR and ECS of the models directly. \citet{tokarska2020past} provide an overview of the TCR and ECS underlying each of the ESMs/GCMs considered in CMIP6 (Table \ref{tab:rank_gcm_us}). Because both the TCR and ECS contribute to the model's response and relative hotness, we create a combined index to rank the set of models. This is done by summing the TCR and ECS and weighting the sum by the mean of each parameter. This results in a weighted sum of the TCR and ECS such that both are given an equal weight in the ranking. A higher number suggest a hotter model, and vice versa. Of the 21 available GCMs, 18 were evaluated in \citet{tokarska2020past}). US EPA's Climate Science and Impacts Branch (CSIB) has also examined the set of CMIP6 ESM/GCM models and identified 5 for use as a minimum set based on three criteria: (1) matching historic temperature and precipitation metrics either globally \citep{brunner2020reduced} or for the CONUS/North America region \citep{ashfaq2022evaluation,zhang2024evaluation}; (2) independence from the other selected models \citep{brunner2020reduced,zhang2024evaluation}; and (3) availability in the LOCA and STAR-ESDM statistical downscaling datasets. Of the five selected by EPA-CSIB, four are available in the model and summarized in Table \ref{tab:rank_gcm_us}.  

\begin{figure}[htbp]
    \caption{The 4 GCMs U.S. CONUS climate projections ranked by GMST in 2100}
    \label{fig:rank_gcm_us}
    \centering
    \includegraphics[width=0.95\linewidth]{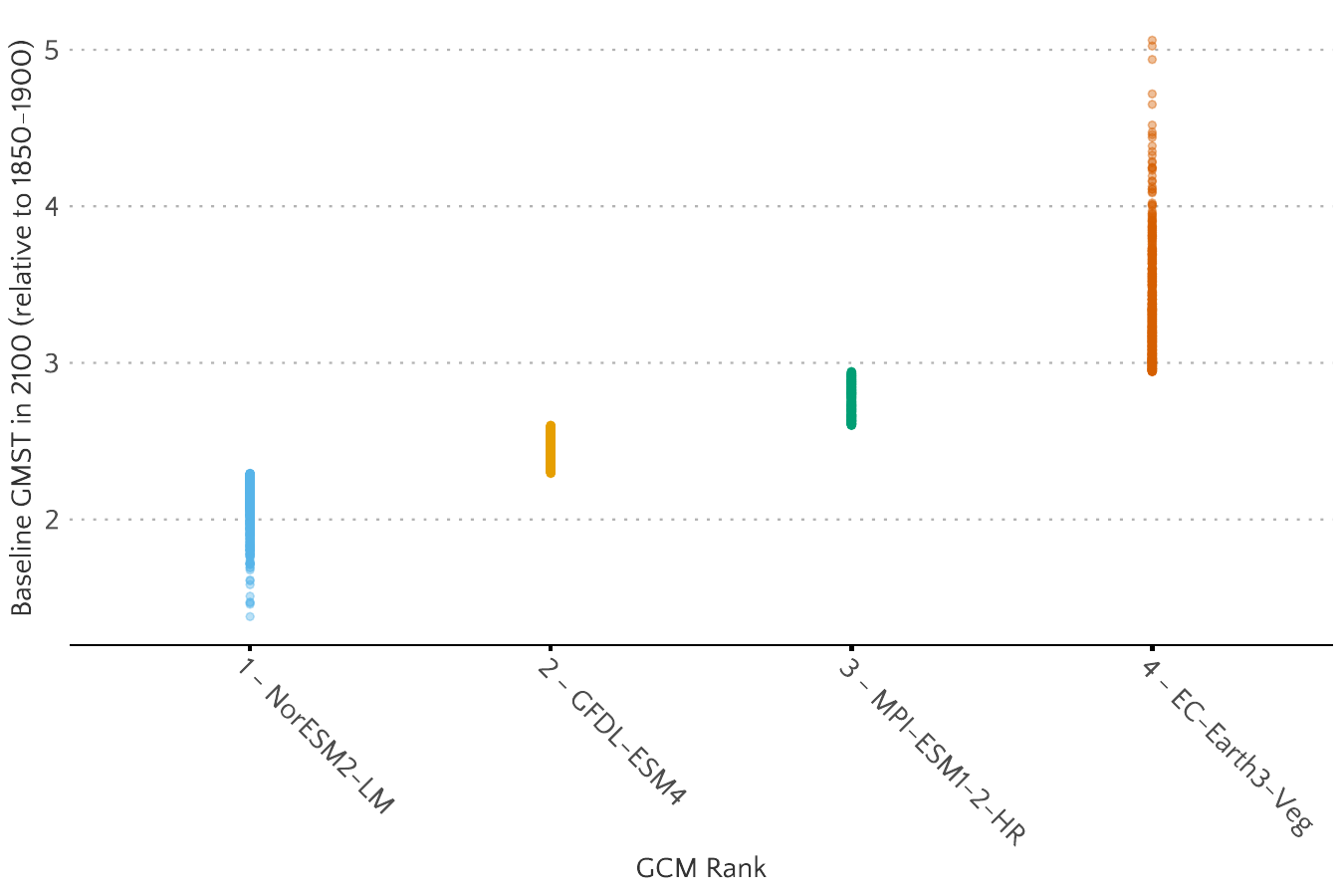}
\end{figure}

\begin{figure}[htbp]
    \caption{Baseline GMST from \FaIR{} under SSP245 and corresponding rankings for 4 CONUS GCMs}
    \label{fig:rank_baseline_gcm_us}
    \centering
    \includegraphics[width=0.95\linewidth]{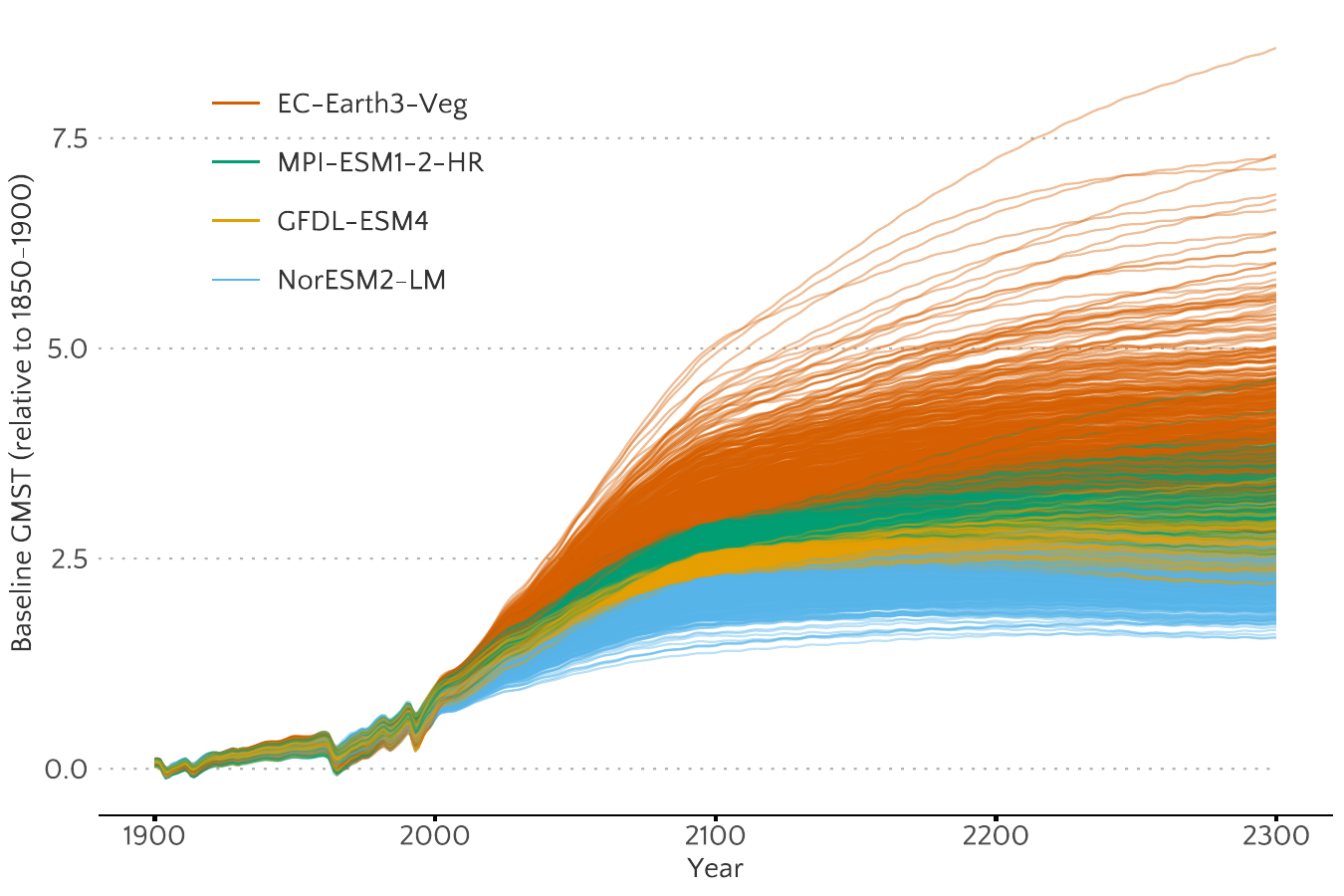}
\end{figure}

Pairing \FaIR{} parameter sets with ESM/GCM patterns was done by simply pairing the relative hotness of each. That is, the 2,237 \FaIR{} parameter sets were ranked into groups from cool to hot based on their resulting GMST in the year 2100. These groups were then paired with the ESM/GCMs based on their ranked weighted sum, from 1 to 4 (Figure \ref{fig:rank_gcm_us}). That is, the 2,237 \FaIR{} trials were ranked into 4 uniformly sized groups (i.e., 559 in the three hottest, and 560 in the coolest group). These relationships are presented in Figures \ref{fig:rank_gcm_us} and \ref{fig:rank_baseline_gcm_us}, showing the full time-series of GMST under SSP245 and colored according to their paired ESM/GCM. 


\begin{table}[htbp]
    \caption{Ranking of GCMs based on the minimum set for U.S.}
    \label{tab:rank_gcm_us}
    \centering
    \begin{threeparttable}
        \begin{tabular}{lcccccc}
            \toprule \toprule
            \textbf{Model (ESM/GCM)} & \textbf{TCR} & \textbf{ECS} & \textbf{Weighted} & \textbf{TCR} & \textbf{ECS} & \textbf{Weighted} \\
            & & & \textbf{Sum} & \textbf{Rank} & \textbf{Rank} & \textbf{Sum Rank} \\
            \midrule	
NorESM2-LM & 1.48 & 2.6 & 1.43 & 1 & 1 & 1 \\
GFDL-ESM4 & 1.61 & 2.62 & 1.5 & 2 & 2 & 2 \\
MPI-ESM1-2-HR & 1.65 & 2.97 & 1.62 & 3 & 3 & 3 \\
EC-Earth3-Veg & 2.61 & 4.3 & 2.45 & 4 & 4 & 4 \\
\bottomrule
        \end{tabular}
        \begin{tablenotes} 		
            \footnotesize
            \item Data on TCR and ECS comes from \citet{tokarska2020past}. Weighted sum is created by weighting the TCR and ECS by their means. Ranking of 1 denotes relatively ``cooler'', while a ranking of 4 means relatively ``hotter''. 
        \end{tablenotes}
	\end{threeparttable}
\end{table}


\newpage

\subsection{Validation of the pattern scaling approach}
\label{sec:pattern_scaling}

\begin{figure}[htbp]
    \begin{center}
        \caption{Historical Global Mean Surface Temperature from \FaIR{} 1.6.2 and Berkeley Earth}
        \label{fig:hist_gmst}
        \includegraphics[width=1\linewidth]{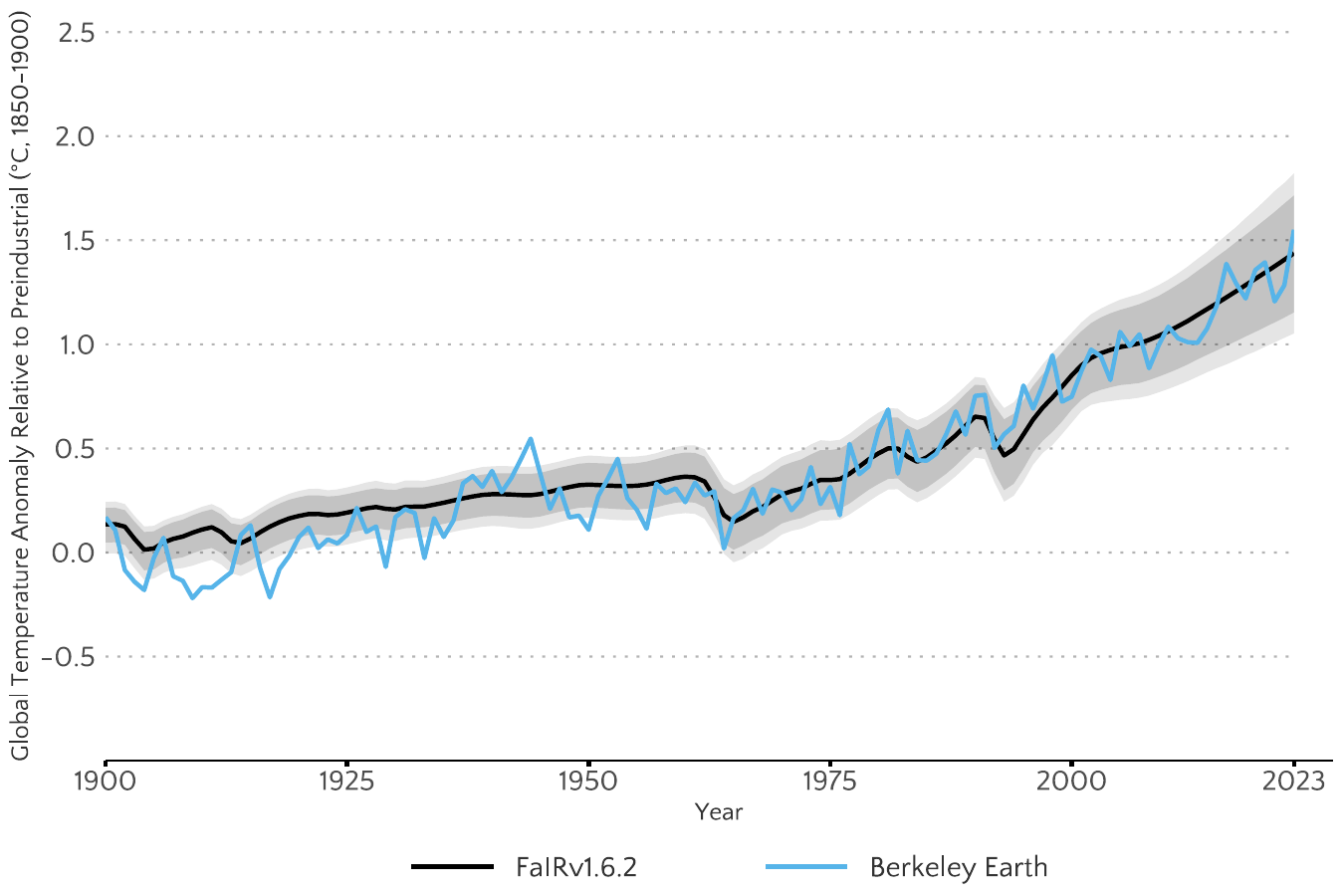}
    \end{center}
    \footnotesize
    Black lines represent mean (solid) GMST from \FaIR{} 1.6.2 along with 5\textsuperscript{th} to 95\textsuperscript{th} (dark shade) and 1\textsuperscript{st} to 99\textsuperscript{th} (light shade) percentile ranges that are a result of the uncertainty underlying \FaIR{} 1.6.2. Blue line represents the historical record from Berkeley Earth.
\end{figure}

\begin{figure}[htbp]
    \begin{center}
        \caption{Historical U.S. mean surface temperature from patterned GMST (\FaIR{} 1.6.2) with CONUS patterns and observations from Berkeley Earth}
        \label{fig:hist_gmst_us}
        \includegraphics[width=1\linewidth]{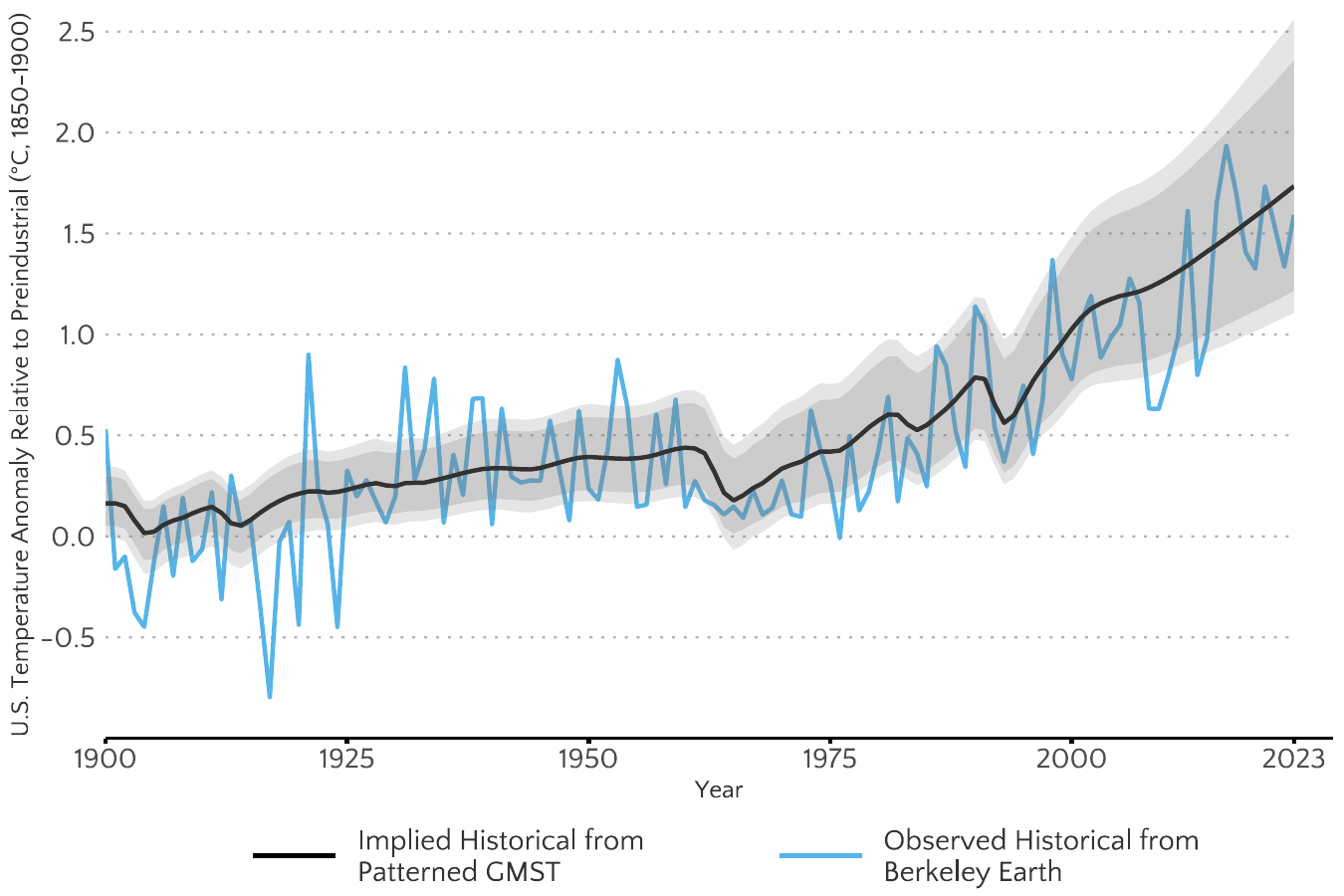}
    \end{center}
    \footnotesize
    Black lines represent mean (solid) U.S. mean surface temperature from combining \FaIR{} 1.6.2 GMST and CONUS population-weighted patterns, along with 5\textsuperscript{th} to 95\textsuperscript{th} (dark shade) and 1\textsuperscript{st} to 99\textsuperscript{th} (light shade) percentile ranges that are a result of the uncertainty underlying \FaIR{} 1.6.2. Blue line represents the population-weighted U.S. historical record from Berkeley Earth.
\end{figure}

The pattern scaling approach described above is used to translate the \FaIR{} model's annual GMST projections to annual average U.S. temperature. To explore how closely the patterns take historical observations of GMST and translate them into historical observations of U.S. mean surface temperature, we use a time-series of global gridded data from Berkeley Earth \citep{berkeley2024global}. Figure \ref{fig:hist_gmst} compares the historical GMST observation from this dataset against the GMST coming from \FaIR{} 1.6.2. Figure \ref{fig:hist_gmst_us} compares the result of using the CONUS patterns discussed above to recover U.S. mean surface temperature from \FaIR{} against the observed historical record from the gridded Berkeley Earth dataset. Both show that the GMST and U.S. temperatures closely match the historical record, suggesting the pattern scaling approach used in this paper reasonably recovers U.S. temperatures from GMST as estimated by \FaIR{} 1.6.2. In \ref{app:sec:td_feedback}, we present the result of our pattern scaling methods on U.S. temperature projections through the year 2300 (Figure \ref{fig:usmst_wo_feedbacks}). 

\subsection{U.S. Ramsey-based discounting}

Following the approach used in EPA \citeyearpar{epa2023report}, the discounting module in this paper uses stochastic discounting based on the Ramsey equation \citep{ramsey1982a}:\footnote{A detailed discussion of the discounting methodology can be found in EPA \citeyearpar{epa2023report}.}

\begin{equation}
    \label{eq:eq_ramsey}
    r_{t} = \rho + \eta g_{t}
\end{equation}

where $r_{t}$ is the consumption discount rate in year t, $\rho$ is the pure rate of time preference, $\eta$ is the elasticity of marginal utility of consumption, and $g_{t}$ is the growth rate of consumption in year $t$. As explained in EPA \citeyearpar{epa2023report}, this approach provides internal consistency within the modeling and a more complete accounting of uncertainty consistent with economic theory (\citealt{arrow2013determining}, \citealt{cropper2014declining}). The approach also follows the National Academies' \citepalias{national2017valuing} recommendation to employ a more structural, Ramsey-like approach to discounting that explicitly recognizes the relationship between economic growth and discounting uncertainty. 

Ramsey-based discounting implies that future states of the world with lower consumption will be discounted less than states of the world with higher consumption. The use of the Ramsey formula in our IAM framework also appropriately addresses consumption risk. That is, the discount rate varies over time consistent with changes in estimated economic growth and, therefore, explicitly accounts for the correlation between consumption growth and climate change damages when discounting.\footnote{The covariance between marginal climate change damages and future consumption growth is often referred to as the climate beta \citep[See][for a discussion]{dietz2018climate,gollier2014}. The climate beta can similarly be viewed as the covariance between returns to investments in climate mitigation and consumption growth. Because each Monte Carlo trial has a separate discount rate path that varies with consumption growth, the climate beta is internally determined within our IAM framework, and therefore, our discounting approach is also internally risk-adjusted.}

The parameters of the Ramsey equation, $\rho$ and $\eta$, are calibrated following the approach of \citet{newell2022a}, so that: (1) the decline in the certainty-equivalent\footnote{The certainty-equivalent discount rate is the discount rate that corresponds to the expectation of uncertain discount factors.} discount rate matches the latest empirical evidence on interest rate uncertainty estimated by \citet{bauer2020interest,bauer2023rising}, and (2) the average of the certainty-equivalent discount rate over the first decade matches a near-term consumption rate of interest. Consistent with EPA \citeyearpar{epa2023report}, we use a 2 percent near-term target rate for this calibration. This is based on multiple lines of evidence on observed the average real return on 10-Year Treasury securities and academic research on real interest rates, which support a risk-free 2 percent rate when discounting climate damages 
\citetext{\citealp{bauer2020interest,bauer2023rising}; \citetalias{cbo2022long,cbo2023long}; \citealp{drupp2018discount,giglio2014very,giglio2021,howard2020wisdom,pindyck2019social}}. See EPA \citeyearpar{epa2023report} for a more detailed discussion of these lines of evidence. For this U.S.-focused paper, we make two minor modifications to the implementation of the \citet{newell2022a} approach used in recent SC-GHG studies \citetext{\citealp{rennert2022comprehensive,rennert2022social}; \citetalias{epa2023report}}: (1) we use U.S. economic growth projections from the RFF-SPs, as opposed to global economic growth projections, in the calibration of $\rho$ and $\eta$, which yields $\rho$=0.41\% and $\eta$=1.02,\footnote{In \citet{rennert2022comprehensive,rennert2022social} and EPA \citeyearpar{epa2023report}, parameter calibrations for a 2 percent near-term target discount rate following this approach but using global consumption growth rates are $\rho$=0.20\% and $\eta$=1.24.} and (2) when discounting climate damages we assume a U.S. representative agent and, therefore, $g_{t}$ in the Ramsey formula is a measure of the U.S. consumption growth rate (net of baseline climate damages to the U.S.) instead of the global consumption growth rate (net of baseline climate damages globally). 

\newpage

\section{Technical Documentation for Estimation of Carbon Feedback Effects}
\label{app:sec:td_feedback}

In this paper, the climate module's temperature projections incorporate two additional carbon feedback effects that are not accounted for in the \FaIR{} 1.6.2 climate model: permafrost thaw and the dieback of the Amazon rainforest. These two feedbacks have been identified in recent reviews of Earth system feedbacks and tipping points as ones not yet represented in \FaIR{} and which have sufficient scientific and methodological basis to be reliably incorporated in our modeling framework at this time \citetext{\citealp{dietz2021economic}; \citetalias{rff2024increasing}}. RFF \citeyearpar{rff2024increasing} assessed the eight feedbacks included in \citet{dietz2021economic} and three others identified by \citet{wang2023mechanisms}. Of these eleven, three are already accounted for in the modeling of temperature in \FaIR{} 1.6.2 and associated SLR in the enumerative damage models described in \ref{app:sec:add_fig_tab}: (1) the rate of disintegration of the Antarctic ice sheet, (2) the rate of disintegration of the Greenland ice sheet, and (3) the surface albedo feedback. RFF \citeyearpar{rff2024increasing} find five additional feedback systems require more physical science and/or economic research to permit an adequate representation in the SC-GHG modeling framework. These include: (1) the slowdown of the Atlantic Meridional Overturning Circulation (AMOC), (2) the release of ocean methane hydrates, (3) shifts in boreal ecosystems, (4) the breakup of the low-latitude stratocumulus cloud deck, and (5) the die-off of tropical coral reefs. It is therefore important to note that our carbon feedback module is only a partial representation of the magnitude of U.S.-specific impacts from earth system feedbacks and tipping points identified in the academic literatures. For example, climate modeling indicates that one of the many global and regional consequences of a weakened AMOC is accelerated SLR on the eastern coast of North America, which would impact highly populated cities of the United States \citep{wang2023mechanisms}.  

\begin{figure}[htbp]
    \begin{center}
        \caption{Global mean surface temperature anomaly with and without additional carbon feedbacks}
        \label{fig:gmst_wo_feedbacks}
        \includegraphics[width=1\linewidth]{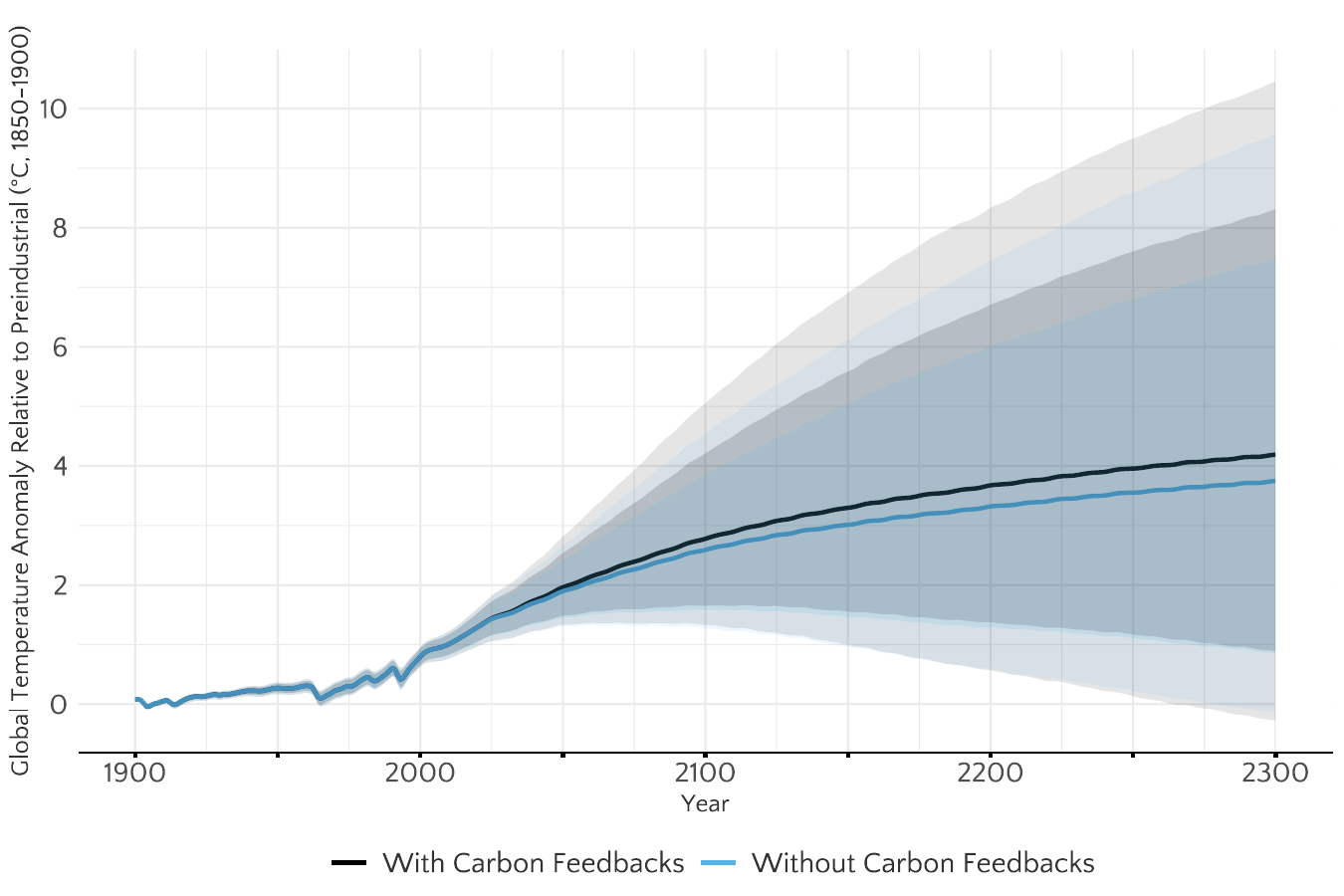}
    \end{center}
    \footnotesize
    Global mean surface temperature anomaly (GMST) (solid line) from \FaIR{} 1.6.2, along with 5\textsuperscript{th} to 95\textsuperscript{th} (dark shade) and 1\textsuperscript{st} to 99\textsuperscript{th} (light shade) percentile ranges that are a result of the joint uncertainty underlying \FaIR{} 1.6.2 and the RFF-SPs.
\end{figure}

\begin{figure}[htbp]
    \begin{center}
        \caption{U.S. mean surface temperature anomaly with and without additional carbon feedbacks}
        \label{fig:usmst_wo_feedbacks}
        \includegraphics[width=1\linewidth]{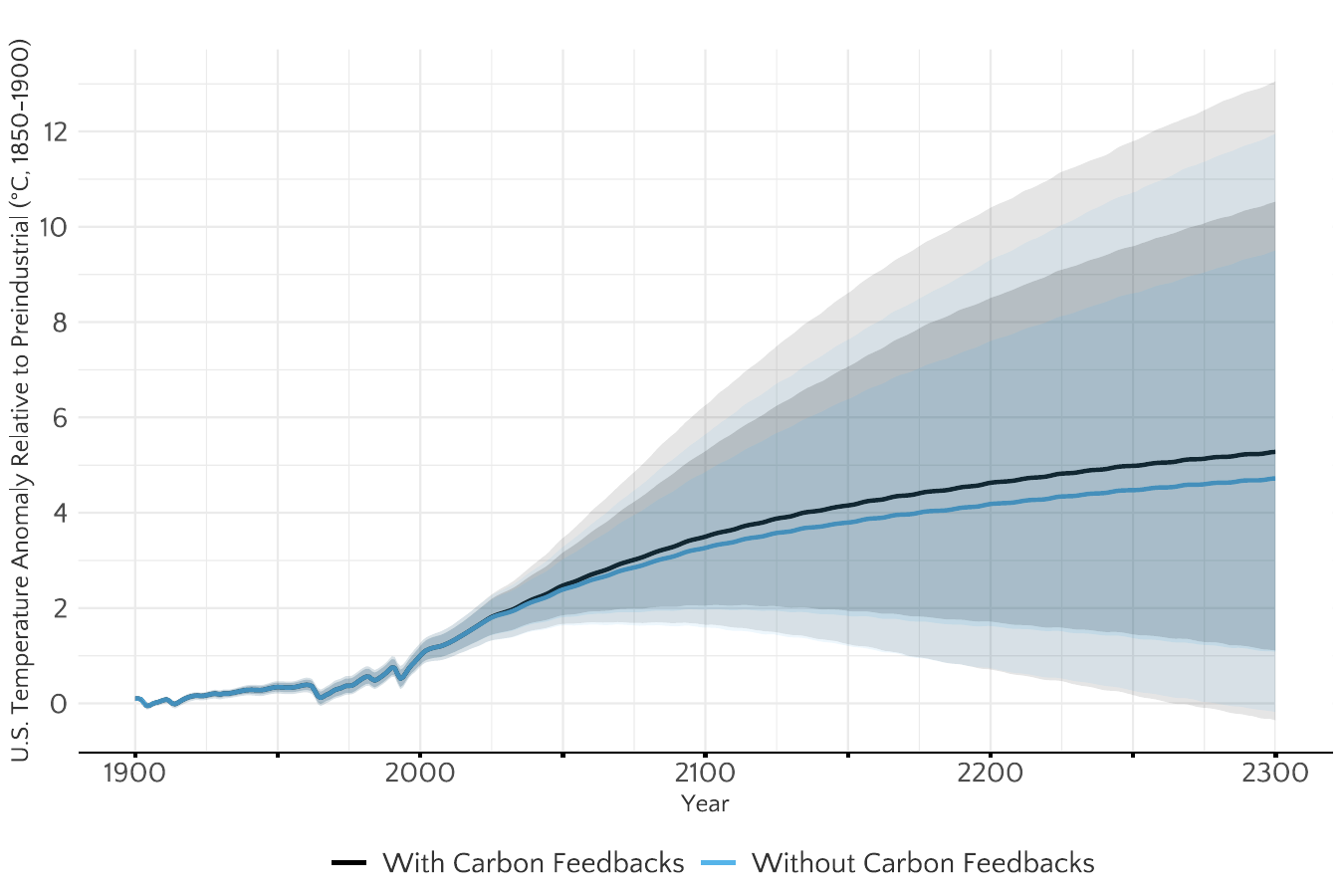}
    \end{center}
    \footnotesize
    U.S. mean surface temperature (solid line) from combining \FaIR{} 1.6.2 GMST and CONUS population-weighted patterns \citepalias{epa2024pattern}, along with 5\textsuperscript{th} to 95\textsuperscript{th} (dark shade) and 1\textsuperscript{st} to 99\textsuperscript{th} (light shade) percentile ranges that are a result of the joint uncertainty underlying \FaIR{} 1.6.2 and the RFF-SPs.
\end{figure}

RFF \citeyearpar{rff2024increasing} develop code to model the three remaining mechanisms using approaches closely tied to those in \citet{dietz2021economic}: (1) the carbon feedback from permafrost thaw, (2) the carbon feedback from Amazon rainforest dieback, and (3) damages to the Indian economy from the disruption of the Indian summer monsoon (ISM). Since \citeauthor{dietz2021economic}'s modeling of the ISM is limited to the direct effect on the Indian economy, it is not relevant to the damage module used in this U.S.-focused paper. Therefore, in this paper we incorporate only the carbon feedbacks from permafrost thaw and Amazon rainforest dieback, based on the methods in \citet{dietz2021economic} as described below. Accounting for these two carbon feedback effects increases the U.S.-specific SC-\ce{CO2} for 2030 (under 2 percent near-term Ramsey discount rate) by 9 percent in the enumerative \GIVE{} and \FrEDI{} models, and by 50 percent in \DSCIM. This increase represents the temperature-driven damages from the additional carbon released from permafrost thaw and Amazon dieback. It does not reflect the value of other physical impacts from these feedbacks, such as lost ecosystem services and biodiversity loss from Amazon rainforest dieback \citep{dietz2018climate,dietz2021economic} and infrastructure damage from permafrost thaw \citep{huntington2023alaska}. We briefly describe the modeling of the Amazon rainforest dieback and permafrost thaw below and then present the impact of these two carbon feedback effects on the GMST and U.S. temperature projections used in this paper.  Specifically, Figures \ref{fig:gmst_wo_feedbacks} and \ref{fig:usmst_wo_feedbacks} show how the inclusion of these carbon feedbacks impact the GMST and the U.S. mean surface temperature, respectively. 

\subsection{Amazon rainforest dieback}
\label{app:sec:amazon_dieback}

Dieback of the Amazon rainforest due to climate change is expected to add billions of additional tons of \ce{CO2} to the atmosphere, which will amplify warming. Other important damages associated with the Amazon rainforest dieback, including the loss of biodiversity and ecosystem services, are not incorporated in this module. To implement this feedback in our framework, we follow the approach of \citet{dietz2021economic}. The input into this module is GMST change, and the output is added \ce{CO2} released, which is added to the emissions inputs to the \FaIR{} climate model in the subsequent year. The probability that the Amazon dieback begins in any year after 2010 increases in global mean surface temperature. A uniform [0,1] random variable is drawn for each Monte Carlo draw and the feedback is triggered if the value is less than this probability. Following \citet{dietz2021economic}, it is assumed that there is a total of 183 gigatons of \ce{CO2} is released over time. The amount of time that the carbon is released is sampled from a triangular distribution with a minimum of 10 years, a maximum of 250 years, and a central value of 50 years.  

\subsection{Permafrost carbon feedback}
\label{app:sec:permafrost}

Similar to the modeling of Amazon rainforest dieback, we represent the permafrost carbon feedback following the approach of \citet{dietz2021economic}. The module uses GMST as an input, and as temperatures rise, thawing of the permafrost is expected to release additional \ce{CH4} and \ce{CO2} into the atmosphere, which are the output variables of the module that are input back into the \FaIR{} climate model in the subsequent year. Other damages associated with permafrost thawing, such as infrastructure damage and soil erosion, are not included in this module. There are five uncertain variables that are sampled according to a normal distribution with the same parameter values as those used in \citet{dietz2021economic} with an adjustment of truncating the normal distribution at zero. The equations for the relationships between these uncertain variables are presented in the supplemental materials of \citet{dietz2021economic} and govern the stock of organic carbon in the permafrost, the proportion of the permafrost that thaws with increased temperatures, how quickly the thawed carbon decomposes into \ce{CO2} and \ce{CH4}, and the proportion of the thawed carbon that is released into a passive reservoir.

\newpage

\section{Technical Documentation for the Implementation of Additional Nonmarket Damages and Combining Estimates from Different Lines of Evidence}
\label{app:sec:td_wildfire}

As discussed at length in EPA \citeyearpar{epa2023report}, there are many categories of climate change impacts and associated damages that are not yet or only partially represented in enumerative models such as \GIVE, \DSCIM, and \FrEDI. Continued progress in filling data gaps and estimating the magnitude of many of these omitted impacts includes a large and growing body of research providing evidence on a wide range of U.S.-specific outcomes, including heat-related kidney disease and other morbidity outcomes \citep[see, e.g.,][]{bell2024climate,yang2024degree}, human capital impacts \citep{park2021learning}, flooding impacts on mortality \citep{mueller2024sunny,lynch2025large} and drinking water quality \citep{austin2025drinking-water}, forestry and ocean impacts \citep{baker2023projecting,bastien2026ocean-scc}, long-term impacts to coastal wetlands \citep{fant2022valua}, net impacts on outdoor recreation \citep{parthum2022winterrecreation,willwerth2023}, climate impacts on mental health \citep{obradovich2022} and crime \citep{harp2020crime}, and the distortionary effects of fiscal impacts \citep{barrage2023fiscal}, to name a few. The U.S. National Climate Assessment offers a periodic overview of much of this literature, including examples of how changes in temperature and other climate variables have affected, and are projected to affect, various U.S. economic outcomes as estimated in particular studies \citepalias[see, e.g., Table 19.1 in ][]{USGCRP2023fifth}.  While the majority of this research has not yet been converted into damage functions of the kind needed for developing U.S. impact-specific SC-GHG estimates, two exceptions include research on climate-driven wildfire-related health impacts \citep{qiu2026wildfire} and monetized damages from biodiversity loss \citep{wingenroth2024accounting}. We discuss our implementation of each within our modeling framework in turn.

\subsection{Mortality Burden from Wildfire Smoke}
\label{sec:wildfire}

One source of climate change damages receiving increasing attention in the academic literature, which may be especially relevant to assessing the climate damages experienced by U.S. citizens and residents, is wildfire. Wildland fires in the U.S., though a naturally occurring disturbance regime important to diverse ecosystems, are increasing in frequency, severity, burned area, and wildfire season length \citetext{\citealp{ostoja2023focus}; \citetalias{epa2024climate}}. Climate change is projected to exacerbate these trends, with increasing temperatures causing earlier spring melts and diminished snowpacks, which allow for drier summer conditions. The increasing frequency and longevity of droughts are predicted to further lengthen fire seasons and augment wildfire size \citepalias{epa2024climate}. At the same time, U.S. populations’ exposure to wildfire risks has increased with the expansion of the wildland-urban interface, as more Americans live near forests, grasslands, and other natural areas affected by fire \citep{mockrin2022growth,guo2024global}. Additionally, U.S. populations exposed to wildfire smoke cover even larger areas of the country as the impact of wildfire smoke on U.S. air quality has increased in recent years \citep{qiu2024mortality,gellman2025wild}. \citet{burke2021changing} estimate that wildfires have accounted for up to 25 percent of \ce{PM_{2.5}} in the U.S. in recent years with even higher percentages in some Western U.S. regions. 

Wildfire events can result in many types of damages. Market-based damages can occur from the fire itself (e.g., property destruction, suppression costs, business productivity impacts, and energy supply disruptions), smoke exposure (e.g., labor market impacts), and spillover effects (e.g., through trade and supply chain networks). Nonmarket damages can also occur through these mechanisms (e.g., mortality and morbidity effects from fire or smoke exposure, recreation impacts, population displacement, and changes to ecosystem services). See \citet{gellman2025wild} and \citet{rosu2026canada} for a recent review of the various types of impacts of wildfire smoke on human health and economic activity. Current methods for estimating these damages for a given wildfire vary by impact or damage category. Estimating future marginal wildfire-related damages from GHG emissions also requires modeling of wildfire starts and severity and smoke exposure under climate change, which is currently not well modeled at the global level, and incompletely even for the U.S. \citepalias[see review in][]{cil2024estimating}. As discussed above, FrEDI incorporates some wildfire damages for the U.S., but currently only includes response costs and mortality and morbidity from wildfire smoke generated in the Western U.S. \citetext{\citealp{neumann2021estim}; \citetalias{epa2024tdfredi}}. Recent research has advanced the climate-wildfire linkage to associated damages in the U.S. for one category of wildfire related damages: climate-driven mortality risks of wildfire smoke exposure \citep{qiu2024mortality,qiu2026wildfire}.

\cite{qiu2026wildfire} develop a damage function for the specific purpose of estimating a partial social cost of carbon from increased wildfire-related \ce{PM_{2.5}}. They implement their damage function, which builds off \cite{qiu2024mortality}, in the \MimiGIVE{} framework. Alongside the manuscript, the authors published the GIVE module which we have adopted as published and implemented in this study.

Table \ref{tab:impacts_bottom_up} includes the U.S. impact-specific SC-\ce{CO2} estimates for 2030 emissions that result from combining the damage function based on \citet{qiu2026wildfire} with the same U.S.-specific socioeconomic and emissions inputs, climate modeling described in Section \ref{app:sec:td_discounting} and and the discounting described in \ref{app:sec:drupp_derivation}. The results show that wildfire PM2.5-related mortality damages as estimated using the \citet{qiu2026wildfire} damage function are considerable. The U.S. impact-specific SC-\ce{CO2} is estimated to be \$18 dollars per metric ton of \ce{CO2} for 2030 emissions (under 2 percent near-term Ramsey discounting). This is of similar magnitude as the analogously estimated U.S. temperature-related mortality SC-\ce{CO2} in DSCIM and GIVE, and significantly greater than the share of the FrEDI based damages attributable to smoke and other impacts from wildfires in U.S. Western states \citetext{\citealp{hartin2023advanc}; \citetalias{epa2024tdfredi}}. \citet{qiu2026wildfire} argue this partial SC-\ce{CO2} is not double counting the effect of temperature itself on mortality because \citet{qiu2024mortality} include temperature control variables in their empirical estimation. The results in Table \ref{tab:impacts_bottom_up} do not reflect any other type of nonmarket or market damages from wildfires, nor are the carbon feedback effects from wildfires (see e.g., \citealp{jerrett2022up}) accounted for in the climate module of our modeling framework. Analogous U.S. impact-specific SC-GHG results for 2030 emissions of \ce{CH4} and \ce{N2O} are presented in Table \ref{tab:impact_categories_us}.  

\subsection{Nonmarket damages from biodiversity loss}
\label{sec:biodiversity}

Another area in which emerging research has advanced our understanding of climate damages to U.S. populations is the economic damages from climate-driven biodiversity loss. A full accounting of the damages from the loss of biodiversity ecosystem services would include consideration of both market impacts (e.g., crop pollination, nutrient cycling, tourism, and pharmaceutical innovations) and nonmarket impacts (e.g., existence and aesthetic values). Developing causal estimates of these types of impacts is challenging due to significant data and methodological limitations given the complexity of ecosystem changes, and biodiversity loss is likely to be associated with other ecological shocks \citep{druckenmiller2022accounting}. A new working paper published by RFF \citep{wingenroth2024accounting} has advanced the estimation of one category of nonmarket damages associated with climate-driven biodiversity loss: the loss of its nonuse value. Nonuse value of biodiversity includes individuals’ WTP for knowing that a species exists, bequest values for future generations, and altruistic values for others' enjoyment of the species. 

\citet{wingenroth2024accounting} develop a biodiversity loss damage function by combining a species loss function (relating GMST change to species loss) with a WTP function (relating species loss to nonuse value) to link changes in temperature to economic impacts from biodiversity loss. The functional form of the species loss function is derived from the Climate Framework for Uncertainty, Negotiation, and Distribution (FUND) model \citep{anthoff2014climate}. \citet{wingenroth2024accounting} re-parameterize the temperature-effect coefficient of the species loss function using data from a more recent and globally comprehensive synthesis of studies on extinction rates \citep{urban2015acceler} and found temperature to have a larger impact on species loss than previous parameterizations. For the WTP function, the authors employ the method developed by \citet{brooks2014updated} for estimating WTP for avoided biodiversity loss from stated preference estimates. The preference for biodiversity parameter of the WTP function is calibrated following \citet{kaushal2023inter}, who estimate this parameter for the 16 regions used in the FUND model. Using country-level population and income forecasts from the RFF-SPs, \citet{wingenroth2024accounting} develop country-level WTP estimates for avoided species loss for 184 countries, including in the U.S. That is, it provides a damage function that reflects U.S. populations’ nonuse damages from global species loss resulting from a change in GMST. 

Table \ref{tab:impacts_bottom_up} presents the U.S. impact-specific SC-\ce{CO2} estimates for 2030 emissions that result from combining the damage function based on \citet{wingenroth2024accounting} with the same U.S.-specific socioeconomic and emissions inputs, climate modeling described in Section \ref{app:sec:td_discounting} and and the discounting described in \ref{app:sec:drupp_derivation}. The results show that the non-use value of biodiversity impacts contributes to monetized U.S. economic damages from GHG emissions. For 2030 emissions, the U.S. impact-specific SC-\ce{CO2} for this endpoint is similar in magnitude to coastal and energy use impacts represented in the \DSCIM{} and \GIVE{} damage modules, and on par with several other impact categories currently represented in \FrEDI{} \citetext{\citealp{hartin2023advanc}; \citetalias{epa2024pattern}}. As discussed above, the biodiversity specific results in Table \ref{tab:impacts_bottom_up} are a subset of the value of biodiversity-related ecosystem services. They do not reflect the value of any market and nonmarket use impacts of species loss (e.g., from hunting, commercial, and recreational activities). Analogous U.S. impact-specific SC-GHG results for 2030 emissions of \ce{CH4} and \ce{N2O} are presented in Table \ref{tab:impact_categories_us}.

\subsection{Combining estimates from different lines of evidence}
\label{sec:combine_details}

In \ref{app:sec:add_fig_tab} Table \ref{tab:combined_evidence} we take a first step in illustrating the impact on U.S.-specific SC-GHG estimates from including two categories of nonmarket damages with the GDP-based market damage function from several macroeconomic studies. Specifically, we combine the heat- and cold-related mortality damage function from \citet{cromar2022global}\footnote{We use the mortality damage function in \GIVE{} \citep{rennert2022comprehensive} due to modeling constraints in incorporating other health damage functions (\citet{carleton2022valuing} from \DSCIM) into the \MimiGIVE{} framework.} and wildfire smoke-related mortality damages from \citet{qiu2026wildfire} with the U.S. GDP damage functions derived from recent macroeconomic studies that project some persistence in the effect of temperature on GDP growth over time. Importantly, in this integrated modeling exercise, we do not simply add GDP and mortality impacts estimated separately, but integrate both impacts into the full Monte Carlo estimation procedure. This allows for feedbacks between the market (GDP) and nonmarket (temperature and wildfire-related mortality) impacts.  The feedbacks occur through (1) the climate-induced changes in net market income overtime and its impact on the VSL used in monetizing the nonmarket impacts and adjusted using the income elasticity of VSL; and (2) the effect of market-only damages on consumption growth rates in the discounting equation used for both market and nonmarket future impacts. The income elasticity of VSL is unit elastic ($\varepsilon = 1$) and consistent with the central value recommended in the literature \citetext{\citealp{viscusi2017income,robinson2019valuing,rennert2022comprehensive}; \citetalias{epa2024guidelines}}.\footnote{Projected changes in premature mortality in the U.S. are monetized using the same value of mortality risk reduction as in the EPA’s regulatory analyses, \$4.8 million in 1990 (1990USD), and adjusted for income growth and inflation following current EPA guidelines and practice \citepalias{epa2024guidelines}, resulting in a 2020 value of \$10.05 million (2020USD). See EPA \citeyearpar{epa2023report} for more discussion. In the case of combining market and nonmarket damages, the adjustment made for future income growth becomes based on the projected exogenous GDP per capita minus the GDP losses projected from the macroeconomic studies' damage functions.} 

The impact of climate damages on the discounting of future damages follows common practice in enumerative studies. In particular, our integrated implementation of the Ramsey discounting procedure described in \ref{app:sec:td_discounting} uses the ``realized'' growth rate of consumption after netting off damages to growth in only market consumption (not nonmarket). This approach preserves an important aspect of the Ramsey discounting approach --- climate damages to income are reflected in the discounting procedure. Despite the intuitive advantage of considering total income in the discounting procedure, this standard approach only accounts for the relative scarcity impacts highlighted by \citet{traeger2011sustain, drupp2018limits,drupp2021relative} in a limited way. In Appendix \ref{app:sec:drupp_derivation}, we show that the procedure can be derived as a special case of the \citet{drupp2018limits} framework paired with the assumption of zero baseline growth in the nonmarket good (zero growth without climate damages). In the context of mortality impacts we view this as a reasonable assumption, and leave a generalization to multiple nonmarket goods with different nonzero baseline growth rates for future research.

\subsection{Discounting with Market and Nonmarket Goods}
\label{app:sec:drupp_derivation}

Consider the following welfare function, which is a function of market consumption, $C(t)$, and nonmarket consumption, $E(t)$:

\begin{equation} \label{eq:b2}
    W = \int_{t=0}^{\infty} u(C(t),E(t),t) e^{-\rho t} dt.
\end{equation}

\noindent where $\rho$ is the pure rate of time preference. Denote the first derivatives of the utility function $u(C,E,t)$ as $u_C, u_E$ and second partial derivatives as $u_{CC}, u_{EE}$, and $u_{CE} $. Following \citet{drupp2018discount} and \citet{traeger2011sustain}, the social discount rate of consumption can be represented as:

\begin{equation} \label{eq:b3}
r^C(t) = \rho + \left[ \frac{- u_{CC}(t) C(t) }{u_{C}(t)} \right] g^C(t) + \left[ \frac{- u_{CE}(t) E(t) }{u_{C}(t)} \right] g^E(t)
\end{equation}

\noindent where $g^X(t) = \frac{\dot{X(t)}}{X(t)}$ is the continuous time growth rate of $X(t)$. The marginal willingness to pay for the environmental good in terms of consumption is

\begin{equation} \label{eq:b4}
MWTP^E(t) = \frac{u_{E}(t) }{u_{C}(t)}
\end{equation}

\noindent
Now, consider the specific utility function:
 
\begin{equation} \label{eq:b5}
     u(C,E,t) = \big[ C(t) E(t)^{\alpha} \big]^{1-\eta}
\end{equation}

\noindent where $\alpha, \eta > 0$ are constants. Collecting the elements necessary to specify equations \ref{eq:b3} and \ref{eq:b4}, we have: 
\begin{align}
    u_C &= C^{-\eta} E^{\alpha (1-\eta)} \\
    u_{CC} &= -\eta C^{-\eta - 1} E^{\alpha (1-\eta)} \\
    u_{CE} &= \alpha (1-\eta) C^{-\eta} E^{\alpha (1-\eta) - 1} \\
    u_E &= \alpha C^{1-\eta} E^{\alpha (1-\eta)-1} 
\end{align}

\noindent The $t$ function argument is suppressed in the arguments of the utility function for simplicity. Plugging these values into equation \ref{eq:b3}, the implied social discount rate of consumption is:

\begin{equation} \label{eq:b10}
r^C(t) = \rho + \eta g^C(t) + \alpha(\eta-1) g^E(t)
\end{equation}

\noindent
And the implied marginal willingness to pay function for the environmental good in equation \ref{eq:b4} becomes:

\begin{equation} \label{eq:b11}
MWTP^E(t) = \alpha \frac{ C(t)}{E(t)}
\end{equation}

We can rewrite the growth rates as an exogenous growth rate minus damages to the growth rate from climate change: $g^X(t) = g0^X (t) - dg^X(t)$ for $X=C$ and $X=E$. Under the assumptions that $g0^E(t) = 0$ and $ \alpha = \eta/(\eta - 1)$, the resulting social discount rate is the standard Ramsey formula from equation \ref{eq:eq_ramsey} with market and environmental damages from climate change subtracted from the exogenous growth rate of market consumption. 

\begin{equation} \label{eq:b12}
r^C(t) = \rho + \eta \left[ g0^C(t) - dg^C(t) - dg^E(t) \right]
\end{equation}


Equation \ref{eq:b12} is the approach that we implement in this paper. Alternatively, under the assumptions that $g^E(t) = 0$, $\eta=1$ or $\alpha =0$, the social discount rate becomes:

\begin{equation} \label{eq:b13}
r^C(t) = \rho + \eta \left[ g0^C(t) - dg^C(t) \right]
\end{equation}

\noindent An assumption that $\alpha =0$ is equivalent to implying that the environmental good is absent from the utility function and therefore that $MWTP^E=0$. Under the assumption that $g^E(t) = 0$ or $\eta=1$ the resulting social discount rate is the standard Ramsey formula from equation \ref{eq:eq_ramsey} with market damages from climate change subtracted from the exogenous growth rate of market consumption. 

\section{Technical Documentation for Implementation of Macroeconomic Econometric Studies}
\label{app:sec:td_macro}

For each of the macroeconomic econometric studies discussed in Section \ref{sec:macro_damage_functions}, we construct a damage component that replicates the econometric results and projection methods in the paper and that can be implemented in the \MimiGIVE{} modeling framework to estimate a U.S. impact-specific SC-\ce{CO2} reflecting the GDP-based market damages from \ce{CO2} emissions changes. For each paper, we focus on the authors' stated ``preferred'', ``main'', or ``central'' specification, or the specification that is used for climate damage projections in the paper if no preferred specification is stated. This Appendix provides (1) a summary of key differences in data used for the econometric estimation and projections presented in each study, (2) a description of central end-of-century damage estimates for the U.S. from our modelling results and from each study, and (3) a description of how we validated that each damage component replicates the authors' empirical results.

\subsection{Summary of data underlying the estimation and projections in each study.}
\label{app:sec:marco_summary}

Tables \ref{tab:macro_summary_temp}, \ref{tab:macro_summary_precip}, and \ref{tab:macro_summary_socio} present a summary of the data used in each macroeconomic econometric study discussed in Section \ref{sec:macro_damage_functions}. It lists both the sources of data and measurement applied in the econometric estimation of the relationship between weather variables and economic growth, as well as the data used for the GDP loss projections presented in each paper. 

\begin{landscape}
\begin{table}[htbp]
    \caption{Source temperature data used in the eight macroeconomic econometric studies}
    \label{tab:macro_summary_temp}
    \centering
    \begin{threeparttable}
    \footnotesize
        \begin{tabular}{llllll}
            \toprule \toprule
			\textbf{Paper} & \textbf{Estimation} & \textbf{Estimation} & \textbf{Estimation} & \textbf{Projection} & \textbf{Projections\tnote{$\alpha$}} \\
            & \textbf{Data} & \textbf{Sample } & \textbf{Weighting and} & \textbf{Baseline} &  \\
            & & & \textbf{Base Year} & &  \\
            \midrule	
\citet{burke2015global} & UDel gridded  & 1960-2010, & Population weighted, & mean 1980-2010 & Ensemble mean warming of  \\
& \citep{matsuura2012terrest} & 166 countries & 2000 & & models in CMIP 5, RCP 8.5 \\
\citet{kalkuhl2020impact} & CRU gridded & 1900-2014, & Area weighted & mean 2015-2019 & Princeton Earth System Model \\
& \citep{harris2014updated} & 77 countries, 1,545 regions & & & GFDL-ESM2M, RCP 8.5 \\
& & & & & \citep{warszawski2014} \\
\citet{newell2021the} & UDel gridded & 1960-2010, & Population weighted, & mean 1980-2010 & Ensemble mean warming of  \\
& \citep{matsuura2012terrest} & 166 countries & 2000 & \citep{burke2015global} & models in CMIP 5, RCP 8.5 \\
\citet{acevedo2020effects} & CRU gridded & 1950-2015, & Population weighted, & 2005 & Earth Exchange Global Daily  \\
& \citep{harris2014updated} & 189 economies & 1950 & & Downscaled Projections  \\
& & & & & (NEX-GDDP) mean of all\\
& & & & & models, RCP 8.5 and 4.5 \\
\citet{kahn2021long} & UDel gridded & 1960-2014, & Population weighted, & 2014 & Author-calculated parametric  \\
& \citep{matsuura2014terrest} & 174 countries & 2010 & & forecasts using estimation \\ 
& & & & & data, RCP 8.5 and 2.6 \\
\citet{casey2023projecting} & UDel gridded & 1960-2010, & Population weighted, & 2010 & World Meteorological \\
& \citep{matsuura2018terrest} & 155 countries & 2000 & \citep{burke2015global} &  Organization (WMO), RCP 8.5 \\ 
\citet{harding2025distribution} & UDel gridded & 1960-2017, & Population weighted, & mean 1980-2010 & Ensemble mean warming of \\
& \citep{matsuura2018terrest} & 166 countries & 2000 & \citep{burke2015global} & models in CMIP 5, RCP 8.5 \\
\citet{nath2024} & Global Meteorological Forcing & 1960-2015, & Population weighted,  & mean 1980-2010 & Ensemble mean warming of  \\
& Dataset \citep{sheffield2006develop} & 206 economies & annual & \citep{burke2015global} & models in CMIP 5, RCP 8.5 \\
\bottomrule
		\end{tabular}
        \begin{tablenotes} 	
            \footnotesize
            \item This table summarizes the source temperature data for each study's stated ``preferred'', ``main'', or ``central'' specification, or the specification that is used for climate damage projections in the paper if no preferred specification is stated. \citet{burke2015global}: Table 1, col 1. \citet{kalkuhl2020impact}: Table 4, col 5. \citet{newell2021the}: Levels version of \citet{burke2015global}. \citet{kahn2021long}: Table 2, spec 2 (m=30). \citet{casey2023projecting}: Table 1, col 2. \citet{harding2025distribution}: Table 1, col 6. \citet{nath2024}: Full dynamics with time FEs (Fig 6c).
            \item[$\alpha$] The Projections column describes the source of RCP-consistent temperature projections used in the papers. Aside from \citet{kalkuhl2020impact}, most papers link directly to a web interface for data request and collection (where citation is relatively clear). Papers using CMIP 5 models and WMO data draw from the 
            \emph{KNMI Climate Explorer}. \citet{acevedo2020effects} draw from the \emph{NASA Center for Climate Simulation}. 
        \end{tablenotes}
	\end{threeparttable}
\end{table}
\end{landscape}

\begin{landscape}
\begin{table}[htbp]
    \caption{Source precipitation data used in the eight macroeconomic econometric studies}
    \label{tab:macro_summary_precip}
    \centering
    \begin{threeparttable}
    \footnotesize
        \begin{tabular}{llllll}
            \toprule \toprule
			\textbf{Paper} & \textbf{Estimation} & \textbf{Estimation} & \textbf{Estimation} & \textbf{Projection} & \textbf{Projections} \\
            & \textbf{Data} & \textbf{Sample } & \textbf{Weighting and} & \textbf{Baseline} &  \\
            & & & \textbf{Base Year} & &  \\
            \midrule
\citet{burke2015global} & UDel gridded & 1960-2010, & Population weighted, & NA   & NA \\ 
& \citep{matsuura2012terrest} & 166 countries & 2000 & & \\
\citet{kalkuhl2020impact} & CRU gridded & 1900-2014, & Area weighted & NA   & NA \\
& \citep{harris2014updated} & 77 countries, 1,545 regions& & & \\
\citet{newell2021the} & UDel gridded & 1960-2010, & Population weighted, & NA   & NA \\
& \citep{matsuura2012terrest} & 166 countries & 2000 & & \\
\citet{acevedo2020effects} & CRU gridded & 1950-2015, & Population weighted, & NA   & NA \\
& \citep{harris2014updated} & 189 economies & 1950 & & \\
\citet{kahn2021long} & UDel gridded & 1960-2014, & Population weighted, & NA   & NA \\
& \citep{matsuura2014terrest} & 174 countries & 2010 & & \\
\citet{casey2023projecting} & UDel gridded & 1960-2010, & Population weighted, & 2010 & Assumed constant   \\
&\citep{matsuura2018terrest} & 155 countries & 2000 & & growth at mean historical  \\ 
& & & & & rate (with and without \\
& & & & & climate change) \\
\citet{harding2025distribution} & UDel gridded & 1960-2017, & Population weighted, & mean 1980-2010, & Ensemble mean of \\
& \citep{matsuura2018terrest} & 166 countries & 2000 & \citep{burke2015global} & models in CMIP 5, RCP 8.5  \\
& & & & & (inferred based on ) \\
& & & & & replication code \\
\citet{nath2024} & NA & NA & NA & NA & NA  \\
& & & & & \\
\bottomrule
		\end{tabular}	
        \begin{tablenotes} 	
            \footnotesize
            \item This table summarizes the source temperature data for each study's stated ``preferred'', ``main'', or ``central'' specification, or the specification that is used for climate damage projections in the paper if no preferred specification is stated. \citet{burke2015global}: Table 1, col 1. \citet{kalkuhl2020impact}: Table 4, col 5. \citet{newell2021the}: Levels version of \citet{burke2015global}. \citet{kahn2021long}: Table 2, spec 2 (m=30). \citet{casey2023projecting}: Table 1, col 2. \citet{harding2025distribution}: Table 1, col 6. \citet{nath2024}: Full dynamics with time FEs (Fig 6c).
        \end{tablenotes}
	\end{threeparttable}
\end{table}
\end{landscape}

\begin{landscape}
\begin{table}[htbp]
    \caption{Source socioeconomic data used in the eight macroeconomic econometric studies}
    \label{tab:macro_summary_socio}
    \centering
    \begin{threeparttable}
    \footnotesize
        \begin{tabular}{llllll}
            \toprule \toprule
            \textbf{Paper} & \textbf{Estimation} & \textbf{Estimation} & \textbf{Estimation} & \textbf{Projection} & \textbf{Projections\tnote{$\alpha$}} \\
            & \textbf{Data} & \textbf{Sample } & \textbf{Weighting and} & \textbf{Baseline} &  \\
            & & & \textbf{Base Year} & &  \\
            \midrule
\citet{burke2015global} & World Bank & 1960-2010, & \textemdash & 2010 & SSPs 1-5 \\
& \emph{World Development Indicators} & 166 countries & & & \citep{o2014new} \\
\citet{kalkuhl2020impact}& Author-compiled  & 1900-2014, & \textemdash & 2015 & SSP 2 \\
& Gross Regional Product & 77 countries, 1,545 regions & & &  \citep{o2014new} \\
\citet{newell2021the} & World Bank & 1960-2010, & \textemdash & 2010 & SSPs 1-5 \\
& \emph{World Development Indicators} & 166 countries & & & \citep{o2014new} \\
\citet{acevedo2020effects} & International Monetary Fund  & 1950-2015, & \textemdash & NA & NA \\
& \emph{World Economic Outlook}, & 189 economies & & & \\
& World Bank & & & & \\
& \emph{World Development Indicators} & & & & \\
\citet{kahn2021long} & World Bank & 1960-2014, & \textemdash & NA & NA \\
& \emph{World Development Indicators} & 174 countries & & & \\
\citet{casey2023projecting} & Penn World Tables 10.0  & 1960-2010, & \textemdash & 2010 & Author-calculated parametric  \\
& \citep{feenstra2015next} & 155 countries & & & projections of socioeconomics \\
\citet{harding2025distribution} & World Bank & 1960-2017, & \textemdash & 2010 & SSPs 1-5 \\
& \emph{World Development Indicators} & 166 countries & & & \citep{o2014new} \\
\citet{nath2024} & World Bank & 1960-2015, & \textemdash & 2010 & SSPs 1-5 \\
& \emph{World Development Indicators} & 206 economies & & & \citep{o2014new} \\
\bottomrule
		\end{tabular}	
        \begin{tablenotes} 	
            \footnotesize
            \item This table summarizes the source temperature data for each study's stated ``preferred'', ``main'', or ``central'' specification, or the specification that is used for climate damage projections in the paper if no preferred specification is stated. \citet{burke2015global}: Table 1, col 1. \citet{kalkuhl2020impact}: Table 4, col 5. \citet{newell2021the}: Levels version of \citet{burke2015global}. \citet{kahn2021long}: Table 2, spec 2 (m=30). \citet{casey2023projecting}: Table 1, col 2. \citet{harding2025distribution}: Table 1, col 6. \citet{nath2024}: Full dynamics with time FEs (Fig 6c).
            \item[$\alpha$] The SSPs projections for GDP use those developed by Organization for Economic Co-operation and Development (OECD).
        \end{tablenotes}
	\end{threeparttable}
\end{table}
\end{landscape}

\subsection{Projected U.S. GDP Loss from \MimiGIVE{} Implementation of Macroeconomic Econometric Damage Functions}

\begin{figure}[htbp]
    \begin{center}
        \caption{Projections of U.S. GDP loss from climate change, \MimiGIVE{} implementation of recent macroeconomic econometric studies}
        \label{fig:macro_summary}
        \includegraphics[width=1\linewidth]{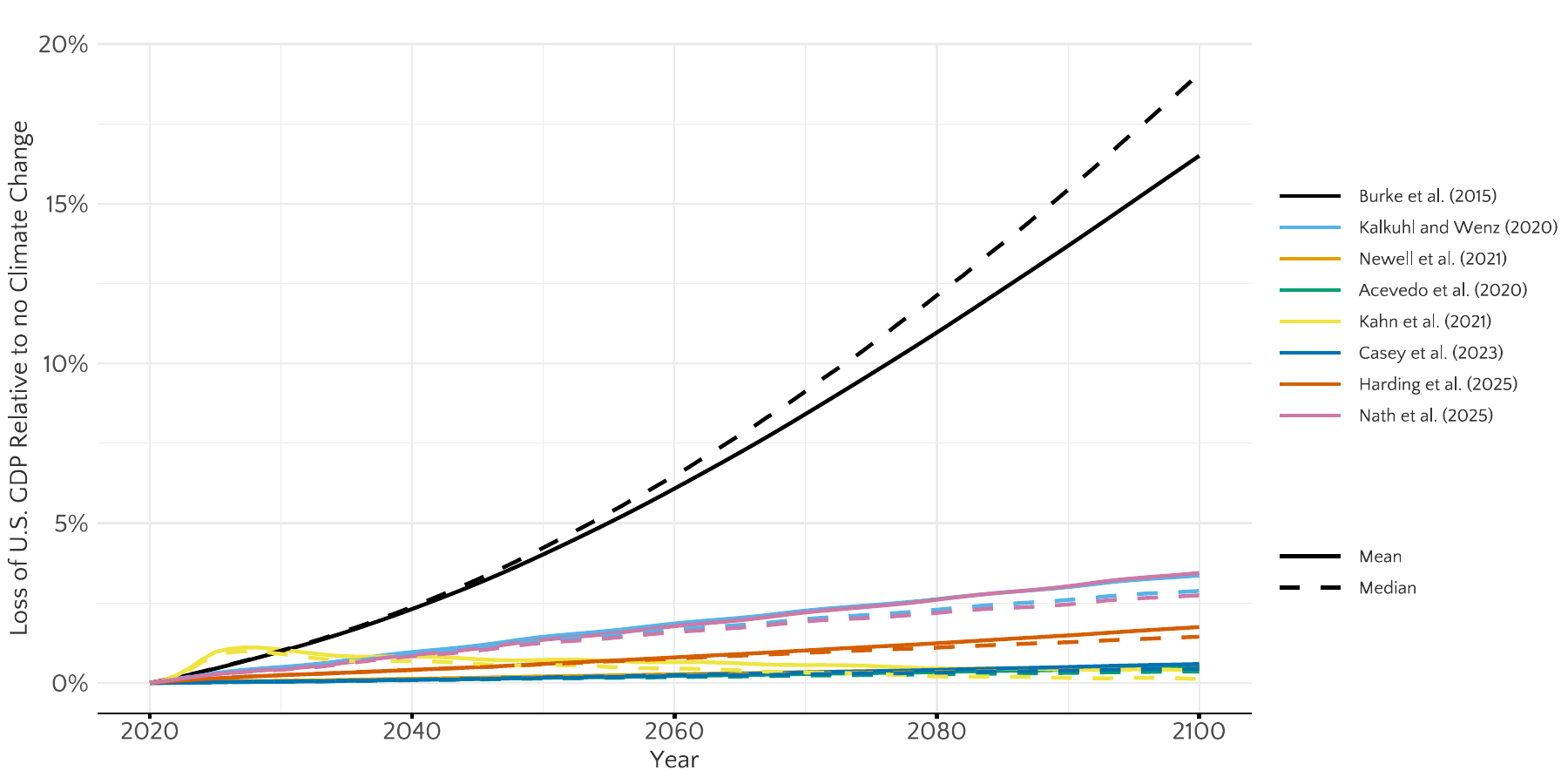}
    \end{center}
    \footnotesize
    This figure shows each study's stated ``preferred'', ``main'', or ``central'' specification, or the specification that is used for climate damage projections in the paper if no preferred specification is stated. Mean (solid line) and median (dashed line) percent losses to U.S. GDP from climate change under each of the eight macroeconomic studies' damage functions as implemented in \MimiGIVE, accounting for U.S. socioeconomic, climate, and damage parameter uncertainty across 10,000 Monte Carlo simulations. Each damage function requires a measure of current climate or baseline temperature; for all damage functions we use country-level population-weighted mean temperatures from 1980 to 2010 drawn from \citet{burke2015global}. All GDP losses are relative to no climate change (i.e., no temperature anomalies above current climate) except \citet{kahn2021long}, which is relative to a continuation of their country-level estimated historic warming trend from 1960 to 2014. U.S. damages include Puerto Rico.
\end{figure}

Figure \ref{fig:macro_summary} presents projections of U.S. GDP loss from climate through 2100 resulting from the \MimiGIVE{} implementation of each study's constructed damage function together with the U.S. projections of socioeconomic variables from the RFF-SPs and the U.S. temperature projections described in \ref{app:sec:td_feedback}. The Figure displays both the mean (solid lines) and the median (dashed lines) of estimated damages across the 10,000 Monte Carlo simulations in each year, where the socioeconomic and climate module parameters are consistent across the studies (i.e., the first trial of \citet{burke2015global} damage function takes the same socioeconomic pathways and climate parameters as the first trial of \citet{kalkuhl2020impact}, \citet{newell2021the} and so forth).

Consistent with the discussion in Section \ref{sec:macro_damage_functions}, earlier papers in the literature such as \citet{burke2015global} that project permanent growth impacts from temperature change imply large damages to the U.S. economy over the next century (i.e., reductions in GDP of more than 15 percent relative to a scenario without climate change). Papers finding temperature to have temporary effects on GDP growth project modest impacts no more than a few percentage point reduction in U.S. GDP, and recent papers modeling varying degrees of persistence reflect an in-between case.

\subsection{Validation of \MimiGIVE{} Implementation of Macroeconomic Econometric Damage Functions}
\label{sec:validate_implementation}

To validate that the damage component constructed for use in \MimiGIVE{} provides an accurate representation of each paper's findings, we confirmed that U.S. GDP per capita loss projections based on our constructed damage functions under an SSP5/RCP8.5 scenario through 2100 approximates the U.S. GDP per capita loss projections from a similar RCP8.5 radiative forcing scenario generated using the replication code for each paper. There are two main advantages to this validation approach. First, the comparison is an intuitive one because all the papers we analyze provide projections consistent with an RCP8.5 scenario through 2100. Second, this approach allows us to focus attention on validating the damage component implementation: the projected GDP per capita loss comparison requires no discounting, marginal analysis, or explicit treatment of uncertainty.

\begin{figure}[htbp]
    \begin{center}
         \caption{Projections of U.S. GDP per capita loss under RCP8.5 climate scenario, from replication code of macroeconomic econometric studies}
        \label{fig:macro_gdp_loss}
        \includegraphics[width=\linewidth]{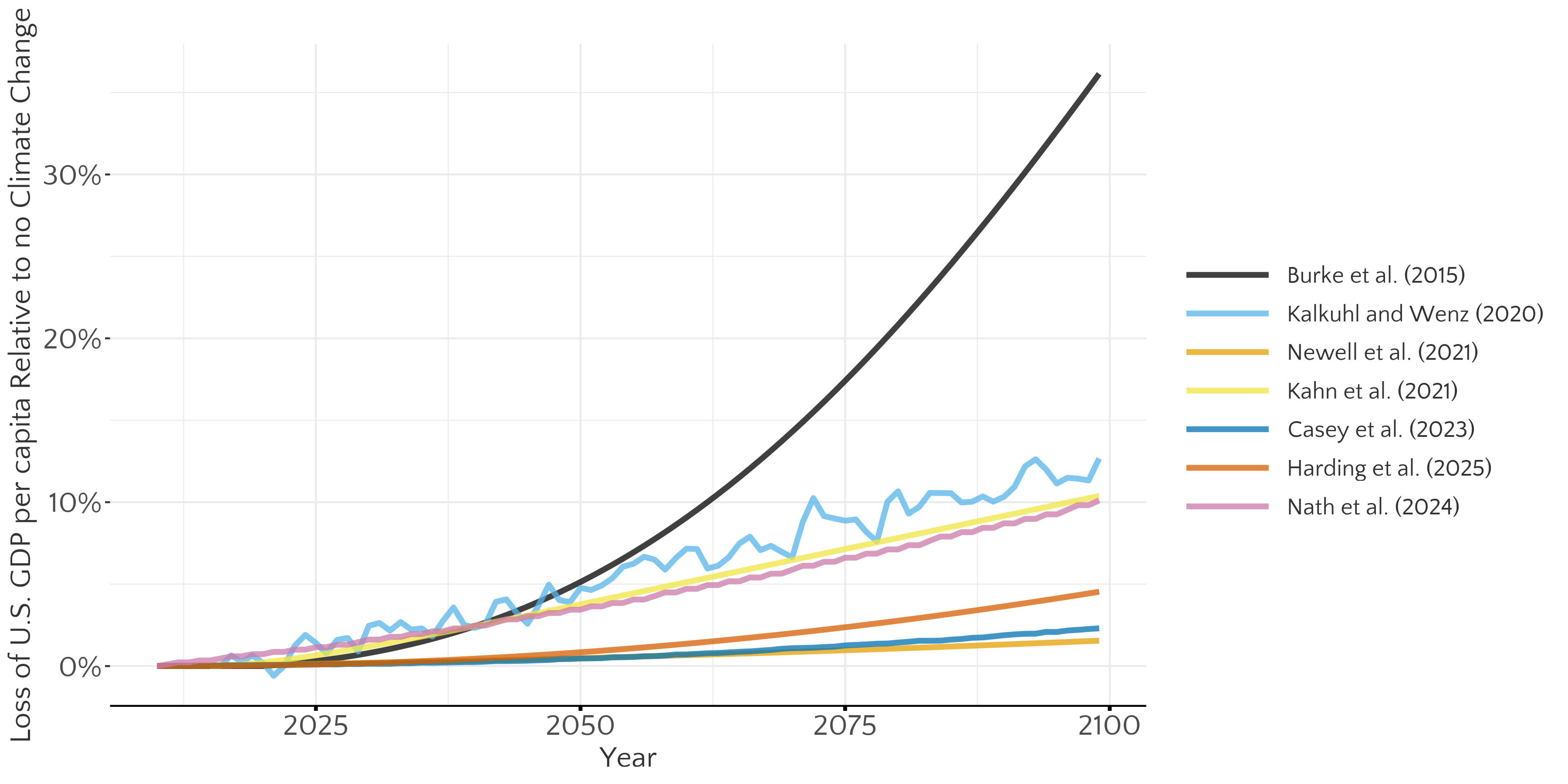}
    \end{center}
    \footnotesize
    This figure shows each study's stated ``preferred'', ``main'', or ``central'' specification, or the specification that is used for climate damage projections in the paper if no preferred specification is stated. \citet{burke2015global}: Table 1, col 1. \citet{kalkuhl2020impact}: Table 4, col 5. \citet{newell2021the}: Levels version of \citet{burke2015global}. \citet{kahn2021long}: Table 2, spec 2 (m=30). \citet{casey2023projecting}: Table 1, col 2. \citet{harding2025distribution}: Table 1, col 6. \citet{nath2024}: Full dynamics with time FEs (Fig 6c). All GDP losses are relative to no climate change (i.e. no temperature anomalies above current climate) except \citet{kahn2021long}, which is relative to a continuation of their country-level estimated historic warming trend from 1960 to 2014. U.S. GDP per capita losses in the figure exclude Puerto Rico.
\end{figure}

Figure \ref{fig:macro_gdp_loss} plots the U.S. GDP per capita losses projected under the RCP8.5-consistent scenario that were extracted from running the replication code available for each paper.\footnote{We obtained replication codes for most of the published papers from publicly available replication packages, listed in the references. For the working papers and \citet{newell2021the}, which do not have publicly available replication packages, we obtained the relevant replication codes and data via correspondence with the authors. Additionally, \citet{acevedo2020effects} is not listed in this appendix – although the code is publicly available, the damage projections are not included in the replication package. We confirmed our projection approach via correspondence with the authors but are unable to show a quantitative comparison with paper results for this study.} For the purposes of the validation exercise, we exclude damages to Puerto Rico, which are commonly estimated separately. The figure shows relative differences across papers under this scenario that are broadly consistent with the patterns found in Figure \ref{fig:macro_summary}. Earlier literature such as \citet{burke2015global} that projects permanent growth impacts from temperature change imply much larger damages to the U.S. economy over the next century than papers finding temperature to have temporary effects on GDP growth \citep[e.g.,][]{newell2021the}. Papers modeling varying degrees of persistence reflect an in-between case \citep[e.g.,][]{nath2024}. 

The projections displayed in Figure \ref{fig:macro_gdp_loss} reflect not only differences in econometric methods used to estimate the effect of climate variables on economic output, but also differences in the underlying data used in the estimation and projection methods. As seen in Tables \ref{tab:macro_summary_temp}, \ref{tab:macro_summary_precip}, and \ref{tab:macro_summary_socio} above, temperature projections consistent with RCP8.5 across the papers are drawn from a variety of sources, projections are made with different base years or measurements of baseline temperatures, or with different treatment of precipitation effects. Therefore, the next steps of the validation procedure required harmonizing the diverse set of year coverage, exogenous inputs, and model assumptions until an apples-to-apples comparison could be made between damage paths from paper replication codes and the \MimiGIVE{} implementation of each paper's damage function.

Figure \ref{fig:validation_procedure} below illustrates these steps. Each panel of the figure shows four (sometimes overlapping) damage paths under an RCP8.5-consistent scenario, plotted with time on the x-axis. The two solid lines are outputs produced with paper replication codes, and the two dashed lines display outputs from damage function implementations in the \MimiGIVE{} IAM. An initial comparison between the paper replication outputs and the \MimiGIVE{} implementation is represented by the lighter-colored lines. The solid light grey line displays the same GDP loss projections as the summary Figure \ref{fig:macro_gdp_loss} above (the raw replication code outputs), while the light-colored dashed line shows results from the \MimiGIVE{} damage component implementation under default SSP5 and RCP8.5 scenarios.\footnote{For more details on SSP and RCP implementation in the Mimi framework see: \url{https://github.com/anthofflab/MimiSSPs.jl}.} For most papers, this initial comparison yields damage paths of similar shape and order of magnitude, with some exceptions. We then explore to what extent the remaining differences are driven by model inputs and assumptions described above.  

The outcome of our input adjustments (required to create an apples-to-applies comparison) is represented by the darker lines in each panel. The solid dark grey line shows paper replication outputs with adjustment to the replication codes. For example, for all papers we dropped any damages prior to 2020, the base year of the \MimiGIVE{} IAM. Other adjustments are described beside each panel. Finally, the dark-colored dashed line displays \MimiGIVE{} damage component results after substituting relevant information from each paper (such as country-level temperature projections), again extracted using replication codes, into the damage components. In general, we considered a damage component to be validated when the solid dark grey line and the dark-colored dashed line coincided, indicating that under equivalent (baseline and RCP8.5) temperature assumptions the \MimiGIVE{} implementation yielded the same U.S. GDP per capita losses as the paper's projections. Substitutions and other issues specific to each paper are explained alongside the relevant panel of Figure \ref{fig:validation_procedure}. 

\begin{figure}[!b]
    \caption{Illustration of validation procedure for \MimiGIVE{} damage function implementations}
    \label{fig:validation_procedure}
    \begin{subfigure}{1\textwidth}
        \begin{minipage}{0.65\linewidth}
            \includegraphics[width=\linewidth]{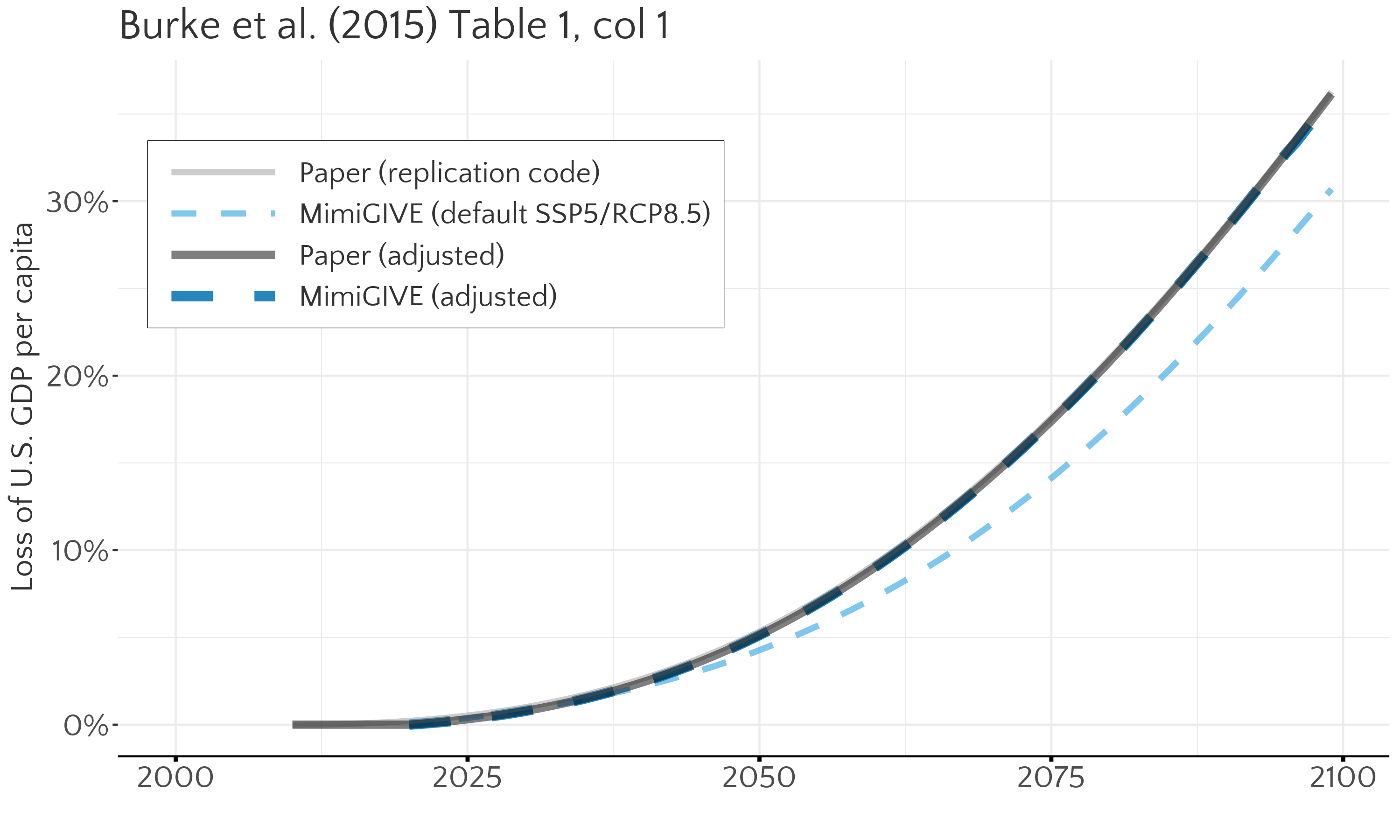}
        \end{minipage}
        \begin{minipage}{0.35\linewidth}
            \subcaption{\citet{burke2015global}}
            \label{fig:val_burke}
            \footnotesize
            \textbf{Paper (adjusted)}: 
            
            Drops damages prior to 2020 from the replication code.
            
            \bigskip
            \textbf{\MimiGIVE{} (adjusted)}:
            
            Substitutes country-level projected temperature paths consistent with RCP8.5 from the paper into the \MimiGIVE{} climate component. 
            \hfill
        \end{minipage}
    \end{subfigure}
    \begin{subfigure}{1\textwidth}
        \begin{minipage}{0.65\linewidth}
            \includegraphics[width=\linewidth]{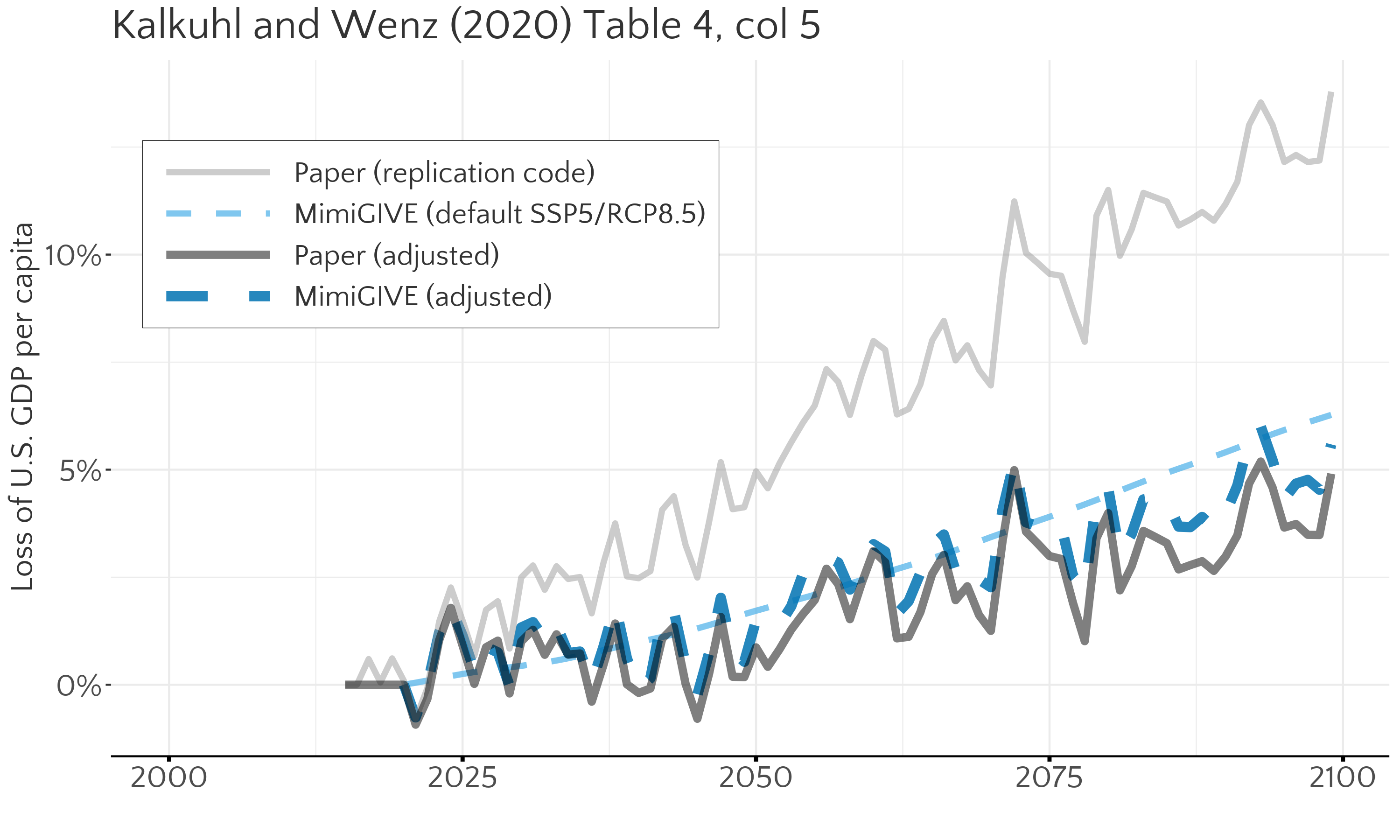}
        \end{minipage}
        \begin{minipage}{0.35\linewidth}    
            \subcaption{\citet{kalkuhl2020impact}}
            \label{fig:val_kalkuhl}
            \footnotesize
            \MimiGIVE{} estimate of U.S. damages was compared with an output-weighted sum of state-level damages from the paper.

            \bigskip
            \textbf{Paper (adjusted)}: 
            
            Drops damages prior to 2020 from replication code and edits the damage computations to be consistent with the econometric lags and empirical estimates presented in the paper text.
            
            \bigskip
            \textbf{\MimiGIVE{} (adjusted)}:
            
            Substitutes population-weighted sum of state-level temperature projections from the paper into \MimiGIVE. (The adjusted lines do not exactly coincide because paper damages for the U.S. are an output-weighted sum of nonlinear damages estimated at the state level, whereas \MimiGIVE{} damages are estimated at the country level.)
            \hfill
        \end{minipage}
    \end{subfigure}
    \begin{subfigure}{1\textwidth}
        \begin{minipage}{0.65\linewidth}
            \includegraphics[width=\linewidth]{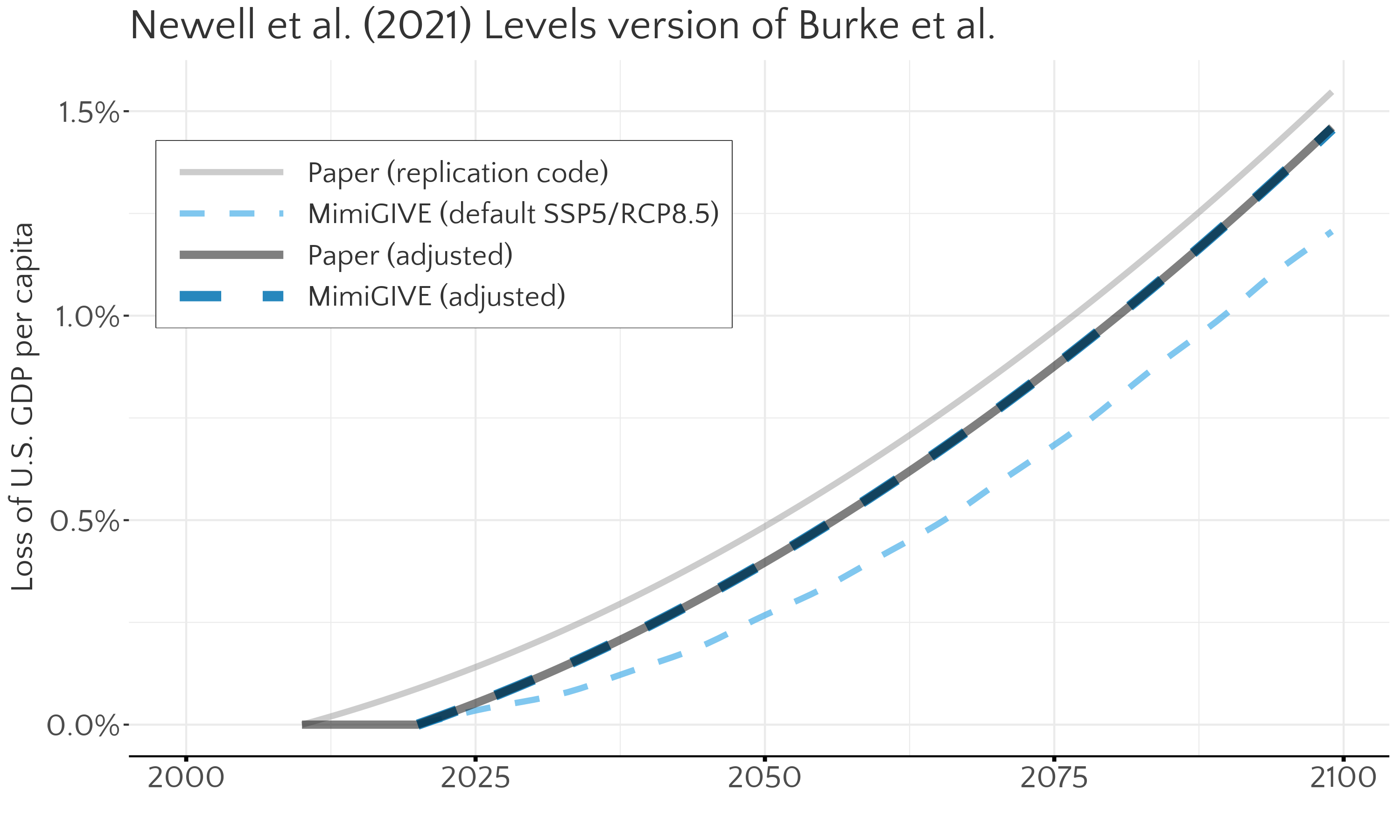}
        \end{minipage}
        \begin{minipage}{0.35\linewidth}    
            \subcaption{\citet{newell2021the}}
            \label{fig:val_newell}
            \footnotesize
            Damage estimation is a variant of the \citet{burke2015global} implementation. 

            \bigskip
            \textbf{Paper (adjusted)}: 
            
            Same as in \citet{burke2015global} above.
            
            \bigskip
            \textbf{\MimiGIVE{} (adjusted)}:
            
            Same as in \citet{burke2015global} above.
            \hfill
        \end{minipage}
    \end{subfigure}
\end{figure}%
\begin{figure}[ht]\ContinuedFloat
    \centering
    \caption[]{Illustration of validation procedure for \MimiGIVE{} damage function implementations, continued...}
    \begin{subfigure}{1\textwidth}
        \begin{minipage}{0.65\linewidth}
            \includegraphics[width=\linewidth]{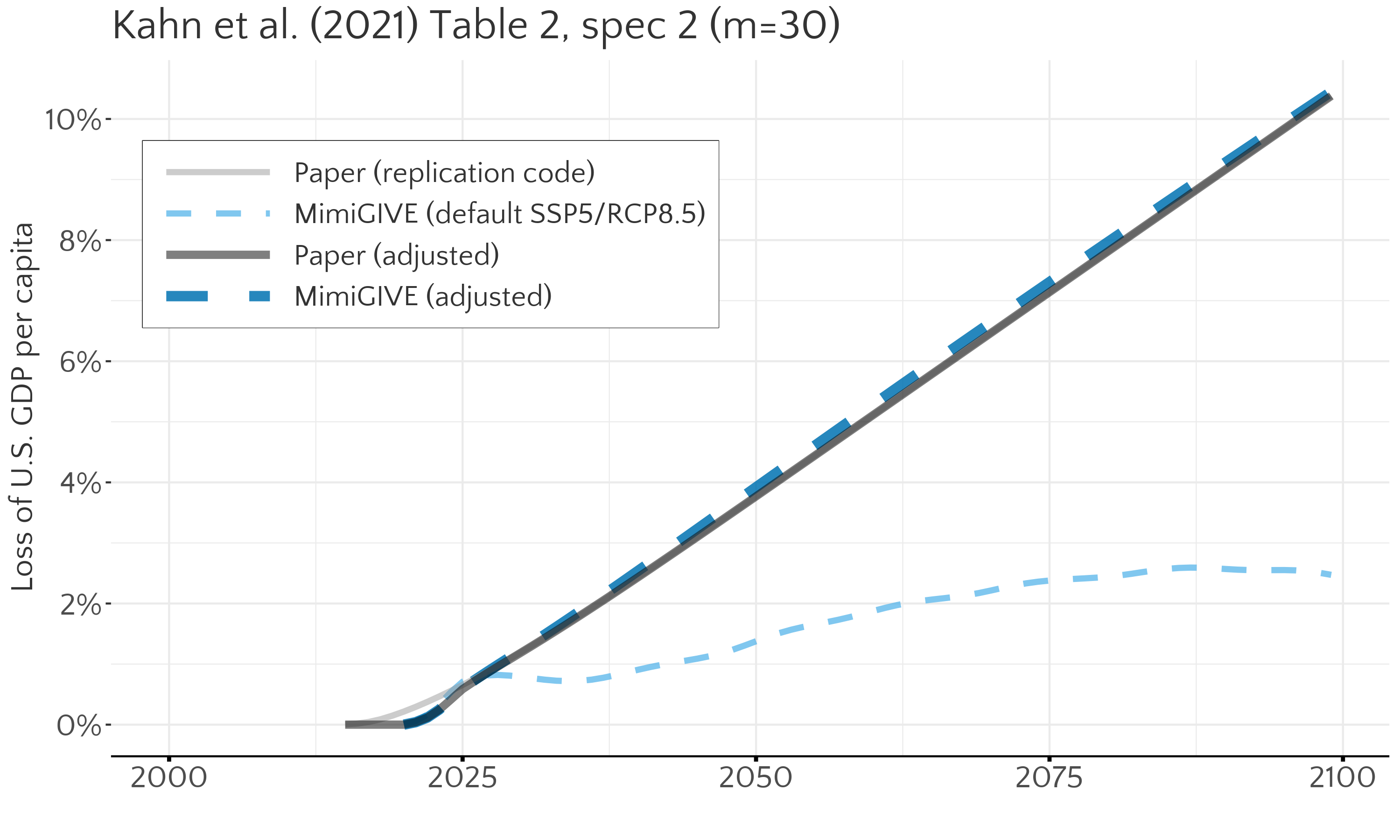}
        \end{minipage}
        \begin{minipage}{0.35\linewidth}    
            \subcaption{\citet{kahn2021long}}
            \label{fig:val_kahn}
            \footnotesize
            Damage estimation uses a recursive computation of long-run damages combined with parametric assumptions on expected future variations in temperature. 

            \bigskip
            \textbf{Paper (adjusted)}: 
            
            Drops damages prior to 2020 from replication code.
            
            \bigskip
            \textbf{\MimiGIVE{} (adjusted)}:
            
            Substitutes paper climate assumptions into MimGIVE. (The paper's parametric assumptions imply a high degree of U.S. warming relative to publicly available country-level projections of future temperatures.)
            \hfill
        \end{minipage}
    \end{subfigure}
    \begin{subfigure}{1\textwidth}
        \begin{minipage}{0.65\linewidth}
            \includegraphics[width=\linewidth]{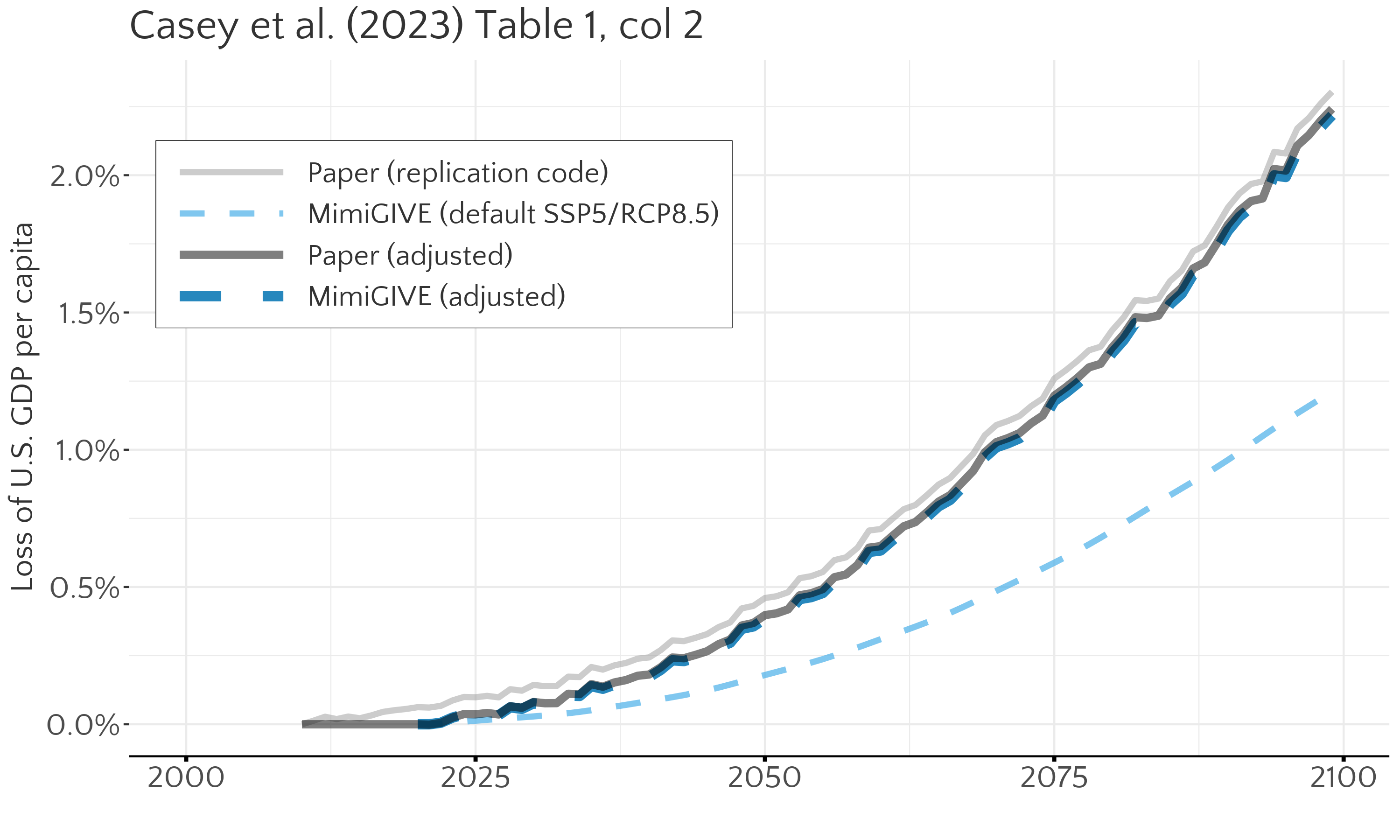}
        \end{minipage}
        \begin{minipage}{0.35\linewidth}    
            \subcaption{\citet{casey2023projecting}}
            \label{fig:val_casey}
            \footnotesize
            Temperature affects TFP and GDP per capita losses are derived from a Solow-model-based computation.  

            \bigskip
            \textbf{Paper (adjusted)}: 
            
            Drops damages prior to 2020 from replication code.
            
            \bigskip
            \textbf{\MimiGIVE{} (adjusted)}:
            
            Substitutes paper's inputs into \MimiGIVE, including the paper's reduced-form socioeconomic projections, baseline capital stocks, temperatures, and projected future temperatures.
            \hfill
        \end{minipage}
    \end{subfigure}
    \begin{subfigure}{1\textwidth}
        \begin{minipage}{0.65\linewidth}
            \includegraphics[width=\linewidth]{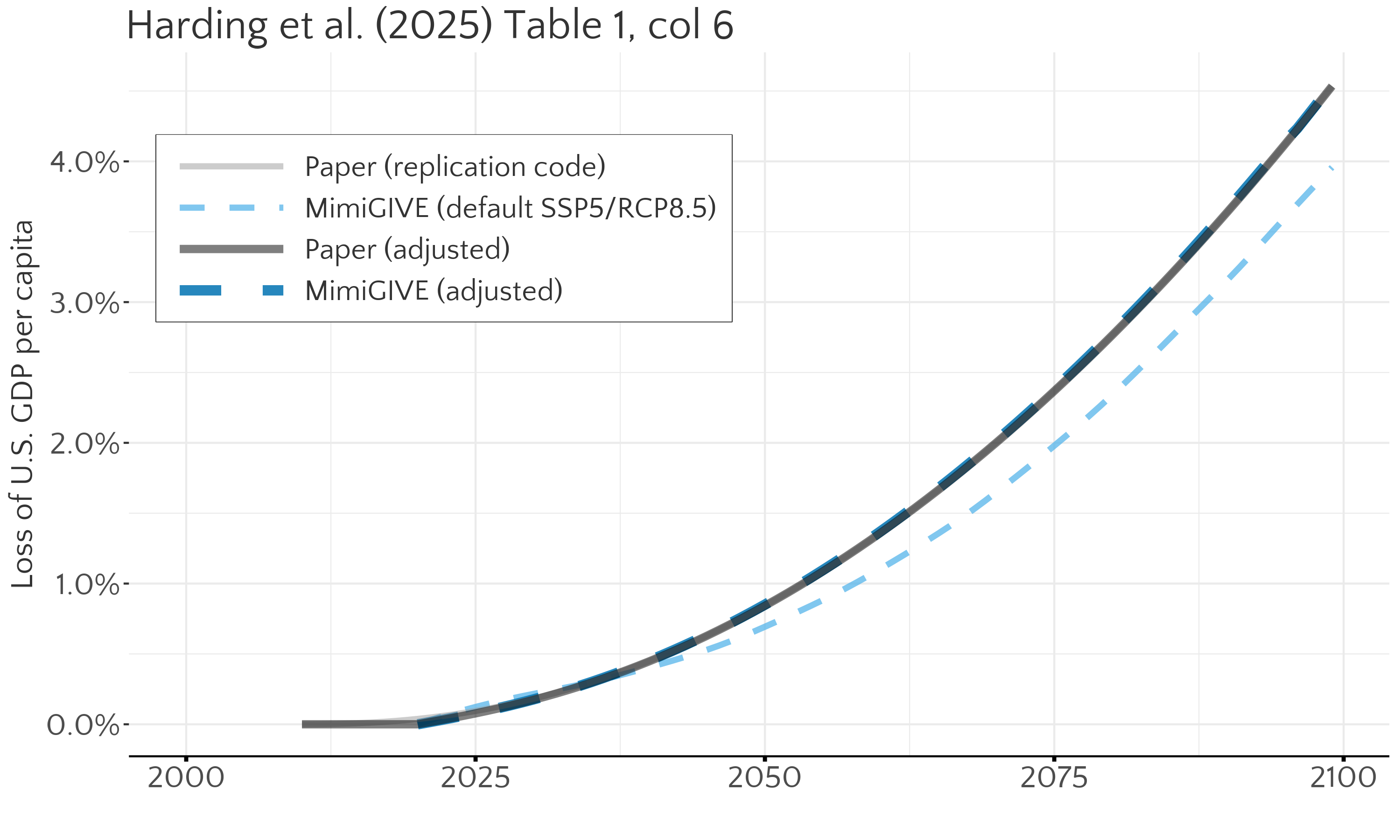}
        \end{minipage}
        \begin{minipage}{0.35\linewidth}    
            \subcaption{\citet{harding2025distribution}}
            \label{fig:val_harding}
            \footnotesize
            Damage estimation is a variant of the \citet{burke2015global} implementation, modified with a convergence term. 

            \bigskip
            \textbf{Paper (adjusted)}: 
            
            Drops damages prior to 2020 from replication code.
            
            \bigskip
            \textbf{\MimiGIVE{} (adjusted)}:
            
            Substitutes paper's temperature and precipitation projections into \MimiGIVE.
            \hfill
        \end{minipage}
    \end{subfigure}
\end{figure}%
\begin{figure}[ht]\ContinuedFloat
    \centering
    \caption[]{Illustration of validation procedure for \MimiGIVE{} damage function implementations, continued...}
    \begin{subfigure}{1\textwidth}
        \begin{minipage}{0.65\linewidth}
            \includegraphics[width=\linewidth]{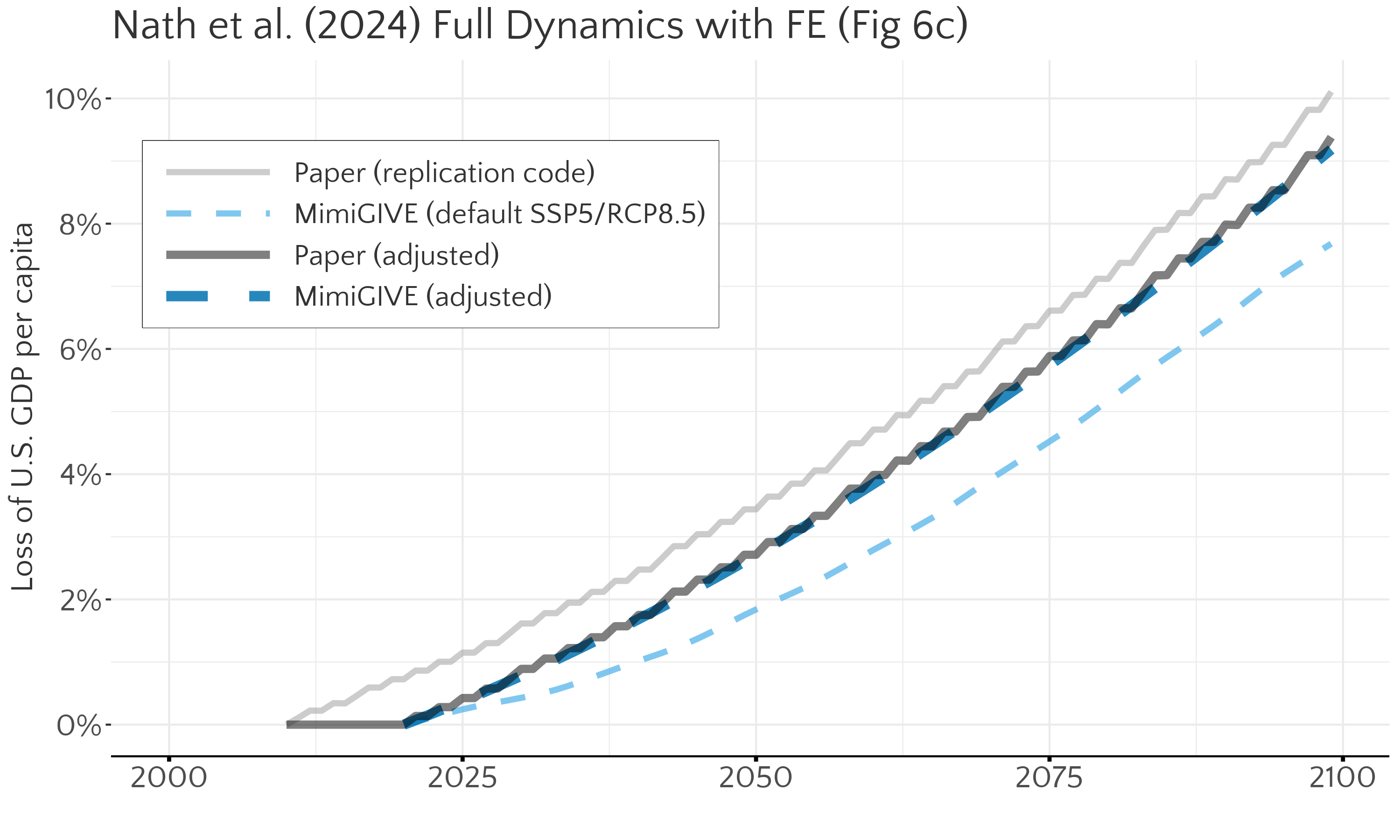}
        \end{minipage}
        \begin{minipage}{0.35\linewidth}    
            \subcaption{\citet{nath2024}}
            \label{fig:val_nath}
            \footnotesize
            \MimiGIVE{} implementation uses a smooth approximation of their state-dependent step function.

            \bigskip
            \textbf{Paper (adjusted)}: 
            
            Drops damages prior to 2020 from replication code.
            
            \bigskip
            \textbf{\MimiGIVE{} (adjusted)}:
            
            Substitutes paper's country-level projected temperature paths into \MimiGIVE.
            \hfill
        \end{minipage}
    \end{subfigure}
\end{figure}%

To summarize, we validated our \MimiGIVE{} implementation of the damage functions from seven macroeconomic econometric studies by comparing U.S. GDP per capita loss projections based on our constructed damage function under an RCP8.5-consistent scenario with the projections generated using the replication code for each paper. We showed that when year coverage, exogenous model inputs, and model assumptions are harmonized, the \MimiGIVE{} implementations yield similar results to the paper replication codes. Table \ref{tab:us_gdppc_loss} summarizes the projected end-of-century U.S. GDP per capita losses from the validation exercise. The first column shows U.S. GDP per capita loss in 2099 drawn directly from unadjusted replication codes, the second column shows paper replication code damages including our adjustments to the replication codes (described above), and the third column shows damages from our \MimiGIVE{} implementations under climate assumptions used in the papers. Thus, the comparison relevant for assessing error in the \MimiGIVE{} implementations is between the second and third columns of the table.

\begin{table}[b]
    \caption{U.S. GDP per capita lost in 2099 under RCP 8.5 (relative to no climate change), from paper replication codes and \MimiGIVE{} results}
    \label{tab:us_gdppc_loss}
    \centering
    \begin{threeparttable}
        \centering
        \begin{tabular}{lccc}
            \toprule \toprule
            \textbf{Paper and} & \textbf{Paper} & \textbf{Paper} & \textbf{\MimiGIVE} \\
            \textbf{Specification} & \textbf{(Replication code)} & \textbf{(Adjusted)} & \textbf{(Adjusted)} \\
            \midrule	
\citet{burke2015global} & 36.28\% & 36.14\% & 36.02\% \\
Table 1, col 1 & & & \\
\citet{kalkuhl2020impact} & 13.79\% & 4.91\% & 5.59\% \\
Table 4, col 5 & & & \\
\citet{newell2021the} & 1.55\% & 1.46\% & 1.46\% \\
Levels version of Burke et al. (2015) & & & \\
\citet{kahn2021long} & 10.38\% & 10.38\% & 10.45\% \\
Table 2, spec 2 (m=30) & & & \\
\citet{casey2023projecting} & 2.30\% & 2.24\% & 2.22\% \\
Table 1, col 2 & & & \\
\citet{harding2025distribution} & 4.53\% & 4.53\% & 4.55\% \\
Table 1, col 6 & & & \\
\citet{nath2024} & 10.11\% & 9.38\% & 9.18\% \\
Full dynamics with FE (Fig 6c) & & & \\
\bottomrule
        \end{tabular}
        \begin{tablenotes} 		
            \footnotesize
            \item This table shows validation results for each study's stated ``preferred'', ``main'', or ``central'' specification, or the specification that is used for climate damage projections in the paper if no preferred specification is stated. All GDP losses are relative to no climate change (i.e. no temperature anomalies above current climate) except \citet{kahn2021long}, which is relative to a continuation of their country-level estimated historic warming trend from 1960 to 2014. U.S. GDP per capita losses in the table exclude Puerto Rico.
        \end{tablenotes}
	\end{threeparttable}
\end{table}

\newpage
\clearpage
\singlespacing
\bibliographystyle{jaere}
\bibliography{references-app}